%% file: HiggsDecay.tex
\documentclass[12pt]{article}
\usepackage{amsmath,amssymb,amsfonts}
\usepackage{color,graphicx,cite,soul}
\usepackage{caption}

\input paperdef

\graphicspath{{figs/}}

\evensidemargin \oddsidemargin
\allowdisplaybreaks
\begin{document}
\thispagestyle{empty}

\def\thefootnote{\fnsymbol{footnote}}

\begin{flushright}
\mbox{}
%CERN--PH--TH/2011--291\\
%FR--PHENO--2011--021\\
%arXiv:yymm.nnnn [hep-ph]
\end{flushright}

\vspace{0.5cm}

\begin{center}

{\large\sc {\bf Heavy Higgs Decays into Sfermions in the Complex MSSM:}}

\vspace{0.4cm}

{\large\sc {\bf A Full One-Loop Analysis}}

\vspace{1cm}

{\sc
S.~Heinemeyer$^{1}$%
\footnote{email: Sven.Heinemeyer@cern.ch}%
~and C.~Schappacher$^{2}$%
\footnote{email: schappacher@kabelbw.de}%
\footnote{former address}%
}

\vspace*{.7cm}

{\sl
$^1$Instituto de F\'isica de Cantabria (CSIC-UC), Santander,  Spain

\vspace*{0.1cm}

$^2$Institut f\"ur Theoretische Physik,
Karlsruhe Institute of Technology, \\
D--76128 Karlsruhe, Germany

}

\end{center}

\vspace*{0.1cm}

\begin{abstract}
\noindent
For the search for additional Higgs bosons in the Minimal
Supersymmetric Standard Model (MSSM) as well as for future precision
analyses in the Higgs sector a precise knowledge of their decay
properties is mandatory.
We evaluate all two-body decay modes of the heavy Higgs bosons into 
sfermions in the MSSM with complex parameters (cMSSM). The evaluation is 
based on a full one-loop calculation of all decay channels, also including 
hard QED and QCD radiation.  The dependence of the heavy Higgs bosons 
on the relevant cMSSM parameters is analyzed numerically.  We find sizable 
contributions to many partial decay widths.  They are roughly of 
\order{15\%} of the tree-level results, but can go up to $30\%$ 
or higher. The size of the electroweak one-loop corrections can be 
as large as the QCD corrections.
The full one-loop contributions are important for the correct
interpretation of heavy  
Higgs boson search results at the LHC and, if kinematically allowed, at
a future linear $e^+e^-$ collider. The 
evaluation of the branching ratios of the heavy Higgs bosons will be 
implemented into the Fortran code {\tt FeynHiggs}.
\end{abstract}
%\pacs{}

\def\thefootnote{\arabic{footnote}}
\setcounter{page}{0}
\setcounter{footnote}{0}

\newpage

%%%%%%%%%%%%%%%%%%%%%%%%%%%%%%%%%%%%%%%%%%%%%%%%%%%%%%%%%%%%%%%%%%%%%%%%%%%%%%%
%%%%%%%%%%%%%%%%%%%%%%%%%%%%%%%%%%%%%%%%%%%%%%%%%%%%%%%%%%%%%%%%%%%%%%%%%%%%%%%

\section{Introduction}
\label{sec:intro}

One of the most important tasks at the LHC is to search for physics effects 
beyond the Standard Model (SM), where the Minimal Supersymmetric Standard 
Model (MSSM)~\cite{mssm,HaK85,GuH86} is one of the leading candidates. 
Supersymmetry (SUSY) predicts two scalar partners for all SM fermions as well
as fermionic partners to all SM bosons.
Another important task is investigating the mechanism of electroweak
symmetry breaking. 
The most frequently investigated
models are the Higgs mechanism within the SM and within the MSSM.
Contrary to the case of the SM, in the MSSM 
two Higgs doublets are required.
This results in five physical Higgs bosons instead of the single Higgs
boson in the SM; three neutral Higgs bosons, $h_n$ ($n = 1,2,3$), 
and two charged Higgs bosons, $H^\pm$. 
The Higgs sector is described at the tree-level by two parameters:
the mass of the charged Higgs boson, $\MHp$, and the ratio of the two
vacuum expectation values, $\tb \equiv \TB = v_2/v_1$.
Often the lightest Higgs boson, $h_1$ is identified with the particle 
discovered at the LHC~\cite{ATLASdiscovery,CMSdiscovery} with a mass 
around $\sim 125 \gev$.
If the mass of the charged Higgs boson is assumed to be larger than 
$\sim 200 \gev$ the four additional Higgs bosons are roughly mass
degenerate, $\MHp \approx \mh2 \approx \mh3$ and referred to as the
``heavy Higgs bosons''. 
Discovering one or more of those additional Higgs bosons would be an
unambiguous sign of physics beyond the SM and could yield important 
information about their supersymmetric origin.

If SUSY is realized in nature and the charged Higgs-boson mass is 
$\MHp \lesssim 1.5 \tev$, then the heavy Higgs 
bosons could be detectable at the LHC (including its high luminosity
upgrade, HL-LHC) and/or at a future linear $e^+e^-$ collider such as the
ILC~\cite{ILC-TDR,teslatdr,ilc} or CLIC~\cite{CLIC}. (Results on the
combination of LHC and LC results can be found in \citere{lhcilc}.)
The discovery potential at the HL-LHC goes up to \order{1 \tev} for
large $\tb$ values and somewhat lower at low $\tb$ values. 
At an $e^+e^-$ linear collider the heavy Higgs
bosons are pair produced, and the reach is limited by the center of mass
energy, $\MHp \lesssim \sqrt{s}/2$, roughly independent of $\tb$.
Details about the discovery process(es) depend strongly
on the cMSSM parameters (and will not be further discussed in this paper). 

In the case of a discovery of additional Higgs bosons a subsequent
precision determination of their properties will be crucial determine
their nature and the underlying (SUSY) parameters. 
In order to yield a sufficient accuracy, one-loop corrections to
the various Higgs-boson decay modes have to be considered.
Decays to SM fermions have been evaluated at
the full one-loop level in the cMSSM in \citere{hff}, see also
\citeres{hff0} as well 
as \citeres{deltab,db2l} for higher-order SUSY corrections.
Decays to (lighter) Higgs bosons have been evaluated at the full
one-loop level in the cMSSM in \citere{hff}, see also \citere{hAA}.
Decays to SM gauge bosons can evaluated to a very high precision using
the full SM one-loop result~\cite{prophecy4f} combined with the
appropriate effective couplings~\cite{mhcMSSMlong}.
The full one-loop corrections in the cMSSM listed here together with
resummed SUSY corrections have been implemented into the code 
{\tt FeynHiggs}~\cite{feynhiggs,mhiggslong,mhiggsAEC,mhcMSSMlong,Mh-logresum}.
Corrections at and beyond the one-loop level in the MSSM with real
parameters (rMSSM) are implemented into the code 
{\tt Hdecay}~\cite{hdecay,hdecay2}.
Both codes were combined by the LHC Higgs Cross Section Working Group to
obtain the most precise evaluation for rMSSM Higgs boson decays to SM
particles and decays to lighter Higgs bosons~\cite{YR3}.

The heavy MSSM Higgs bosons can also decay to SUSY particles, \ie to
charginos, neutralinos and scalar fermions.
In \citere{benchmark4} it was demonstrated that the SUSY particle modes
can dominate the decay of the heavy Higgs bosons.
In this work we calculate all two-body decay modes of the heavy Higgs
bosons to scalar fermions in the cMSSM.%
\footnote{
  We neglect flavor violation effects and resulting decay channels.
}
More specifically, we calculate the full one-loop corrections to the 
partial decay widths
\begin{align}
\label{eq:hnsfsf}
\Ga(h_n \to \Sf_i^{} \Sf_j^\dagger) \qquad (n = 2,3;\, i,j = 1,2)\,,
\end{align}
\begin{align}
\label{eq:Hsfsf}
\Ga(\Hpdecay) \quad \text{and} \quad \Ga(\Hmdecay) \qquad (i,j = 1,2)\,, 
\end{align}
where $H^\pm$ denotes the charged, $h_n$ the mixed neutral Higgs bosons
and $\tilde{f}$ $(\tilde{f^\dagger})$ denotes the scalar (anti-) fermions.%
\footnote{
  In the text and figures below we omit the $\mbox{}^\dagger$ 
  (denoting anti-particles) for simplification.
}
The total decay width is defined as the sum of the partial decay
widths (\ref{eq:hnsfsf}) or (\ref{eq:Hsfsf}), the SM decay channels as
described above and the decays to charginos/neutralinos (at the
tree-level, supplemented with effective couplings~\cite{mhcMSSMlong}).

The evaluation of the channels \refeqs{eq:hnsfsf}, (\ref{eq:Hsfsf}) is
based on a full one-loop calculation, \ie including (S)QCD and
electroweak (EW) corrections, as well as soft and hard QCD and QED
radiation. For ``mixed'' decay modes, we evaluate in addition the 
two ``$\CP$-versions'' $(i \ne j)$ of \refeq{eq:hnsfsf} and the two 
`$\CP$-versions'' of \refeq{eq:Hsfsf}, which give different results for 
non-zero complex phases.
While our calculation comprises the decay to {\em all} sfermionic decay
modes of the cMSSM Higgs bosons, in our numerical analysis we will focus
on the decay to the third generation sfermions, scalar top and bottom 
quarks, scalar tau and tau neutrinos.

Higher-order contributions to MSSM Higgs decays to scalar fermions have
been evaluated in various analyses over the last decade. However, they
were in most cases restricted to few specific channels. 
In many cases only parts of a
one-loop calculation has been performed, and no higher-order corrections
in the cMSSM are available so far. 
More specifically, the available literature comprises the following. 
First, \order{\als} corrections to partial decay widths of various squark 
decay channels in the rMSSM were derived: 
to the decay of a charged Higgs to stops and sbottoms in
\citere{Hpstsb_als}, of a heavy Higgs boson to third generation squarks
in \citere{Phisqsq_als}, supplemented later by an effective resummation
of the trilinear Higgs-sbottom coupling in \citere{Phisqsq_als_1}.
First full one-loop corrections in the rMSSM were calculated in the
decays of the $\CP$-odd Higgs boson to scalar quarks~\cite{Asqsq_1L} and
to scalar fermions~\cite{Asfsf_1L}. The full one-loop corrections in the
rMSSM to Higgs decays to squarks was published in
\citere{Phisqsq_1L}. While there results constitute a full one-loop
correction (although not for complex parameters), it differs from our
calculation in the renormalization of the SUSY particles and parameters.
It was shown in \citeres{SbotRen,Stop2decay,Stau2decay} that our
renormalization leads to stable results over nearly the full cMSSM
parameters space.

The full \order{\als} corrections to Higgs decays to scalar quarks were
also evaluated by a different group in \citere{Phisqsq_als_2}, using a
renormalization more similar to ours, but also restricting to the case
of real parameters. Finally, in \citere{Phisqsq_als_3} the \order{\als}
corrections to Higgs decays to scalar quarks were re-analyzed, where the
emphasis was put on the connection of the MSSM squark sector and Higgs
sector couplings to \MSbar\ input parameters. The latter corrections
in particular differ from our treatment of the renormalization of the
scalar quark sector. They have been included into the code {\tt Hdecay}.

In this paper we present for the first time a full one-loop calculation 
for all two-body sfermionic decay channels of the Higgs bosons in the 
cMSSM (with no generation mixing), taking into account soft and hard QED 
and QCD radiation.  In \refse{sec:renorm} we review the renormalization of 
all relevant sectors of the cMSSM.  Details about the calculation can be 
found in \refse{sec:calc}, and the numerical results for all decay 
channels are presented in \refse{sec:numeval} 
(including comments on comparisons with results from other groups).  
The conclusions can be found in \refse{sec:conclusions}.  The results will 
be implemented into the Fortran code 
{\tt FeynHiggs}~\cite{feynhiggs,mhiggslong,mhiggsAEC,mhcMSSMlong,Mh-logresum}.

%%%%%%%%%%%%%%%%%%%%%%%%%%%%%%%%%%%%%%%%%%%%%%%%%%%%%%%%%%%%%%%%%%%%%%%%%%%%%%%
%%%%%%%%%%%%%%%%%%%%%%%%%%%%%%%%%%%%%%%%%%%%%%%%%%%%%%%%%%%%%%%%%%%%%%%%%%%%%%%

\section{The complex MSSM}
\label{sec:renorm}

The channels (\ref{eq:hnsfsf}) and (\ref{eq:Hsfsf}) are calculated at the
one-loop level, including hard QED and QCD radiation. This requires the
simultaneous renormalization of several sectors of the cMSSM, including
the colored sector with top and bottom quarks and their scalar partners
as well as the gluon and the gluino, the Higgs and gauge boson sector with 
all the Higgs bosons as well as the $Z$ and the $W$~boson and the
chargino/neutralino sector. 
In the following subsections we briefly review these sectors and their 
renormalization.

%%%%%%%%%%%%%%%%%%%%%%%%%%%%%%%%%%%%%%%%%%%%%%%%%%%%%%%%%%%%%%%%%%%%%%%%%%%%%%%

\subsection{The Higgs- and gauge-boson sector}
\label{sec:higgs}

The Higgs- and gauge-boson sector follow strictly \citere{MSSMCT} and 
references therein (see especially \citere{mhcMSSMlong}).
This defines in particular the counterterm $\de\tb \equiv \dTB$,
as well as the counterterms for the $Z$~boson mass, $\de\MZ^2$, and the
sine of the weak mixing angle, $\de\SW$.

%%%%%%%%%%%%%%%%%%%%%%%%%%%%%%%%%%%%%%%%%%%%%%%%%%%%%%%%%%%%%%%%%%%%%%%%%%%%%%%

\subsection{The chargino/neutralino sector}
\label{sec:chaneu}

The chargino/neutralino sector is also described in detail in 
\citere{MSSMCT} and references therein.  In this paper we use the so 
called CCN scheme, \ie on-shell conditions for two charginos 
and one neutralino, which we chose to be the lightest one. In the
notation of \citere{MSSMCT} we used:

\Code{\$InoScheme = CCN[1]} --- fixed CCN scheme with on-shell $\neu1$.

\noindent
This defines in particular the counterterm $\de\mu$, where $\mu$
denotes the Higgs mixing parameter.

%%%%%%%%%%%%%%%%%%%%%%%%%%%%%%%%%%%%%%%%%%%%%%%%%%%%%%%%%%%%%%%%%%%%%%%%%%%%%%%

\subsection{The fermion sector}
\label{sec:fermion}

The fermion sector is described in detail in \citere{MSSMCT} and references 
therein.  For simplification we use her the \DRbar\ renormalization for all 
three generations of down-type quarks \textit{and} leptons, again in the 
notation of \citere{MSSMCT}:

\Code{UVMf1[4,\,\uscore]}~ \Code{= UVDivergentPart} \qquad
 \text{\DRbar\ renormalization for $m_d$, $m_s$, $m_b$}

\Code{UVMf1[2,\,\uscore]}~ \Code{= UVDivergentPart} \qquad
 \text{\DRbar\ renormalization for $m_e$, $m_\mu$, $m_\tau$}

%%%%%%%%%%%%%%%%%%%%%%%%%%%%%%%%%%%%%%%%%%%%%%%%%%%%%%%%%%%%%%%%%%%%%%%%%%%%%%%

\subsection{The scalar fermion sector}
\label{sec:sfermion}

The sfermion sector which we use here differ slightly form the one 
described in \citere{MSSMCT}.  For the squark sector we follow 
\citeres{SbotRen,Stop2decay} and for the slepton sector we created an 
additional \DRbar\ type version in full analogy to the squark sector.  
In the following we list all these formulas we used in this analysis.

In the absence of non-minimal flavor violation, the sfermion mass matrix
is given by \cite{HaK85,GuH86}
\begin{equation}
\label{eq:Sfermmassmatrix}
\matr{M}^2_{\Sf_{tg}} = \begin{pmatrix}
  \bigl(\matr{M}^2_{L,f_t}\bigr)_{gg} + \mf{tg}^2
	& \mf{tg} \bigl(\matr{X}_{f_t}\bigr)_{gg}^* \\
  \mf{tg} \bigl(\matr{X}_{f_t}\bigr)_{gg}
	& \bigl(\matr{M}^2_{R,f_t}\bigr)_{gg} + \mf{tg}^2
\end{pmatrix}
\end{equation}
where
\begin{align*}
\matr{M}^2_{L,f_t} &= \MZ^2 (I_3^{f_t} - Q_{f_t}\SW^2) \CBB + 
\begin{cases}
\matr{M}^2_{\tilde L}
	& \text{for left-handed sleptons ($t = 1, 2$)}\,, \\
\matr{M}^2_{\tilde Q}
	& \text{for left-handed squarks ($t = 3, 4$)}\,,
\end{cases} \\
\matr{M}^2_{R,f_t} &= \MZ^2 Q_{f_t}\SW^2 \CBB + 
\begin{cases}
\matr{M}^2_{\tilde E}
	& \text{for right-handed sleptons ($t = 2$)}\,, \\
\matr{M}^2_{\tilde U}
	& \text{for right-handed $u$-type squarks ($t = 3$)}\,, \\
\matr{M}^2_{\tilde D}
	& \text{for right-handed $d$-type squarks ($t = 4$)}\,, \\
\end{cases} \\
\matr{X}_{f_t} &= \matr{A}_{f_t} - \mu^*
\begin{cases}
1/\TB & \text{for isospin-up sfermions ($t = 3$)}\,, \\
\TB & \text{for isospin-down sfermions ($t = 2, 4$)}\,.
\end{cases}
\end{align*}
The soft-SUSY-breaking parameters $\matr{M}^2_{\tilde L,\tilde Q,
\tilde E,\tilde U,\tilde D}$ and $\matr{A}_f$ are $3\times 3$ matrices 
in flavor space whose off-diagonal entries are zero in the minimally
flavor-violating MSSM. 
$Q_f$ and $I_3^f$ denote the charge and the weak iso-spin of the
corresponding fermion, and $\CBB \equiv \cos 2\beta$.

The mass matrix is diagonalized by a unitary transformation
${\matr{U}}_\Sf$,
\begin{align}
\matr{U}_{\!\Sf}^{}\, \matr{M}^2_\Sf\, \matr{U}_{\!\Sf}^\dagger =
\begin{pmatrix}
  \msf1^2 & 0 \\
  0 & \msf2^2
\end{pmatrix}, \qquad
\matr{U}_{\!\Sf} = \begin{pmatrix}
  U^\Sf_{11} & U^\Sf_{12} \\
  U^\Sf_{21} & U^\Sf_{22}
\end{pmatrix}.
\end{align}

We renormalize the up-type squarks 
($\Su_{\{g=1,2,3\}} = \{\tilde u, \tilde c, \tilde t\}$) 
and the sneutrinos
($\Sn_{\{g=1,2,3\}} = \{\tilde \nu_e, \tilde \nu_\mu, \tilde \nu_\tau\}$) 
on-shell (OS).  For the down-type squarks 
($\Sd_{\{g=1,2,3\}} = \{\tilde d, \tilde s, \tilde b\}$) 
and the electron-type sleptons
($\Se_{\{g=1,2,3\}} = \{\tilde e, \tilde\mu, \tilde\tau\}$)
we follow the discussion in Sect.~4 
(option $\mathcal{O}2$) of \citere{SbotRen} and renormalize them 
on-shell.  They then have to be computed from a mass matrix with 
shifted $M_{\tilde L}^2, M_{\tilde Q}^2, M_{\tilde E}^2$ and $M_{\tilde D}^2$, 
see below.  

We apply the ``$m_b, A_b$ \DRbar'' scheme of \citeres{SbotRen,Stop2decay}.
The scheme affecting sfermions $\Se_g$, $\Sd_g$ is chosen with the variable 
\Code{\$SfScheme[\Vt,\,\Vg]}:
\begin{subequations}
\begin{alignat}{2}
\Code{\$SfScheme[2,\,\Vg]}~ &\Code{= DR[2]} &\qquad
	& \text{mixed scheme with $\mse{2g}$ OS, $A_{\Fe_g}$ \DRbar}\,, \\
\Code{\$SfScheme[4,\,\Vg]}~ &\Code{= DR[2]} &\qquad
	& \text{mixed scheme with $\msd{2g}$ OS, $A_{\Fd_g}$ \DRbar}\,.
\end{alignat}
\end{subequations}

In the following, the sfermion index \Vs\ runs over both values 1, 2.
All sfermions are on-shell,
\begin{subequations}
\begin{alignat}{2}
\Code{dMSfsq1[1,\,1,\,1,\,\Vg]}     &\equiv {}& \delta\msn{1g}^2
	&= \ReTilde\mati{\SE{\Sn_g}(\msn{1g}^2)}_{11}\,, \\
\Code{dMSfsq1[\Vs,\,\Vs,\,2,\,\Vg]} &\equiv {}& \delta\mse{sg}^2
	&= \ReTilde\mati{\SE{\Se_g}(\mse{sg}^2)}_{ss}\,, \\
\Code{dMSfsq1[\Vs,\,\Vs,\,3,\,\Vg]} &\equiv {}& \delta\msu{sg}^2
	&= \ReTilde\mati{\SE{\Su_g}(\msu{sg}^2)}_{ss}\,, \\
\Code{dMSfsq1[\Vs,\,\Vs,\,4,\,\Vg]} &\equiv {}& \delta\msd{sg}^2
	&= \ReTilde\mati{\SE{\Sd_g}(\msd{sg}^2)}_{ss}\,.
\end{alignat}
\end{subequations}
The up-type off-diagonal mass-matrix entries receive counterterms 
\cite{mhcMSSM2L,dissHR,SbotRen}
\begin{subequations}
\begin{alignat}{2}
\Code{dMSfsq1[1,\,2,\,3,\,\Vg]} &\equiv {}& \delta Y_{\Fu_g} 
	&= \frac 12\ReTilde\mati{\SE{\Su_g}(\msu{1g}^2) +
	                         \SE{\Su_g}(\msu{2g}^2)}_{12}\,, \\
\Code{dMSfsq1[2,\,1,\,3,\,\Vg]} &\equiv {}& \delta Y_{\Fu_g}^*
	&= \frac 12\ReTilde\mati{\SE{\Su_g}(\msu{1g}^2) +
	                         \SE{\Su_g}(\msu{2g}^2)}_{21}\,.
\end{alignat}
\end{subequations}
For clarity of notation we furthermore define the auxiliary constants
\begin{align}
\Code{dMsq12Sf1[2,\,\Vg]} \equiv \delta M_{\Se_g,12}^2
	&= \mfe{g} (\delta A_{\Fe_g}^* - \mu\,\dTB - \TB\,\delta\mu) +
	   (A_{\Fe_g}^* - \mu\,\TB)\,\delta\mfe{g}\,, \\
\Code{dMsq12Sf1[4,\,\Vg]} \equiv \delta M_{\Sd_g,12}^2
	&= \mfd{g} (\delta A_{\Fd_g}^* - \mu\,\dTB - \TB\,\delta\mu) +
	   (A_{\Fd_g}^* - \mu\,\TB)\,\delta\mfd{g}\,.
\end{align}
The electron/down-type off-diagonal mass counterterms are related as
\begin{subequations}
\begin{align}
\Code{dMSfsq1[1,\,2,\,2,\,\Vg]} &\equiv \delta Y_{\Fe_g}
	= \frac 1{|U^{\Se_g}_{11}|^2 - |U^{\Se_g}_{12}|^2}
	\Bigl\{
	  U^{\Se_g}_{11} U^{\Se_g*}_{21}
	  \bigl(\delta\mse{1g}^2 - \delta\mse{2g}^2\bigr) + {} \\[-1ex]
&\kern 12em
	  U^{\Se_g}_{11} U^{\Se_g*}_{22} \delta M_{\Se_g,12}^2 -
	  U^{\Se_g}_{12} U^{\Se_g*}_{21} \delta M_{\Se_g,12}^{2*}
	\Bigr\}\,, \notag \\
\Code{dMSfsq1[2,\,2,\,2,\,\Vg]} &= \delta Y_{\Fe_g}^*\,, \\
\Code{dMSfsq1[1,\,2,\,4,\,\Vg]} &\equiv \delta Y_{\Fd_g}
	= \frac 1{|U^{\Sd_g}_{11}|^2 - |U^{\Sd_g}_{12}|^2}
	\Bigl\{
	  U^{\Sd_g}_{11} U^{\Sd_g*}_{21}
	  \bigl(\delta\msd{1g}^2 - \delta\msd{2g}^2\bigr) + {} \\[-1ex]
&\kern 12em
	  U^{\Sd_g}_{11} U^{\Sd_g*}_{22} \delta M_{\Sd_g,12}^2 -
	  U^{\Sd_g}_{12} U^{\Sd_g*}_{21} \delta M_{\Sd_g,12}^{2*}
	\Bigr\}\,, \notag \\
\Code{dMSfsq1[2,\,1,\,4,\,\Vg]} &= \delta Y_{\Fd_g}^*\,.
\end{align}
\end{subequations}
The trilinear couplings $A_{f_{tg}}\equiv\mati{\matr{A}_{f_t}}_{gg}$ 
are renormalized by
\begin{subequations}
\begin{alignat}{2}
\Code{dAf1[2,\,\Vg,\,\Vg]} &\equiv \delta A_{\Fe_g} &
	&= \biggl\{\frac 1{\mfe{g}}\Bigl[
	\begin{aligned}[t]
	& U^{\Se_g}_{11} U^{\Se_g*}_{12}
	  (\delta\mse{1g}^2 - \delta\mse{2g}^2) + {} \\
	& U^{\Se_g}_{11} U^{\Se_g*}_{22} \delta Y_{\Fe_g}^* +
	  U^{\Se_g*}_{12} U^{\Se_g}_{21} \delta Y_{\Fe_g} -
	  (A_{\Fe_g} - \mu^*\TB)\,\delta\mfe{g}
	\rlap{$\Bigr] + {}$}
	\end{aligned} \\
&&	&\qquad \delta\mu^*\TB + \mu^*\dTB
        \biggr\}_{\mathrm{div}}\,, \notag \\[1ex]
\Code{dAf1[3,\,\Vg,\,\Vg]} &\equiv \delta A_{\Fu_g} &
	&= \frac 1{\mfu{g}} \Bigl[
	\begin{aligned}[t]
	& U^{\Su_g}_{11} U^{\Su_g*}_{12}
	    (\delta\msu{1g}^2 - \delta\msu{2g}^2) + {} \\
	& U^{\Su_g}_{11} U^{\Su_g*}_{22} \delta Y_{\Fu_g}^* +
	  U^{\Su_g*}_{12} U^{\Su_g}_{21} \delta Y_{\Fu_g} -
	  \bigl(A_{\Fu_g} - \mu^*/\TB\bigr)\,\delta\mfu{g}
	\rlap{$\Bigr] + {}$}
	\end{aligned} \\
&&	&\qquad \delta\mu^*/\TB - \mu^* \dTB/\TB^2\,, \notag \\[1ex]
\Code{dAf1[4,\,\Vg,\,\Vg]} &\equiv \delta A_{\Fd_g} &
	&= \biggl\{\frac 1{\mfd{g}} \Bigl[
	\begin{aligned}[t]
	& U^{\Sd_g}_{11} U^{\Sd_g*}_{12}
	    (\delta\msd{1g}^2 - \delta\msd{2g}^2) + {} \\
	& U^{\Sd_g}_{11} U^{\Sd_g*}_{22} \delta Y_{\Fd_g}^* +
	  U^{\Sd_g*}_{12} U^{\Sd_g}_{21} \delta Y_{\Fd_g} -
	  \bigl(A_{\Fd_g} - \mu^*\TB\bigr)\,\delta\mfd{g}
	\rlap{$\Bigr] + {}$}
	\end{aligned} \\
&&	&\qquad\, \delta\mu^*\TB + \mu^*\dTB
	\biggr\}_{\mathrm{div}}\,, \notag
\end{alignat}
\end{subequations}
where the subscripted ``div'' means to take the divergent part, 
to effect \DRbar\ renormalization of $A_{\Fe_g}$ and $A_{\Fd_g}$ 
\cite{SbotRen}.

As now all the sfermion masses are renormalized as on-shell an explicit 
restoration of the $SU(2)$~relation is needed. 
This is performed in requiring that the left-handed (bare) soft
SUSY-breaking mass parameter is the same in the $\Sd_g$ as in the 
$\Su_g$ squark sector at the one-loop level
(see also \citeres{squark_q_V_als,stopsbot_phi_als,dr2lA}),
\begin{subequations}
\label{eq:MSfShift}
\begin{align}
M_{\tilde L}^2(\Se_g) &= M_{\tilde L}^2(\Sn_g) + 
                     \de M_{\tilde L}^2(\Sn_g) - 
                     \de M_{\tilde L}^2(\Se_g)\,, \\
M_{\tilde Q}^2(\Sd_g) &= M_{\tilde Q}^2(\Su_g) + 
                     \de M_{\tilde Q}^2(\Su_g) - 
                     \de M_{\tilde Q}^2(\Sd_g)
\end{align}
\end{subequations}
with
\begin{align}
\de M_{\tilde L, \tilde Q}^2(\Sf_g) &= 
         |U_{11}^{\Sf_g}|^2 \de\msf{1g}^2 +
         |U_{12}^{\Sf_g}|^2 \de\msf{2g}^2 -
          U_{22}^{\Sf_g} U_{12}^{\Sf_g*} \de Y_{f_g} -
          U_{12}^{\Sf_g} U_{22}^{\Sf_g*} \de Y_{f_g}^* - 2 \mf{g} \de\mf{g} \notag \\
&\quad  + \MZ^2\, \CBB\, Q_{f_g}, \de \SW^2 
        - (I_3^{f_g} - Q_{f_g} \SW^2) (\CBB\, \de \MZ^2 + \MZ^2\, \de \CBB)\,.
\end{align} 
Now $M_{\tilde L}^2(\Se_g)$ and $M_{\tilde Q}^2(\Sd_g)$ are used in the scalar 
mass matrix instead of the parameters $M_{\tilde L, \tilde Q}^2$ in 
\refeq{eq:Sfermmassmatrix} when calculating the values of $\mse{sg}$ and 
$\msd{sg}$.  However, with this procedure, also $\mse{2g}$ and $\msd{2g}$ 
are shifted, which contradicts our choice of independent parameters. 
To keep this choice, also the right-handed soft SUSY-breaking mass 
parameters $M_{\tilde E, \tilde D}^2$ receive a shift%
\footnote{
  If the mass of the $\Sd_{1g}$ squark is chosen as independent mass as 
  $\Sd_{2g}\approx \Sd_{Lg}$ then the shift of $M_{\tilde D}^2$ has to 
  be performed with respect to $\msd{1g}$.
}:
\begin{subequations}
\label{eq:MSfBackshift}
\begin{align}
M_{\tilde E}^2(\Se_g) &= \frac{\mfe{g}^2\, |X_{\Fe_g}|^2}
  {M_{\tilde L}^2(\Se_g) + \mfe{g}^2 
   + \CBB \MZ^2 (I_3^{\Fe} - Q_{\Fe} \SW^2) - \mse{2g}^2} 
   - \mfe{g}^2 - \CBB \MZ^2 Q_{\Fe} \SW^2 + \mse{2g}^2\,, \\
M_{\tilde D}^2(\Sd_g) &= \frac{\mfd{g}^2\, |X_{\Fd_g}|^2}
  {M_{\tilde Q}^2(\Sd_g) + \mfd{g}^2 
   + \CBB \MZ^2 (I_3^{\Fd} - Q_{\Fd} \SW^2) - \msd{2g}^2} 
   - \mfd{g}^2 - \CBB \MZ^2 Q_{\Fd} \SW^2 + \msd{2g}^2\,.
\end{align}
\end{subequations}
Taking into account this shift in $M_{\tilde E, \tilde D}^2$, 
up to one-loop order%
\footnote{
  In the case of a pure OS scheme 
  (see \eg \cite{hr,mhiggsFDalbals} for the rMSSM) the shifts 
  \refeqs{eq:MSfShift} and (\ref{eq:MSfBackshift}) result in a mass 
  parameter $\msd{1g}$ and $\mse{1g}$, which is exactly the same as 
  in \refeq{eq:msfOS}.  This constitutes an important consistency 
  check of these two different methods.
}%
, the resulting mass parameter $\mse{1g}$ and $\msd{1g}$ are the same as 
the on-shell mass
\begin{subequations}
\label{eq:msfOS}
\begin{align}
\bigl(\mse{sg}^\OS\bigr)^2
&= \mse{sg}^2 + \delta\mse{sg}^2 -
  \ReTilde\mati{\SE{\Se_g}(\mse{sg}^2)}_{ss}\,, \\
\bigl(\msd{sg}^\OS\bigr)^2
&= \msd{sg}^2 + \delta\msd{sg}^2 -
  \ReTilde\mati{\SE{\Sd_g}(\msd{sg}^2)}_{ss}\,.
\end{align}
\end{subequations}

\bigskip

The input parameters in the $b/\sbot$ sector have to correspond to the
chosen renormalization. We start by defining the bottom mass, where the
experimental input is the SM \MSbar\ mass \cite{pdg},
\begin{align}
\mb^{\MSbar}(\mb) & = 4.18 \gev\,.
\end{align}
To convert to the \DRbar\ mass the following procedure is taken.
The value of $\mb^{\MSbar}(\mu_R)$ (at the renormalization scale $\mu_R$) 
is calculated from $\mb^{\MSbar}(\mb)$ at the three-loop level 
\begin{align}
\mb^{\MSbar}(\mu_R) &= \mb^{\MSbar}(\mb) \;
  \frac{c(\als^{\MSbar,(n_f)}(\mu_R)/\pi)}{c(\als^{\MSbar,(n_f)}(\mb)/\pi)} 
\end{align}
via the function $c(x)$ following the prescription given in 
\citere{RunDec}. $n_f$ denotes the number of active flavors.
The ``on-shell'' mass is connected to the \MSbar\ mass via
\begin{align}
\mb^{\os} &= \mb^{\MSbar}(\mu_R) \; 
   \LB 1 + \frac{\als^{\MSbar,(n_f)}(\mu_R)}{\pi} 
   \LP \frac 43 + 2\, \ln \frac{\mu_R}{\mb^{\MSbar}(\mu_R)} \RP 
         + \ldots \RB\,,
\end{align}
where the ellipsis denote the two- and three-loop contributions which 
can also be found in \citere{RunDec}.
The $\DRbar$ bottom quark mass at the scale $\mu_R$ is 
calculated iteratively from \cite{deltab2,dissHR,mhiggsFDalbals}
\begin{align}
\label{eq:mbDR}
\mb^{\DRbar} &= \mb^{\os} + \frac{\de\mb^{\OS} - \de\mb^{\DRbar}}{|1 + \db|}
\end{align}
with an accuracy of 
$|1 - (\mb^{\DRbar})^{(n)}/(\mb^{\DRbar})^{(n-1)}| < 10^{-5}$
reached in the $n$th step of the iteration.

The quantity $\db$~\cite{deltab2,deltab1,deltabc} resums the 
\order{(\als\TB)^n} and \order{(\alt\TB)^n} terms and is given by 
\begin{align}
\db = -\frac{\mu^*\, \TB}{\pi} 
  \LB \frac 23\, \als(\mt)\, M_3^*\,
       I(\msbot1^2, \msbot2^2, \mgl^2) \;
     + \frac 14\, \alt(\mt)\, \At^*\, 
       I(\mstop1^2, \mstop2^2, |\mu|^2) 
  \RB
\end{align}
with
\begin{align}
I(a, b, c) = C_0(0,0,0,a,b,c) =
             -\frac{a b \ln(b/a) + a c \ln(a/c) + b c \ln(c/b)}
                   {(c - a) (c - b) (b - a)}\,.
\end{align}
Here $\alt$ is defined in terms of the top Yukawa coupling 
$y_t(\mt) = \sqrt{2} \mt(\mt)/v$ as
$\alt(\mt) = y_t^2(\mt)/(4\pi)$ with 
$v = 1/\sqrt{\sqrt{2}\, G_F} = 246.218 \gev$
and 
$\mt(\mt)\approx \mt/(1-\frac{1}{2\,\pi} \alt(\mt) 
+\frac{4}{3\,\pi}\als(\mt))$.
Setting in the evaluation of $\db$ the scale to $\mt$ was shown to
yield in general a more stable result~\cite{db2l} as long as 
two-loop corrections to $\db$ are not included.%
\footnote{
  It should be noted that in Ref.~\cite{db2l} a different scale has been 
  advocated due to the emphasis on the two-loop contributions presented 
  in this paper. The plots, however, show that $\mt$ is a good scale 
  choice if only one-loop corrections are included.
}
$M_3$ is the soft SUSY-breaking parameter for the gluinos. 
We have neglected any CKM mixing of the quarks.

\medskip

The $Z$~factors of the squark fields are derived in the OS scheme.
They can be found in \citere{MSSMCT}.

%%%%%%%%%%%%%%%%%%%%%%%%%%%%%%%%%%%%%%%%%%%%%%%%%%%%%%%%%%%%%%%%%%%%%%%%%%%%%%%

\subsection{The gluon/gluino sector and the strong coupling constant}
\label{sec:alphas}

The gluon and gluino sector follow strictly \citere{MSSMCT};
see also the references therein.

The decoupling of the heavy particles and the running is taken into 
account in the definition of $\als$: starting point is~\cite{pdg}
\begin{align}\label{eq:alsMS}
  \als^{\MSbar,(5)}(\MZ) &= 0.1184\,,
\end{align}
where the running of $\als^{\MSbar,(n_f)}(\mu_R)$ can be found in \citere{pdg}.
$\mu_R$ denotes the renormalization scale which is typically of the order 
of the energy scale of the considered process.

From the \MSbar\ value the \DRbar\ value is obtained at the two-loop level 
via the phenomenological one-step formula~\cite{alsDRbar}
\begin{align}\label{eq:alsDR}
\als^{\DRbar,(n_f)}(\mu_R) &= \als^{\MSbar,(n_f)}(\mu_R)\, \Bigg\{
1 + \frac{\als^{\MSbar,(n_f)}(\mu_R)}{\pi} \LP \frac{1}{4} - \zeta^{(n_f)}_1 \RP + 
\notag \\ & \quad 
\LP \frac{\als^{\MSbar,(n_f)}(\mu_R)}{\pi} \RP^2 
\LB \frac{11}{8} - \frac{n_f}{12} - \frac{1}{2} \zeta^{(n_f)}_1 + 
2\, (\zeta^{(n_f)}_1)^2 - \zeta^{(n_f)}_2 \RB \Bigg\}\,,
\end{align}
where (for $n_f = 6$)
\begin{align}
\zeta^{(6)}_1 = - \ln\frac{\mu_R^2}{\Mt^2}\,, \qquad
\zeta^{(6)}_2 = - \frac{65}{32} 
               - \frac 52\, \ln\frac{\mu_R^2}{\Mt^2} 
               + \LP \ln\frac{\mu_R^2}{\Mt^2} \RP^2\,, 
\end{align}
with $\Mt^2$ being defined as the geometric average of all 
squark masses multiplied with the gluino mass%
\footnote{
  $\Mt$ is chosen such that
  $\ln\frac{\mgl^2}{\Mt^2} + \frac{1}{6} \sum_{\Sq} \ln\frac{\msq1\msq2}{\Mt^2} = 0$,
  which means that the corresponding diagrams vanish at zero momentum 
  transfer.  Under this condition $\als(\mu_R)$ is well-defined.
}, 
$\Mt^2 = \mgl\, \prod_{\Sq}(\msq1\msq2)^{\frac{1}{12}}$.
The log~terms originates from the decoupling of the SQCD particles 
from the running of $\als$ at lower scales $\mu_R \leq \mu_{\text{dec.}}$.
For simplification we have chosen the energy scale of the considered 
processes (as a typical SUSY scale) also as decoupling scale.

%%%%%%%%%%%%%%%%%%%%%%%%%%%%%%%%%%%%%%%%%%%%%%%%%%%%%%%%%%%%%%%%%%%%%%%%%%%%%%%
%%%%%%%%%%%%%%%%%%%%%%%%%%%%%%%%%%%%%%%%%%%%%%%%%%%%%%%%%%%%%%%%%%%%%%%%%%%%%%%

\section{Calculation of loop diagrams}
\label{sec:calc}

\begin{figure}[htb!]
\begin{center}
\framebox[15cm]{\includegraphics[width=0.8\textwidth]{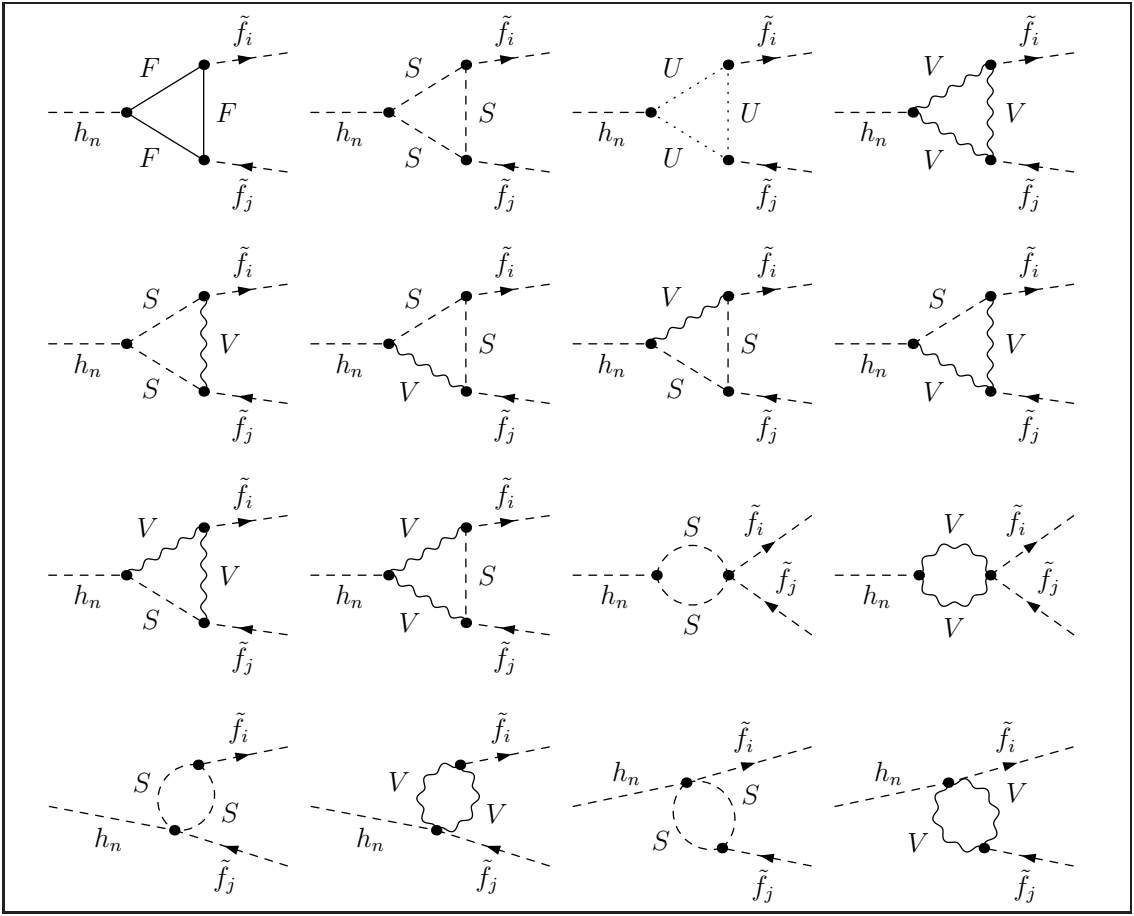}}
\caption{
  Generic Feynman diagrams for the decay $\hndecay$ ($n = 2,3;\, i,j=1,2$).
  $F$ can be a SM fermion, chargino, neutralino, or gluino; 
  $S$ can be a sfermion or a Higgs boson; $U$ denotes the ghosts;
  $V$ can be a $\ga$, $Z$, $W^\pm$, or $g$. 
  Not shown are the diagrams with a $h_n$--$Z$ or $h_n$--$G$ transition 
  contribution on the external Higgs boson leg.
}
\label{fig:hnsfisfj}
\vspace{1em}
\end{center}
\end{figure}

\begin{figure}[htb!]
\begin{center}
\framebox[15cm]{\includegraphics[width=0.8\textwidth]{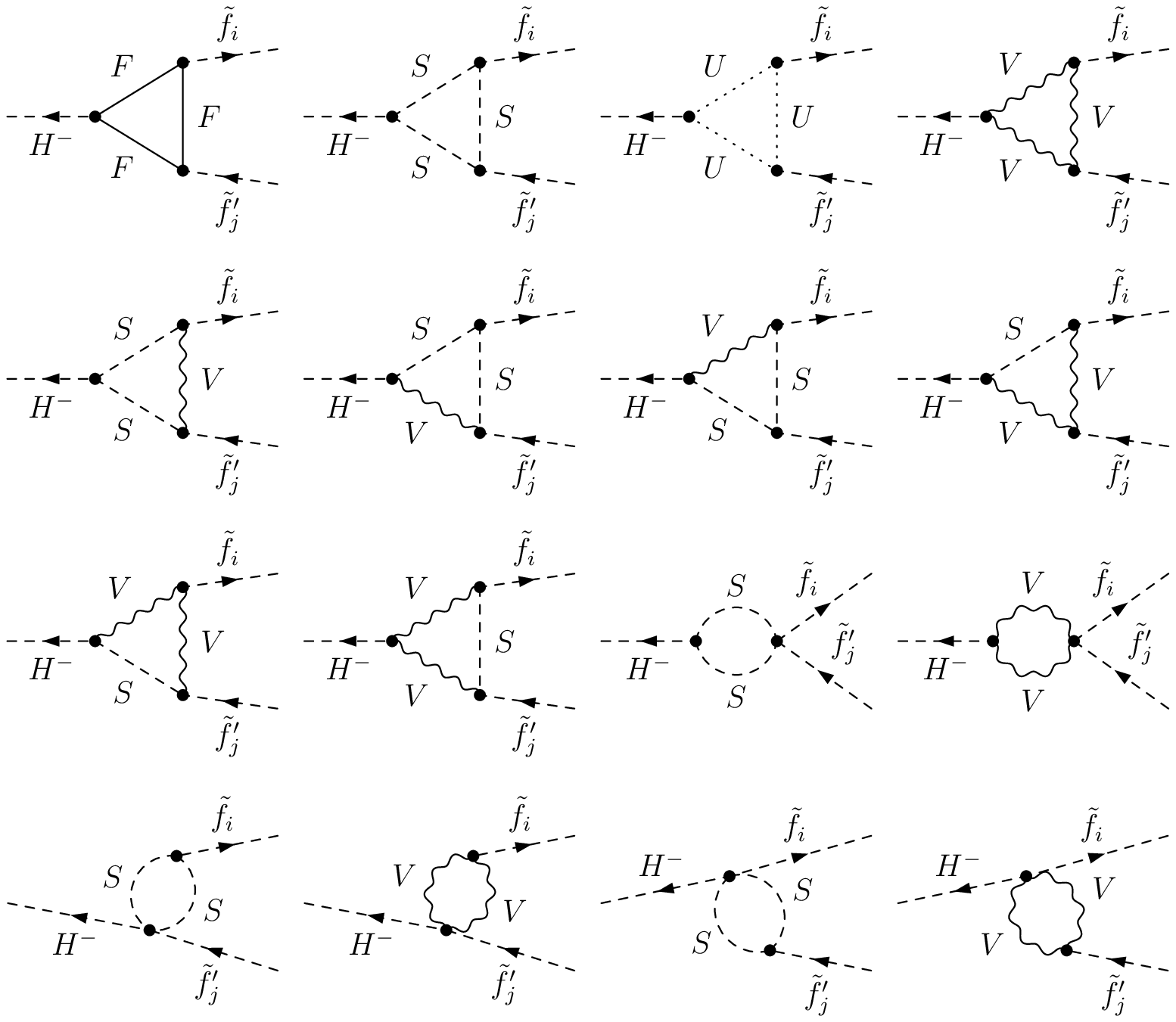}}
\caption{
  Generic Feynman diagrams for the decay $H^+ \to \Sf_i^{} \Sf_j^{\prime}$
  ($i,j = 1,2$).
  (It should be noted that all arrows are inverted in case of a $H^-$ decay.)
  $F$ can be a SM fermion, chargino, neutralino, or gluino;
  $S$ can be a sfermion or a Higgs boson; $U$ denotes the ghosts;
  $V$ can be a $\ga$, $Z$, $W^\pm$, or $g$. 
  Not shown are the diagrams with a $H^\pm$--$W^\pm$ or $H^\pm$--$G^\pm$ 
  transition contribution on the external Higgs boson leg.
}
\label{fig:Hpsfisfj}
\vspace{1em}
\end{center}
\end{figure}

In this section we give some details about the calculation of the
higher-order corrections to the partial decay widths of Higgs bosons. 
Sample diagrams are shown in \reffis{fig:hnsfisfj} and \ref{fig:Hpsfisfj}. 
Not shown are the diagrams for real (hard and soft) photon and gluon
radiation. They are obtained from the corresponding tree-level diagrams
by attaching a photon (gluon) to the electrically (color) charged
particles. The internal generically depicted particles in
\reffis{fig:hnsfisfj} and \ref{fig:Hpsfisfj} are labeled as follows:
$F$ can be a SM fermion $f$, chargino $\cha{j}$, neutralino 
$\neu{k}$, or gluino $\gl$; 
$S$ can be a sfermion $\Sf_i$ or a Higgs boson $h_n$; 
$U$ denotes the ghosts $u_V$;
$V$ can be a photon $\ga$, gluon $g$, or a massive SM gauge boson, 
$Z$ or $W^\pm$. 
For internally appearing Higgs bosons no higher-order
corrections to their masses or couplings are taken into account; 
these corrections would correspond to effects beyond one-loop order.%
\footnote{
  We found that using loop corrected Higgs boson masses 
  in the loops leads to a UV divergent result.
}
For external Higgs bosons, as described in
\refse{sec:higgs}, the appropriate $\hat{Z}$~factors are applied and
on-shell masses (including higher-order corrections) are
used~\cite{mhcMSSMlong}, obtained with 
\FH~\cite{feynhiggs,mhiggslong,mhiggsAEC,mhcMSSMlong,Mh-logresum}.

%%%%%%%%%%%%%%%%%%%%%%%%% F I G U R E %%%%%%%%%%%%%%%%%%%%%%%%%%%%%%%%%%%%%%%%%
%%%%%%%%%%%%%%%%%%%%%%%%% F I G U R E %%%%%%%%%%%%%%%%%%%%%%%%%%%%%%%%%%%%%%%%%

%%%%%%%%%%%%%%%%%%%%%%%%% F I G U R E %%%%%%%%%%%%%%%%%%%%%%%%%%%%%%%%%%%%%%%%%
%%%%%%%%%%%%%%%%%%%%%%%%% F I G U R E %%%%%%%%%%%%%%%%%%%%%%%%%%%%%%%%%%%%%%%%%

Also not shown are the diagrams with a Higgs boson--gauge/Goldstone
self-energy contribution on the external Higgs boson leg. 
They appear in the decay $\hndecay$, \reffi{fig:hnsfisfj}, 
with a $h_n$--$Z/G$ transition and in the decay $\Hpmdecay$, 
\reffi{fig:Hpsfisfj}, with a $H^\pm$--$W^\pm$/$G^\pm$ transition.%
\footnote{
  From a technical point of view, the $H^\pm$--$W^\pm$/$G^\pm$ transitions 
  have been absorbed into the respective counterterms, while the 
  $h_n$--$Z/G$ transitions have been calculated explicitly.
}

Furthermore, in general, in \reffis{fig:hnsfisfj} and \ref{fig:Hpsfisfj} 
we have  omitted diagrams with self-energy type corrections of external 
(on-shell) particles.  While the contributions from the real parts of 
the loop functions are taken into account via the renormalization 
constants defined by on-shell renormalization conditions, the 
contributions coming from the imaginary part of the loop functions can 
result in an additional (real) correction if multiplied by complex 
parameters (such as $A_f$).  In the analytical and numerical evaluation, 
these diagrams have been taken into account via the prescription 
described in \citere{MSSMCT}. 

Within our one-loop calculation we neglect finite width effects that 
can help to cure threshold singularities.  Consequently, in the close 
vicinity of those thresholds our calculation does not give a reliable
result.  Switching to a complex mass scheme \cite{complexmassscheme} 
would be another possibility to cure this problem, but its application 
is beyond the scope of our paper.

The diagrams and corresponding amplitudes have been obtained with \FA\ 
\cite{feynarts}.  The model file, including the MSSM counterterms, 
is largely based on \citere{MSSMCT}, however adjusted to match exactly 
the renormalization prescription described in \refse{sec:renorm}. 
The further evaluation has been performed with \FC\ and \LT\ 
\cite{formcalc}.

\subsubsection*{Ultraviolet divergences}

As regularization scheme for the UV divergences we
have used constrained differential renormalization~\cite{cdr}, 
which has been shown to be equivalent to 
dimensional reduction~\cite{dred} at the \onel\ level~\cite{formcalc}. 
Thus the employed regularization scheme preserves SUSY~\cite{dredDS,dredDS2}
and guarantees that the SUSY relations are kept intact, \eg that the 
gauge couplings of the SM vertices and the Yukawa couplings of the 
corresponding SUSY vertices also coincide to \onel\ order in the SUSY limit. 
Therefore no additional shifts, which might occur when using a different 
regularization scheme, arise.
All UV divergences cancel in the final result.

\subsubsection*{Infrared divergences}

The IR divergences from diagrams with an internal photon or gluon have
to cancel with the ones from the corresponding real soft radiation. 
In the case of QED we have included the soft photon contribution
following the description given in \citere{denner}. 
In the case of QCD we have modified this prescription by replacing the
product of electric charges by the appropriate combination of color
charges (linear combination of $C_A$ and $C_F$ times $\als$).
The IR divergences arising from the diagrams involving a $\ga$ 
(or a $g$) are regularized by introducing a photon (or gluon) 
mass parameter, $\la$. 
While for the QED part this procedure always works, in the QCD part 
due to its non-Abelian character this method can fail. 
However, since no triple or quartic gluon vertices appear, $\la$ can 
indeed be used as a regulator.
All IR divergences, \ie all divergences in the limit $\la \to 0$, 
cancel once virtual and real diagrams for one decay channel are added.

\subsubsection*{Tree-level formulas}

For completeness we show here also the formulas that have been
used to calculate the tree-level decay widths:
\begin{align}
\Gamma^{\rm tree}(\hndecay) &= 
\frac{|C(h_n, \Sf_i, \Sf_j)|^2\,
  \la^{1/2}(\mh{n}^2, \msf{i}^2, \msf{j}^2)}
     {16\, \pi\, \mh{n}^3} \qquad (n = 2,3;\, i,j = 1,2)\,, \\
\Gamma^{\rm tree}(\Hpmdecay) &= 
\frac{|C(H^\pm, \Sf_i, \Sfp_j)|^2\,
  \la^{1/2}(\MHp^2, \msf{i}^2, \msfp{j}^2)}
     {16\, \pi\, \MHp^3}\qquad (i,j = 1,2)\,,
\end{align}
where $\la(x,y,z) = (x - y - z)^2 - 4yz$ and the couplings $C(a, b, c)$ 
can be found in the \FA\ model files~\cite{feynarts-mf}.

%%%%%%%%%%%%%%%%%%%%%%%%%%%%%%%%%%%%%%%%%%%%%%%%%%%%%%%%%%%%%%%%%%%%%%%%%%%%%%%
%%%%%%%%%%%%%%%%%%%%%%%%%%%%%%%%%%%%%%%%%%%%%%%%%%%%%%%%%%%%%%%%%%%%%%%%%%%%%%%

\section{Numerical analysis}
\label{sec:numeval}

In this section we present the comparisons with results from other groups
and our numerical analysis of all heavy Higgs boson decay channels into 
the third generation sfermions in the cMSSM. In the various figures below 
we show the partial decay widths and their relative correction at the 
tree-level (``tree'') and at the one-loop level (``full'').  In addition 
we show the SQCD corrections (``SQCD'') for comparison with the full 
one-loop result.

%%%%%%%%%%%%%%%%%%%%%%%%%%%%%%%%%%%%%%%%%%%%%%%%%%%%%%%%%%%%%%%%%%%%%%%%%%%%%

\subsection{Comparisons}
\label{sec:comparisons}

We performed exhaustive comparisons with results from other groups
for heavy Higgs boson decays. Since loop corrections in the MSSM
with complex parameters have been evaluated in this work for the 
first time, these comparisons were restricted to the MSSM with 
real parameters.

\begin{itemize}

\item
% Hollik: hep-ph/9702426: Phi -> sq sq at O(als)
We calculated the decays $\Phi \to \Sq_i \Sq_j$ at $\order{\als}$ 
($\Phi$ denotes any heavy MSSM Higgs boson) and found good agreement 
with \citere{Phisqsq_als_2}, where only a small difference remains 
due to the slightly different renormalization schemes. 
We successfully reproduced their figures, except their Fig.~4 
($H^+ \to \Stop1 \Sbot1$) which differs substantially.
Unfortunately, seventeen years after publication, the source code of 
\citere{Phisqsq_als_2} is unavailable for a direct 
comparison~\cite{abdesslam}.
On the other hand our results for $H^+ \to \Stop1 \Sbot1$ are in 
good qualitative agreement with \citere{Phisqsq_als_1}, see below.

\item
% Spira: arXiv:1103.4283: h/H/A -> sq sq at O(als)
A comparison with \citere{Phisqsq_als_3} at $\order{\als}$ was rather
difficult.  \citere{Phisqsq_als_3} used running \MSbar\ input 
parameters and significant differences exist w.r.t.\ our treatment 
of the renormalization of the scalar quark sector.  Nevertheless, 
using their input parameters as far as possible, we found 
qualitative agreement.

\item
% Wiener: hep-ph/9508283: H+ -> st sb at O(als)
We performed a detailed comparison with \citere{Hpstsb_als} for the 
decay $H^+ \to \Stop{i} \Sbot{j}$ at $\order{\als}$.  They also differ 
in the renormalization of the scalar quark sector, leading to 
different loop corrections.  Furthermore they used tree/\DRbar/pole 
squark masses in tree/loop/phase space.  Despite these complications
we found rather good qualitative agreement with their Fig.~2.

\item
% Wiener: hep-ph/9701398: H/A -> st st, sb sb, H+ -> st sb at O(als)
% Wiener: hep-ph/9912463: Phi -> qq, sq sq at O(als)
A check with \citeres{Phisqsq_als} and \cite{Phisqsq_als_1} at 
$\order{\als}$ gave good qualitative agreement, although an effective 
resummation of the trilinear Higgs-sbottom coupling was used in 
\citere{Phisqsq_als_1}.

\item
% Wiener: hep-ph/0305250: A -> sq sq full one-loop (rMSSM)
Decays of the $\CP$-odd Higgs boson $A$ to scalar quarks (in the rMSSM) 
have been compared with \citere{Asqsq_1L}. Again, using their input 
parameters as far as possible, we found good (qualitative) agreement 
with their Figs.~4 -- 7.

\item
% Wiener: hep-ph/0308146: A -> sf sf full one-loop (rMSSM)
$A$ boson decays into sfermions in the rMSSM have been analyzed in 
\citere{Asfsf_1L}.  As in the latter item we found good (qualitative) 
agreement, especially for the decay into scalar taus after using our 
new \DRbar\ type version for the slepton sector, as described above 
in \refse{sec:sfermion}. 

\end{itemize}

\noindent
Finally it should be noted that
\citeres{Hpstsb_als,Phisqsq_als,Phisqsq_als_1,Asqsq_1L,Asfsf_1L}
subsequently had been recompiled in \citere{Phisqsq_1L}.

%%%%%%%%%%%%%%%%%%%%%%%%%%%%%%%%%%%%%%%%%%%%%%%%%%%%%%%%%%%%%%%%%%%%%%%%%%%%%

\subsection{Parameter settings}
\label{sec:paraset}

The renormalization scale $\mu_R$ has been set to the mass of the 
decaying Higgs boson.  The SM parameters are chosen as follows; 
see also \cite{pdg}:
\begin{itemize}

\item Fermion masses (on-shell masses, if not indicated differently)
\index{leptonmasses}:
\begin{align}
m_e    &= 0.510998928\mev\,, & m_{\nu_e}    &= 0\mev\,, \notag \\
m_\mu  &= 105.65837515\mev\,, & m_{\nu_{\mu}} &= 0\mev\,, \notag \\
m_\tau &= 1776.82\mev\,,      & m_{\nu_{\tau}} &= 0\mev\,, \notag \\
m_u &= 68.7\mev\,,           & m_d         &= 68.7\mev\,, \notag \\ 
m_c &= 1.275\gev\,,          & m_s         &= 95.0\mev\,, \notag \\
m_t &= 173.07\gev\,,         & m_b(m_b)    &= 4.18\gev\,.
\end{align}
According to \citere{pdg}, $m_s$ is an estimate of a so-called 
"current quark mass" in the \MSbar\ scheme at the scale $\mu \approx 2\gev$. 
$m_c$ and $m_b$ are the "running" masses in the \MSbar\ scheme.
The top quark mass as well as the lepton masses are defined OS.
$m_u$ and $m_d$ are effective parameters, calculated through the hadronic
contributions to
\begin{align}
\Delta\alpha_{\text{had}}^{(5)}(M_Z) &= 
      \frac{\alpha}{\pi}\sum_{f = u,c,d,s,b}
      Q_f^2 \Bigl(\ln\frac{M_Z^2}{m_f^2} - \frac 53\Bigr) \sim 0.027723\,.
\end{align}

\item The CKM matrix has been set to unity.

\item Gauge boson masses\index{gaugebosonmasses}:
\begin{align}
M_Z = 91.1876\gev\,, \qquad M_W = 80.385\gev\,.
\end{align}

\item Coupling constants\index{couplingconstants}:
\begin{align}
\alpha(0) = 1/137.0359895\,, \qquad \alpha_s(M_Z) = 0.1184\,,
\end{align}
where the running and decoupling of $\alpha_s$ is described in 
\refse{sec:alphas}.
\end{itemize}

The Higgs sector quantities (masses, mixings, etc.) have been
evaluated using \FH\ (version 2.10.2)
\cite{feynhiggs,mhiggslong,mhiggsAEC,mhcMSSMlong,Mh-logresum}.

We emphasize again that the analytical calculation has been done for 
{\em all} decays into sfermions, but in the numerical analysis we
concentrate on the decays to third generation sfermions.  Results are 
shown for some representative numerical examples.  The parameters are 
chosen according to the scenarios, \Sce, \Scz\ and \Scd, shown in 
\refta{tab:para}.  The scenarios are defined such that a maximum number 
of (third generation) decay modes are open simultaneously to permit an 
analysis of all channels, \ie not picking specific parameters for each 
decay.  
For the same reason we do not demand that the lightest Higgs boson
has a mass around $\sim 125 \gev$, although for most of the parameter
space this is given.
We will show the variation of $\MHp$ and $\phia$, where the
latter denotes the phase of any trilinear coupling.

%%%%%%%%%%%%%%%%%%%%% T A B L E %%%%%%%%%%%%%%%%%%%%%%%%%%%%%%%%%%%%%%%%%%%%%%
\begin{table}[t!]
\caption{\label{tab:para}
  MSSM parameters for the initial numerical investigation; all parameters 
  (except of $\TB$) are in GeV.  In our analysis $M_{\tilde Q_3}$, 
  $M_{\tilde U_3}$, $M_{\tilde D_3}$, $M_{\tilde L_3}$ and $M_{\tilde E_3}$
  are chosen such that the values of $\mstop1$, $\mstop2$, 
  $\msbot2$, $\mtausneu$ and $\mstau2$ are realized.  For the $\Sd_g$ 
  and $\Se_g$~sector the shifts in $M_{\tilde Q, \tilde D}(\Sd_g)$ and 
  $M_{\tilde L, \tilde E}(\Se_g)$ as defined in \refeqs{eq:MSfShift} and 
  \eqref{eq:MSfBackshift} are taken into account, concerning $\mstau1$ 
  and $\msbot1$ (rounded to 1 MeV).  The values for $\At$, $\Ab$ and 
  $A_{\tau}$ are chosen such that charge- and/or color-breaking minima are 
  avoided~\cite{ccb}.  It should be noted that for the first and second 
  generation of sfermions we chose instead $M_{\tilde L, \tilde E} = 1500\gev$ 
  and $M_{\tilde Q, \tilde U, \tilde D} = 2000\gev$.
}
\renewcommand{\arraystretch}{1.5}
\begin{center}
\begin{tabular}{lrrrrrrrrrrrrrr}
\hline
Scen. & $\TB$ & $\mstop1$ & $\mstop2$ & $\msbot2$ & $\mtausneu$ 
& $\mstau2$ & $\mu$ & $|\At|$ & $|\Ab|$ & $|\Atau|$ & $M_1$ & $M_2$ & $M_3$ \\ 
\hline
\Sce/\Scz/\Scd & 10 & 394 & 771 & 582 & 280 
& 309 & 500 & 1200 & 600 & 1000 & 300 & 600 & 1500 \\
\hline
\end{tabular}

\vspace{5mm}

\begin{tabular}{llrrrrr}
\hline
Scen. & $\MHp$ & $\mh1$ & $\mh2$ & $\mh3$ & $\mstau1$ & $\msbot1$ \\ 
\hline 
\Sce & 1000 & 123.405 &  996.766 &  996.813 & 282.517 & 513.289 \\ 
\Scz & 1400 & 123.428 & 1397.299 & 1398.596 & 282.337 & 513.167 \\
\Scd & 1600 & 123.436 & 1597.174 & 1597.524 & 282.265 & 513.124 \\
\hline
\end{tabular}
\end{center}
\renewcommand{\arraystretch}{1.0}
\end{table}
%%%%%%%%%%%%%%%%%%%%% T A B L E %%%%%%%%%%%%%%%%%%%%%%%%%%%%%%%%%%%%%%%%%%%%%%

The numerical results we will show in the next subsections are of
course dependent on choice of the SUSY parameters. Nevertheless, they
give an idea of the relevance of the full one-loop corrections.
Channels (and their respective one-loop corrections) that may look 
unobservable due to the smallness of their decay width in the plots 
shown below, could become important if other channels are kinematically 
forbidden.

%%%%%%%%%%%%%%%%%%%%%%%%%%%%%%%%%%%%%%%%%%%%%%%%%%%%%%%%%%%%%%%%%%%%%%%%%%%%%%

\subsection{Full one-loop results for varying \boldmath{$\MHp$}
  and \boldmath{$\phia$}}
\label{sec:full1L}

The results shown in this and the following subsections consist of 
``tree'', which denotes the tree-level value and of ``full'', which 
is the partial decay width including {\em all} one-loop corrections 
as described in \refse{sec:calc}.  Also shown are the pure SUSY-QCD 
one-loop corrections (``SQCD'') for colored decays.
We restrict ourselves to the analysis of the decay widths themselves, 
since the one-loop effects on the branching ratios are strongly 
parameter dependent, as discussed in the previous subsection.

When performing an analysis involving complex parameters it should be 
noted that the results for physical observables are affected only 
by certain combinations of the complex phases of the parameters $\mu$, 
the trilinear couplings $A_{t,b,\tau}$ and the gaugino mass parameters 
$M_{1,2,3}$~\cite{MSSMcomplphasen,SUSYphases}.
It is possible, for instance, to rotate the phase $\phiMz$ away.
Experimental constraints on the (combinations of) complex phases 
arise, in particular, from their contributions to electric dipole 
moments of the electron and the neutron (see \citeres{EDMrev2,EDMPilaftsis} 
and references therein), of the deuteron~\cite{EDMRitz} and of heavy 
quarks~\cite{EDMDoink}.
While SM contributions enter only at the three-loop level, due to its
complex phases the MSSM can contribute already at one-loop order.
Large phases in the first two generations of sfermions can only be 
accommodated if these generations are assumed to be very heavy 
\cite{EDMheavy} or large cancellations occur~\cite{EDMmiracle};
see, however, the discussion in \citere{EDMrev1}. 
A review can be found in \citere{EDMrev3}.
Accordingly (using the convention that $\phiMz = 0$, as done in this paper), 
in particular, the phase $\phimu$ is tightly constrained~\cite{plehnix}, 
while the bounds on the phases of the third generation trilinear couplings 
are much weaker.
Setting $\phimu = \phiMe = \phigl = 0$ leaves us with 
$\phiAt$, $\phiAb$ and $\phiAtau$ as the only complex valued parameters. 
It should be noted that the tree-level prediction depends on $\phia$ 
via the sfermion mixing matrix.

Since now complex trilinear $A_f$ parameters can appear in the couplings,  
contributions from absorptive parts of self-energy type corrections on 
external legs can arise.  The corresponding formulas for an inclusion of 
these absorptive contributions via finite wave function correction factors 
can be found in \cite{MSSMCT,Stop2decay}.

We start the numerical analysis with partial decay widths of $H^\pm$
evaluated as a function of $\MHp$, starting at $\MHp = 600\gev$ 
up to $\MHp = 1.6 \tev$, which roughly coincides with the reach of 
the LHC for high-luminosity running as well as an $e^+e^-$ collider
with a center-of-mass energy up to $\sqrt{s} \sim 3 \tev$~\cite{CLIC}.
Then we turn to the $h_n$ ($n = 2,3$) decays.

%%%%%%%%%%%%%%%%%%%%%%%%%%%%%%%%%%%%%%%%%%%%%%%%%%%%%%%%%%%%%%%%%%%%%%%%%%%%%%

\subsubsection{\boldmath{$H^\pm$} decays into sfermions}
\label{Hpdecays}

\begin{figure}[htb!]
\begin{center}
\begin{tabular}{c}
\includegraphics[width=0.49\textwidth,height=7.5cm]{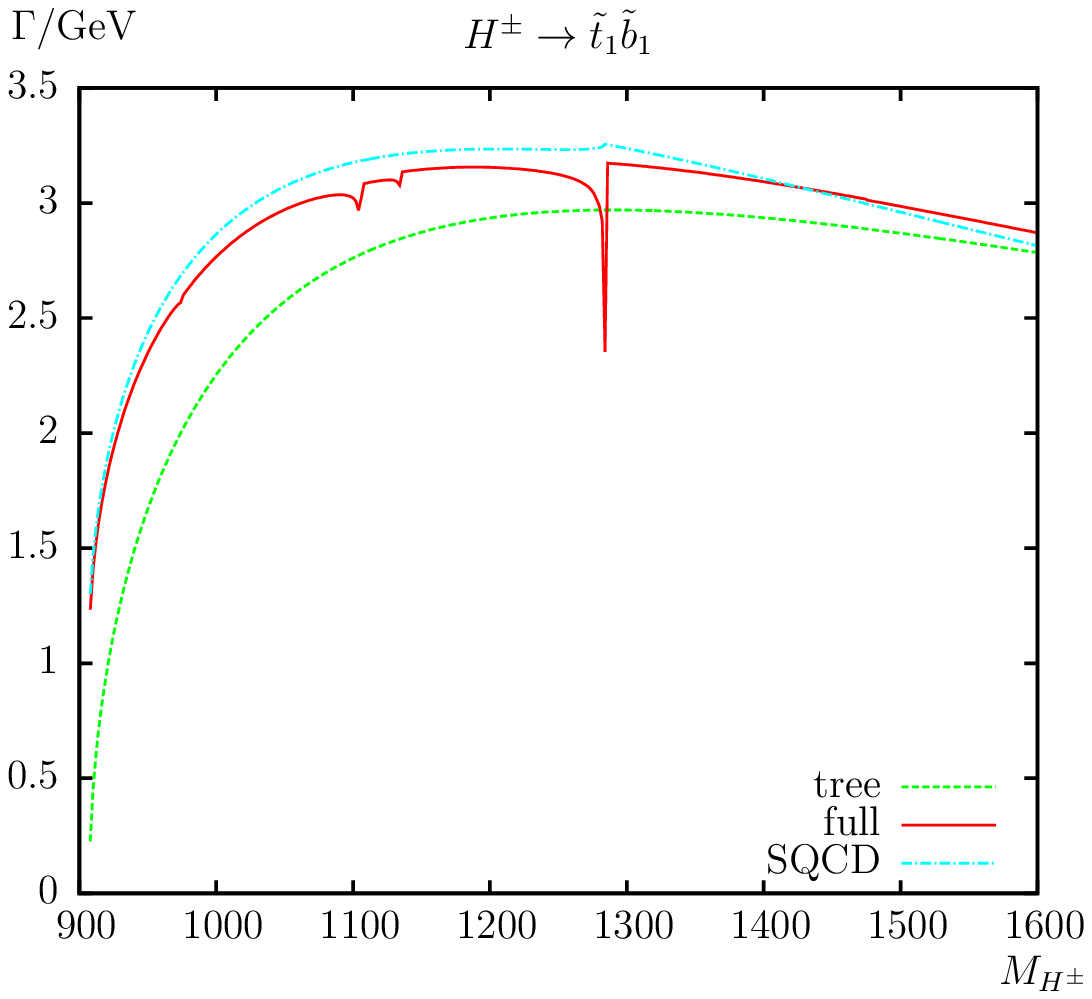}
%\hspace{-4mm}
%\includegraphics[width=0.49\textwidth,height=7.5cm]{} 
\\[4em]
\includegraphics[width=0.49\textwidth,height=7.5cm]{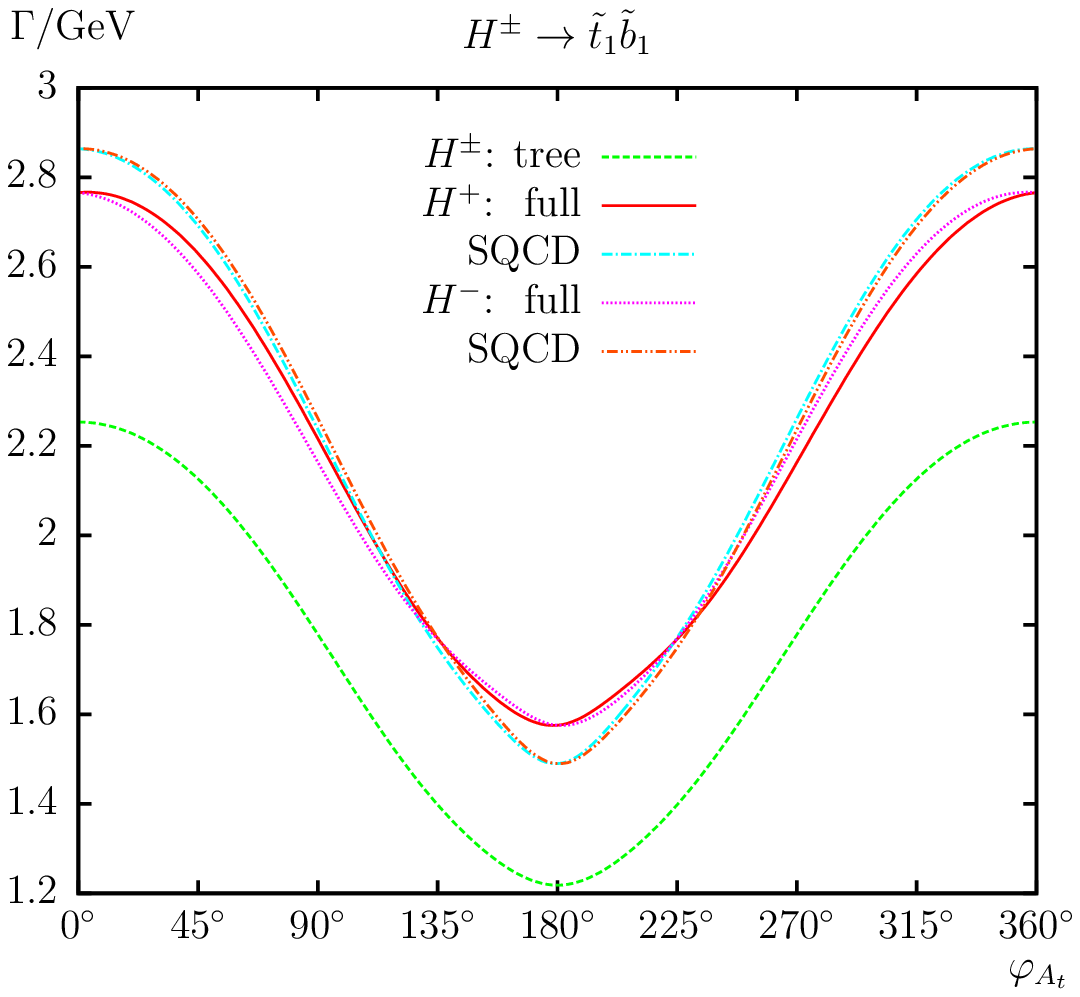}
\hspace{-4mm}
\includegraphics[width=0.49\textwidth,height=7.5cm]{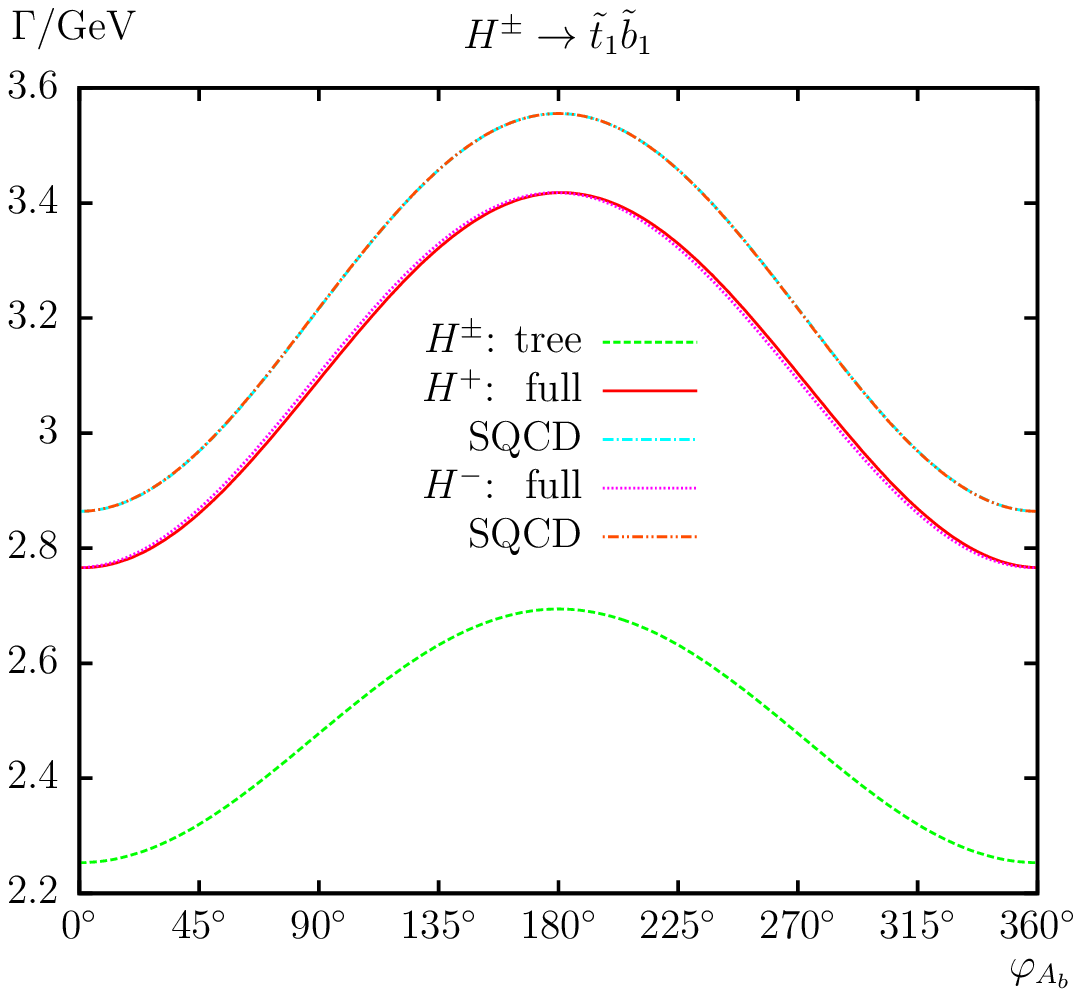}
\end{tabular}
\vspace{1em}
\caption{
  $\Ga(H^\pm \to \tilde{t}_1 \tilde{b}_1)$. 
  Tree-level, full and SQCD one-loop corrected partial decay widths are 
  shown.  The upper plot shows the partial decay width with $\MHp$ varied.
  The lower plots show the complex phases $\phiAt$ (left) and 
  $\phiAb$ (right) varied with parameters chosen according to \Sce\ 
  (see \refta{tab:para}).
}
\label{fig:Hpst1sb1}
\end{center}
\end{figure}

\begin{figure}[htb!]
\begin{center}
\begin{tabular}{c}
\includegraphics[width=0.49\textwidth,height=7.5cm]{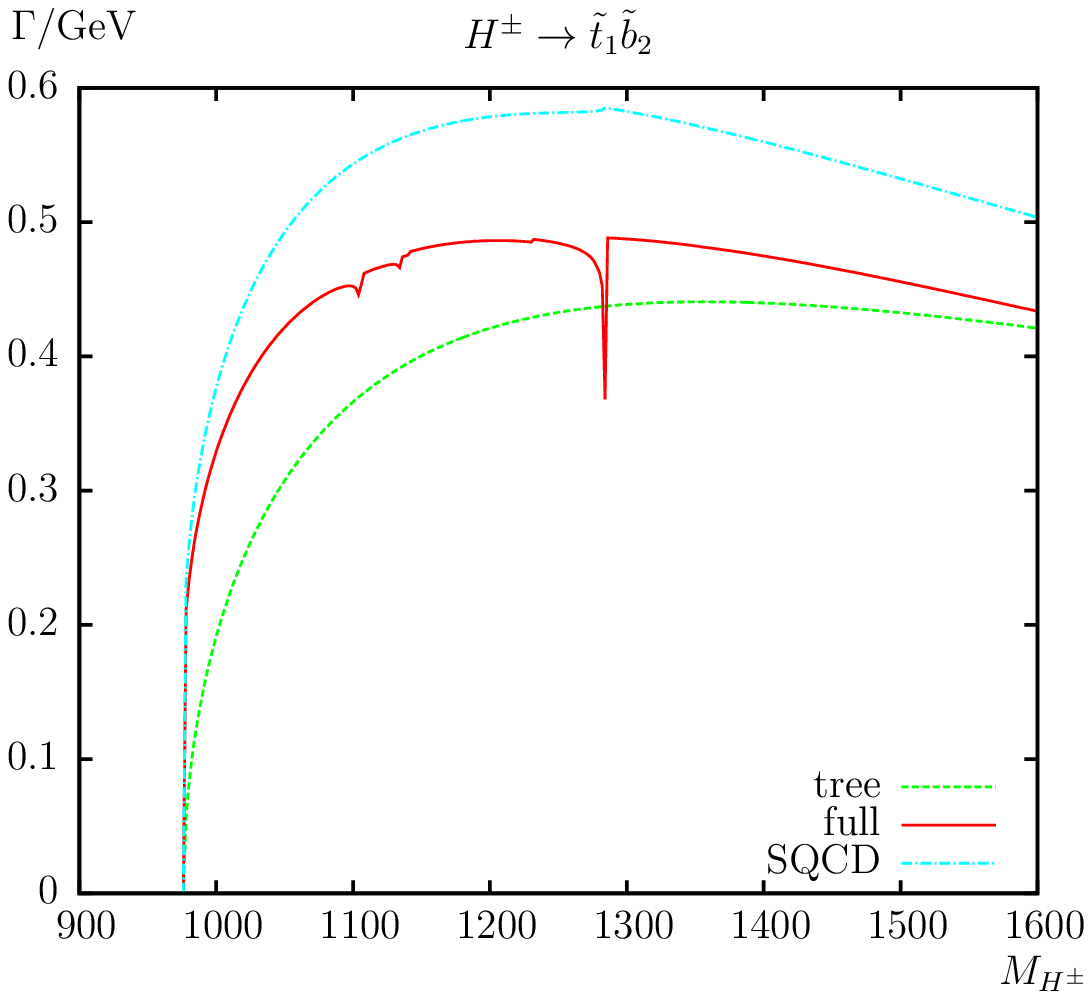}
%\hspace{-4mm}
%\includegraphics[width=0.49\textwidth,height=7.5cm]{} 
\\[4em]
\includegraphics[width=0.49\textwidth,height=7.5cm]{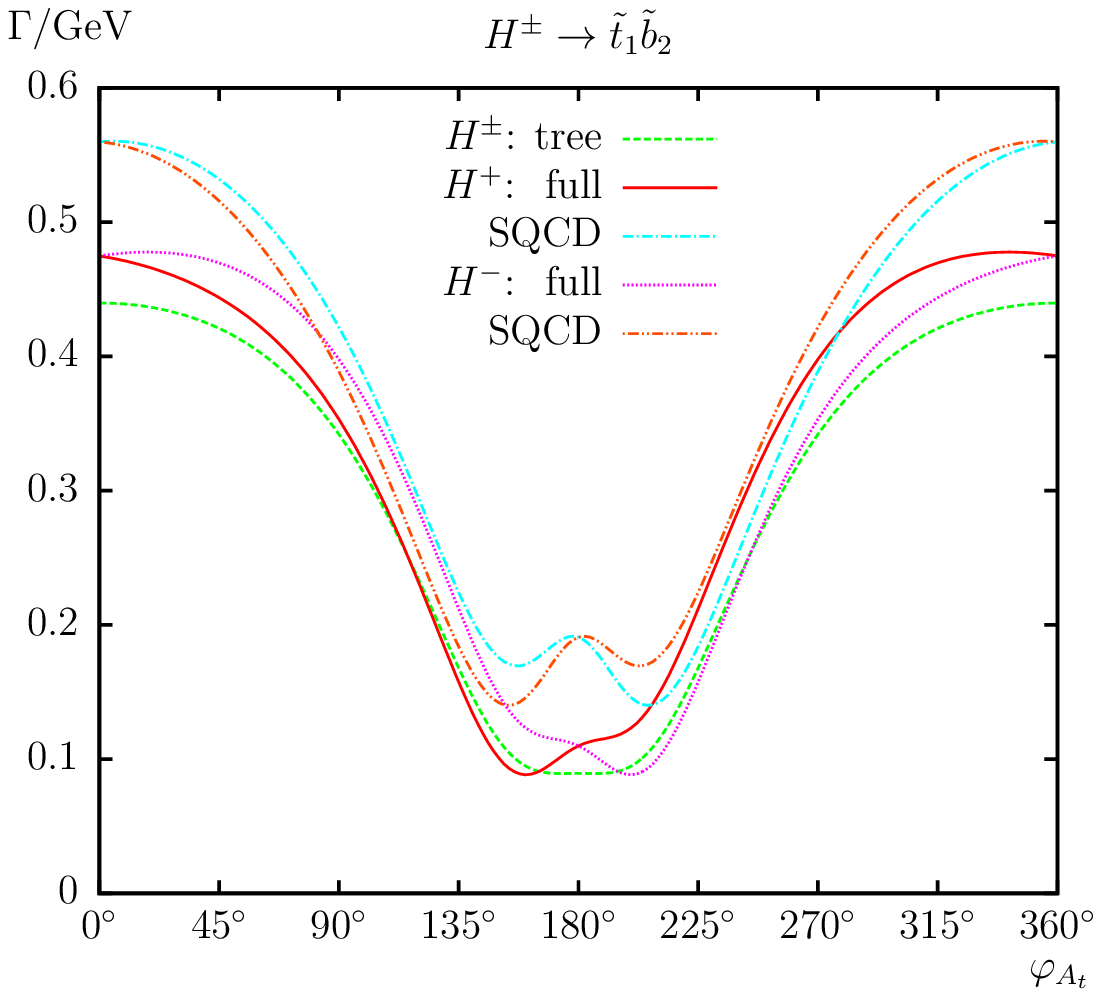}
\hspace{-4mm}
\includegraphics[width=0.49\textwidth,height=7.5cm]{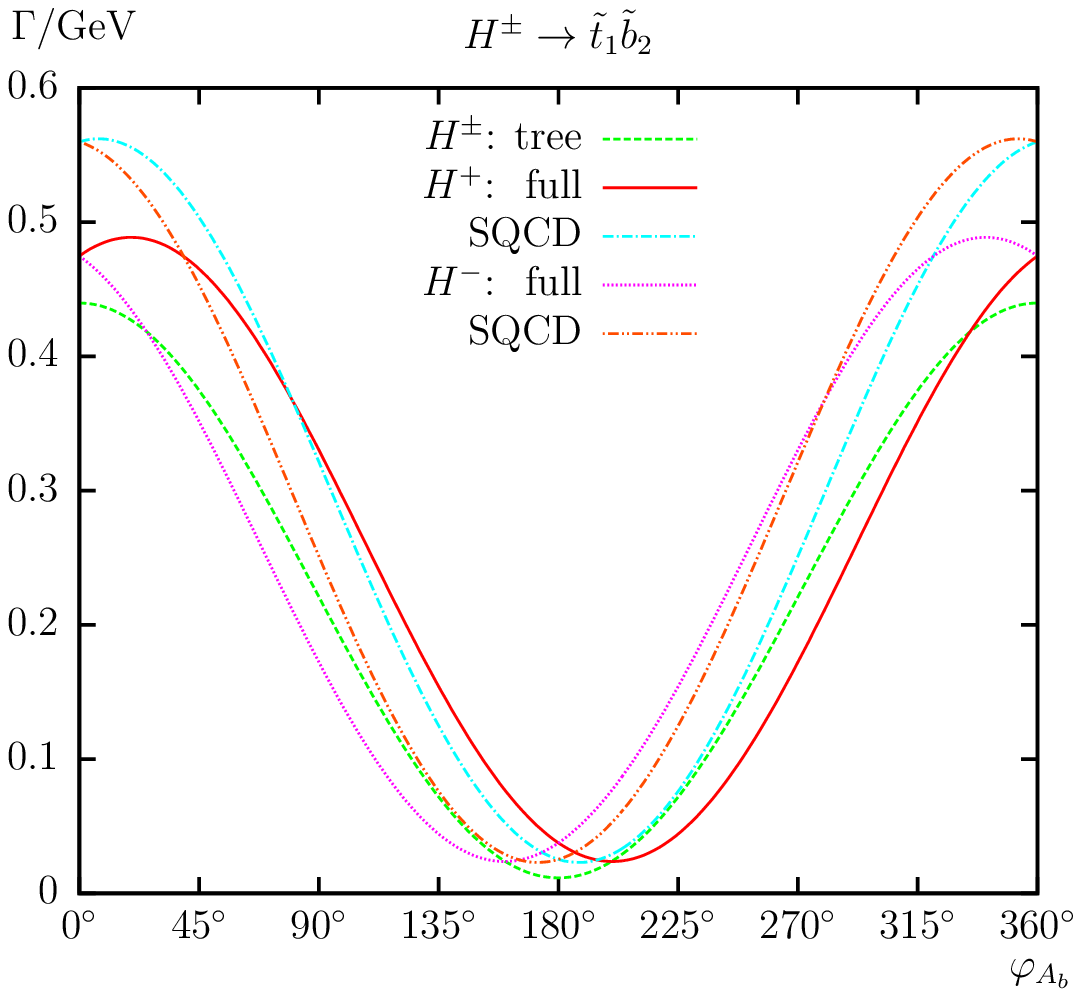}
\end{tabular}
\vspace{1em}
\caption{
  $\Ga(H^\pm \to \tilde{t}_1 \tilde{b}_2)$. 
  Tree-level, full and SQCD one-loop corrected partial decay widths are 
  shown.  The upper plot shows the partial decay width with $\MHp$ varied.
  The lower plots show the complex phases $\phiAt$ (left) and 
  $\phiAb$ (right) varied with parameters chosen according to \Scz\ 
  (see \refta{tab:para}).
}
\label{fig:Hpst1sb2}
\end{center}
\end{figure}

\begin{figure}[htb!]
\begin{center}
\begin{tabular}{c}
\includegraphics[width=0.49\textwidth,height=7.5cm]{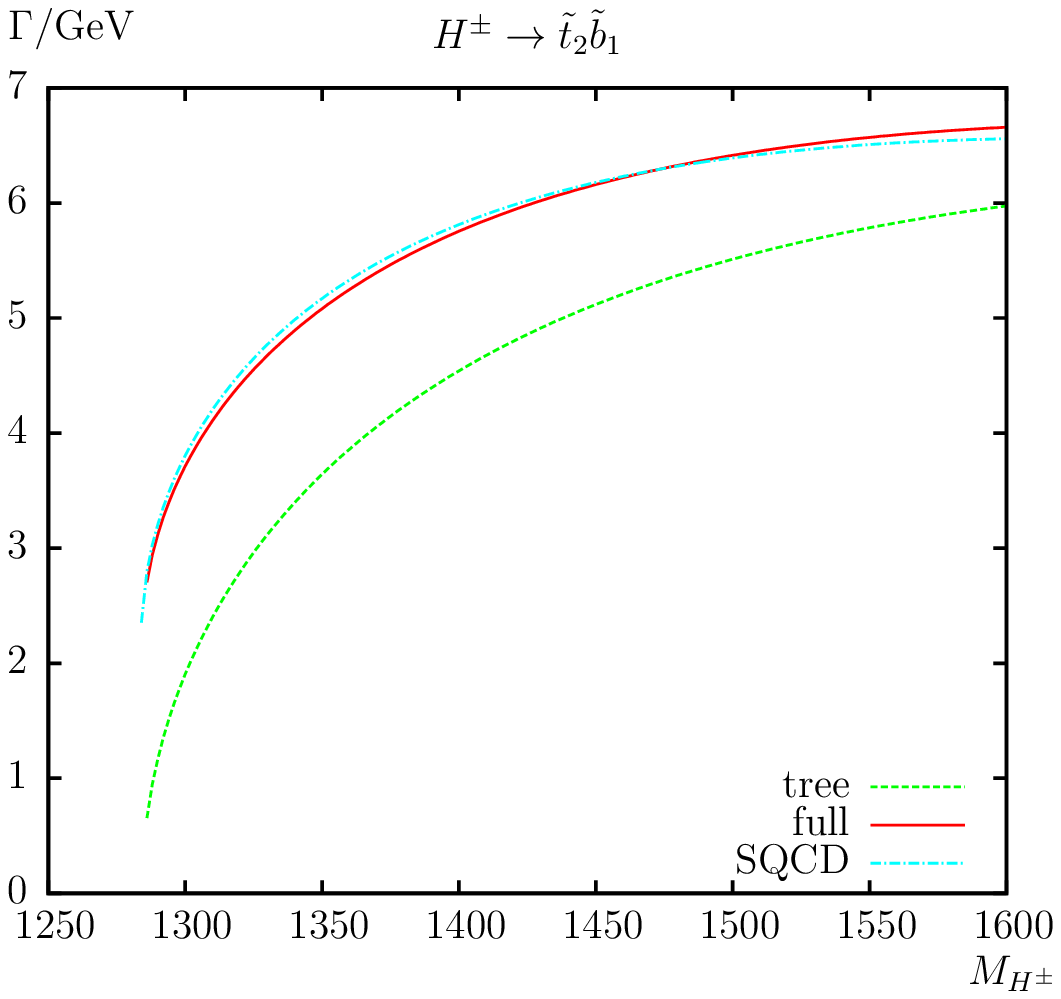}
%\hspace{-4mm}
%\includegraphics[width=0.49\textwidth,height=7.5cm]{} 
\\[4em]
\includegraphics[width=0.49\textwidth,height=7.5cm]{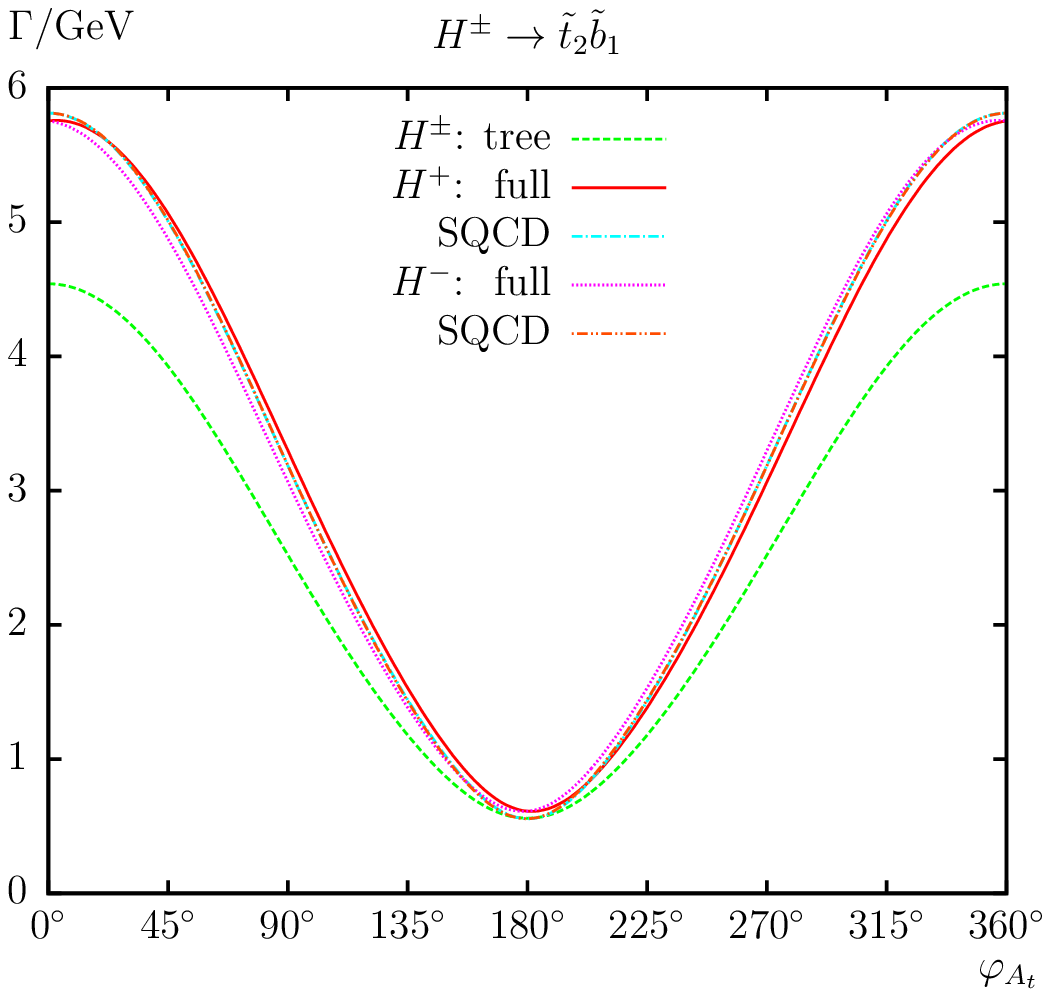}
\hspace{-4mm}
\includegraphics[width=0.49\textwidth,height=7.5cm]{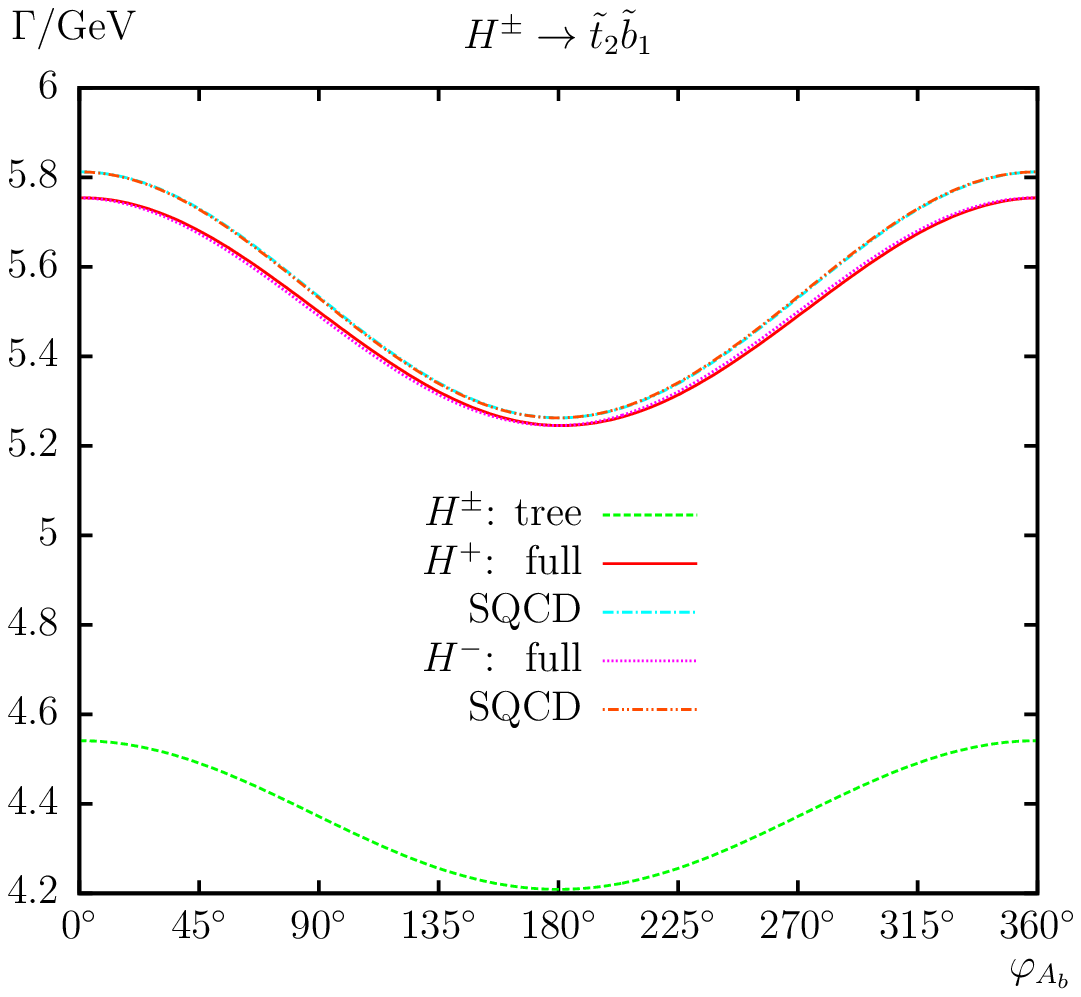}
\end{tabular}
\vspace{1em}
\caption{
  $\Ga(H^\pm \to \tilde{t}_2 \tilde{b}_1)$. 
  Tree-level, full and SQCD one-loop corrected partial decay widths are 
  shown.  The upper plot shows the partial decay width with $\MHp$ varied.
  The lower plots show the complex phases $\phiAt$ (left) and 
  $\phiAb$ (right) varied with parameters chosen according to \Scz\ 
  (see \refta{tab:para}).
}
\label{fig:Hpst2sb1}
\end{center}
\end{figure}

\begin{figure}[htb!]
\begin{center}
\begin{tabular}{c}
\includegraphics[width=0.49\textwidth,height=7.5cm]{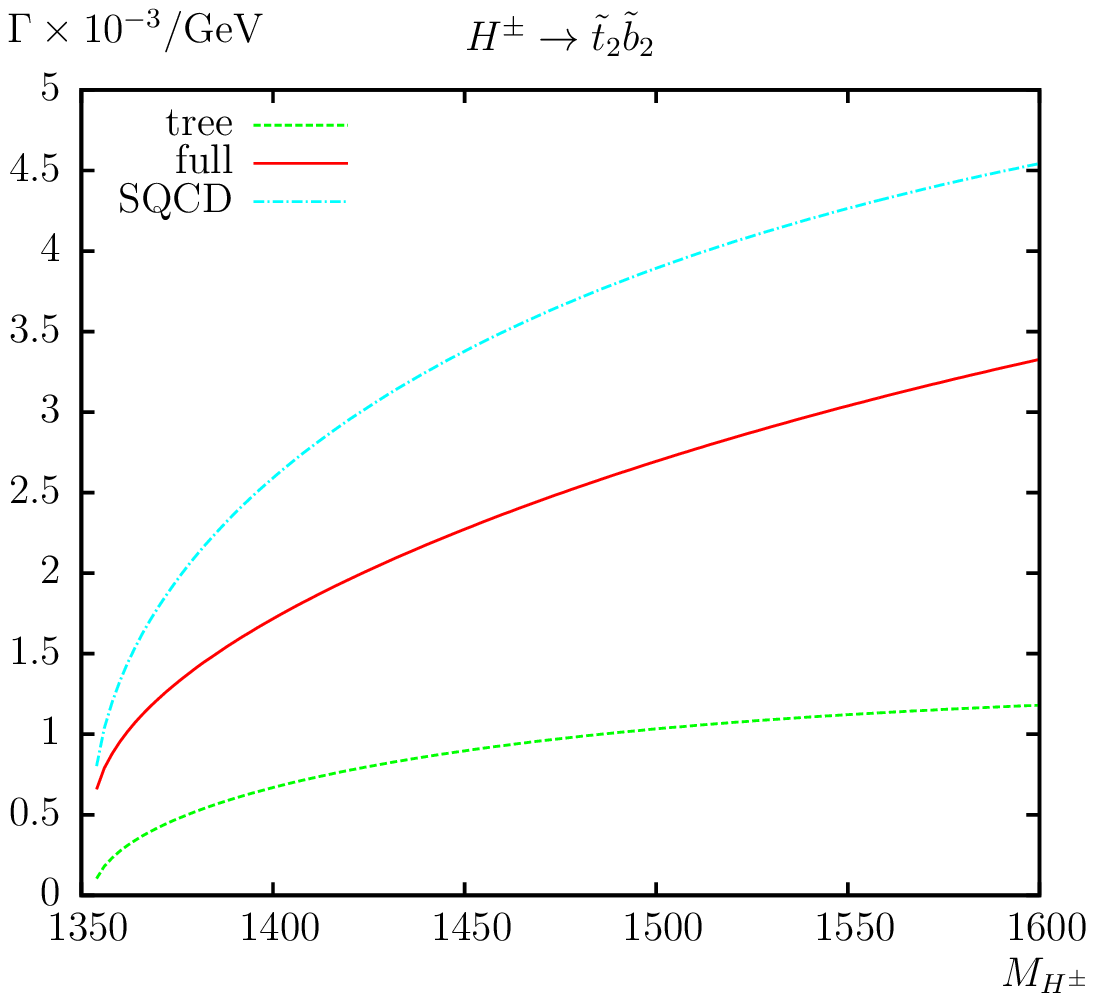}
%\hspace{-4mm}
%\includegraphics[width=0.49\textwidth,height=7.5cm]{} 
\\[4em]
\includegraphics[width=0.49\textwidth,height=7.5cm]{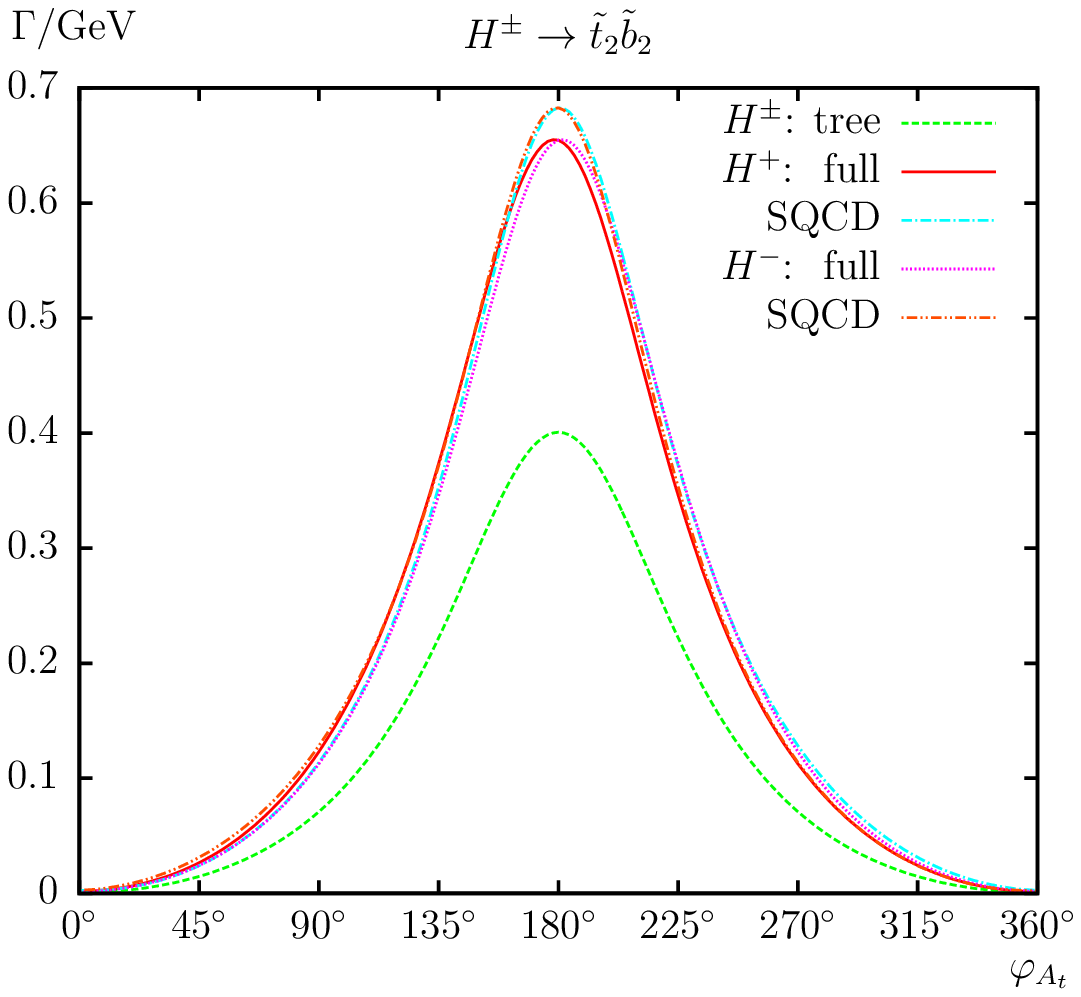}
\hspace{-4mm}
\includegraphics[width=0.49\textwidth,height=7.5cm]{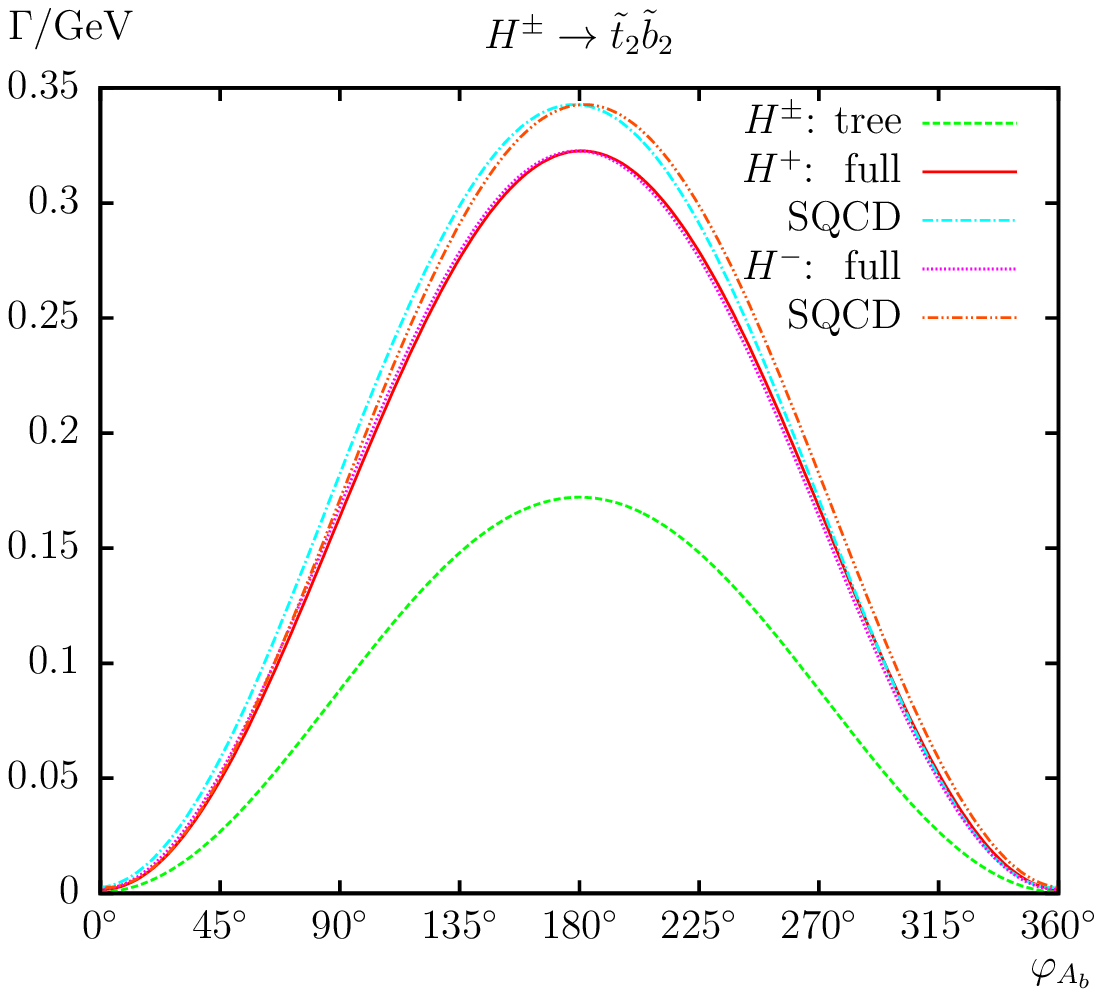}
\end{tabular}
\vspace{1em}
\caption{
  $\Ga(H^\pm \to \tilde{t}_2 \tilde{b}_2)$. 
  Tree-level, full and SQCD one-loop corrected partial decay widths are 
  shown.  The upper plot shows the partial decay width with $\MHp$ varied.
  The lower plots show the complex phases $\phiAt$ (left) and 
  $\phiAb$ (right) varied with parameters chosen according to \Scz\ 
  (see \refta{tab:para}).
}
\label{fig:Hpst2sb2}
\end{center}
\end{figure}

\begin{figure}[htb!]
\begin{center}
\begin{tabular}{c}
\includegraphics[width=0.49\textwidth,height=7.5cm]{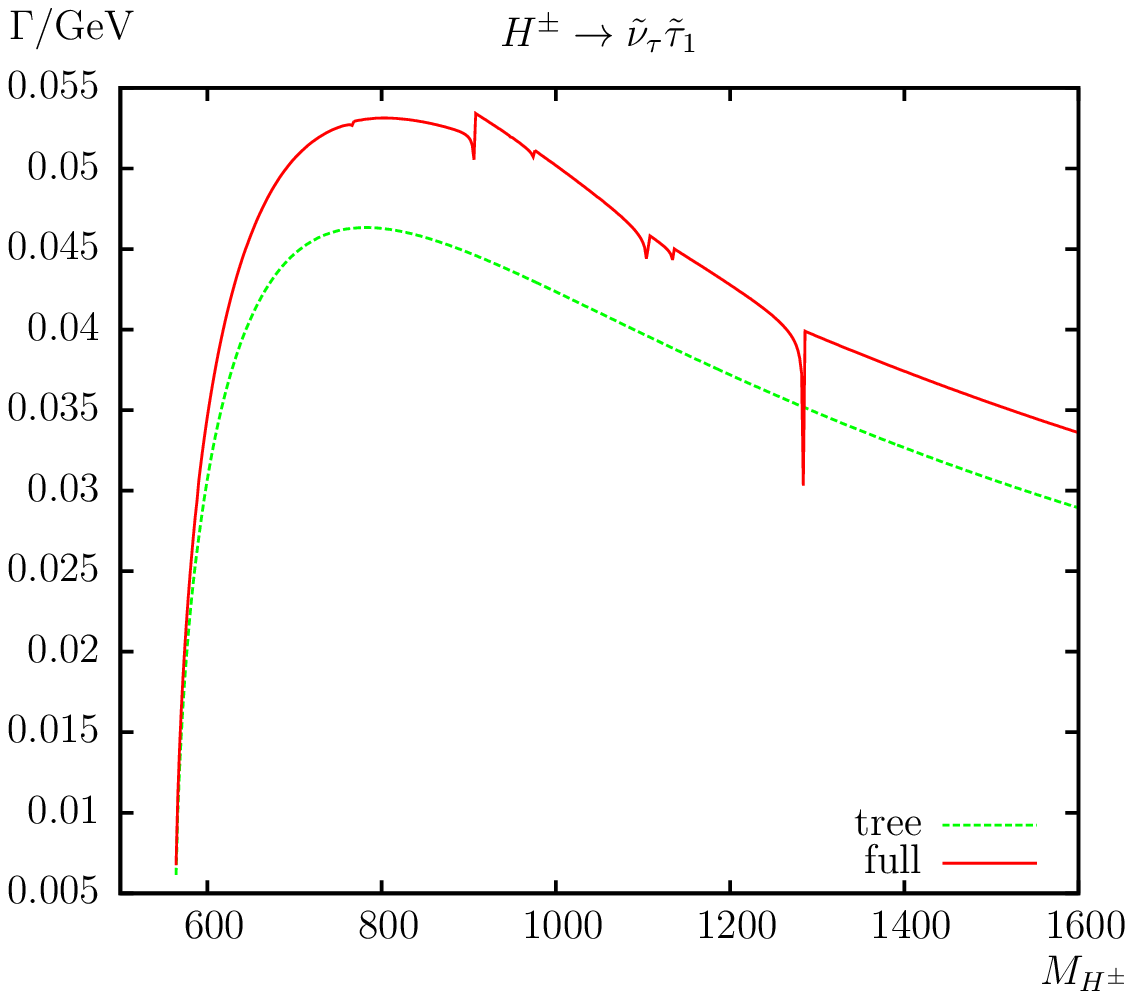}
\hspace{-4mm}
\includegraphics[width=0.49\textwidth,height=7.5cm]{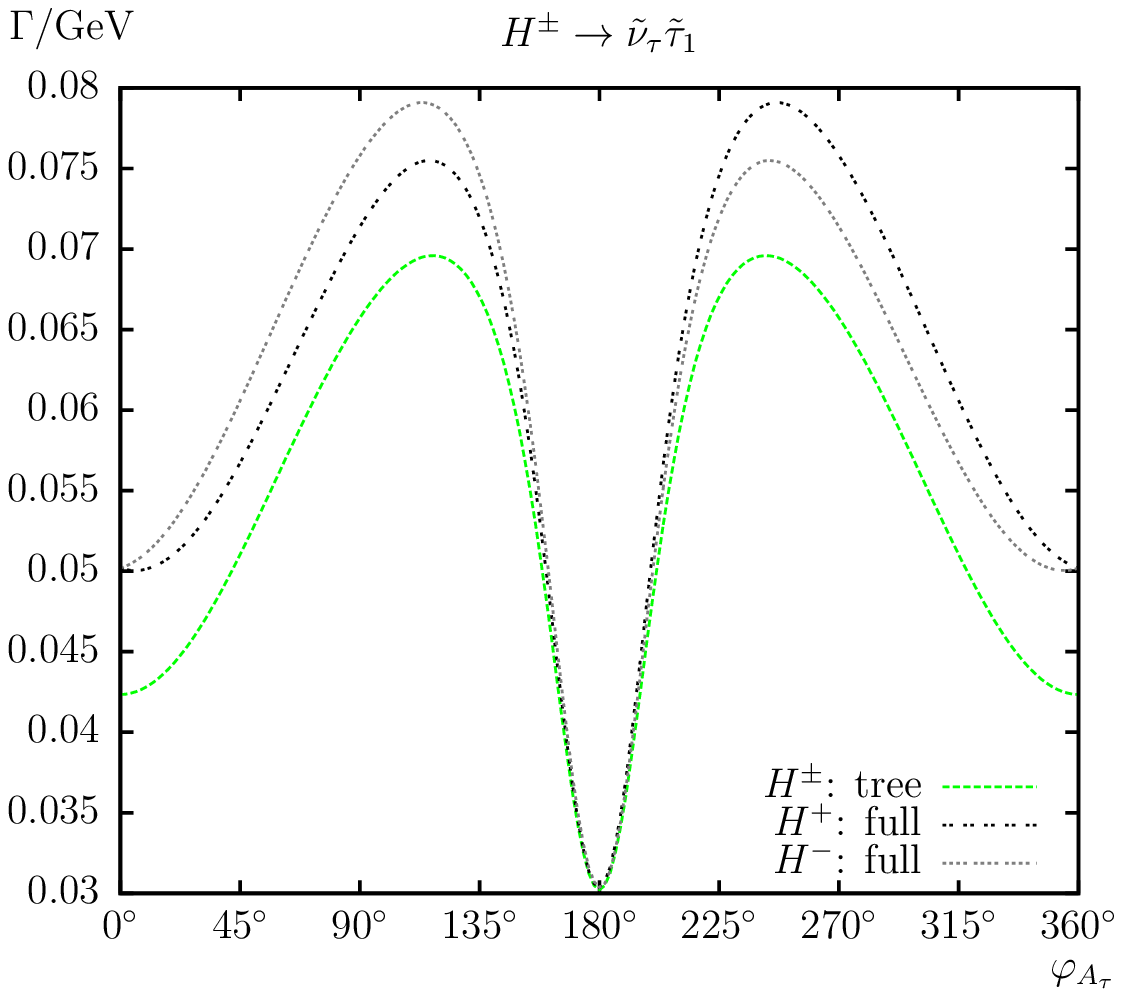}
\end{tabular}
\vspace{1em}
\caption{
  $\Ga(H^\pm \to \tilde{\nu}_{\tau} \tilde{\tau}_1)$. 
  Tree-level and full one-loop corrected partial decay widths are shown.
  The left plot shows the partial decay width with $\MHp$ varied; 
  the right plot shows the complex phase $\varphi_{A_{\tau}}$ varied with 
  parameters chosen according to \Sce\ (see \refta{tab:para}).
}
\label{fig:Hpsnstau1}
\end{center}
\end{figure}

\begin{figure}[htb!]
\begin{center}
\begin{tabular}{c}
\includegraphics[width=0.49\textwidth,height=7.5cm]{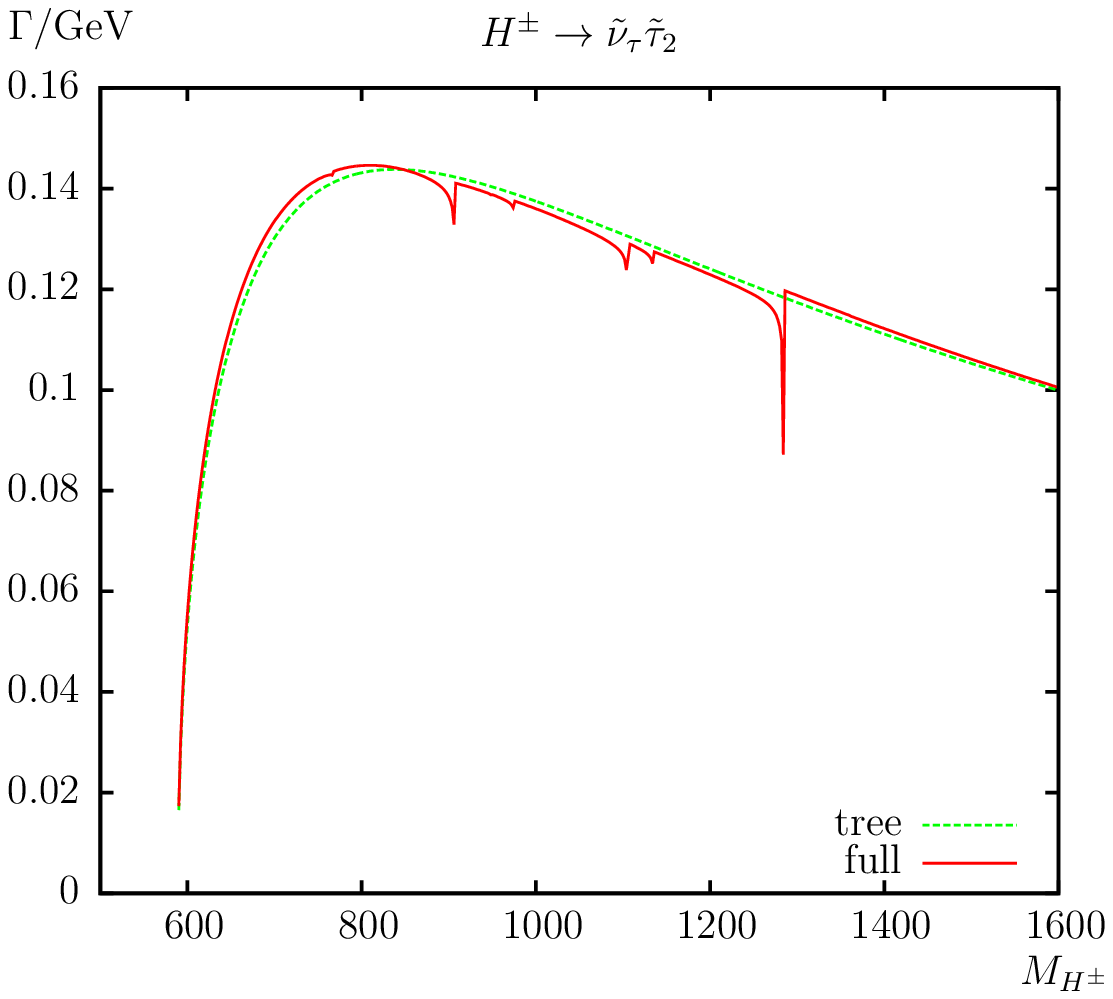}
\hspace{-4mm}
\includegraphics[width=0.49\textwidth,height=7.5cm]{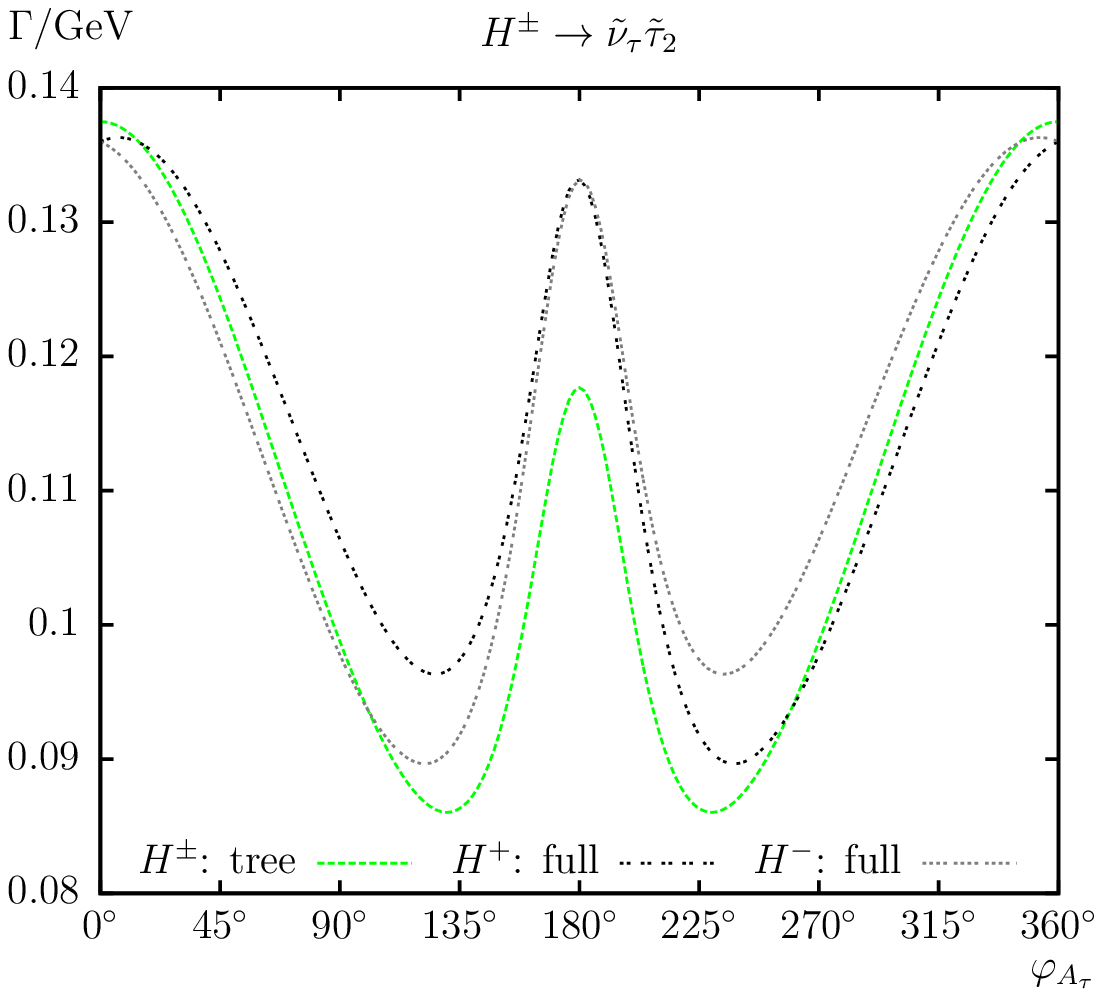}
\end{tabular}
\vspace{1em}
\caption{
  $\Ga(H^\pm \to \tilde{\nu}_{\tau} \tilde{\tau}_2)$. 
  Tree-level and full one-loop corrected partial decay widths are shown. 
  The left plot shows the partial decay width with $\MHp$ varied; 
  the right plot shows the complex phase $\varphi_{A_{\tau}}$ varied with 
  parameters chosen according to \Sce\ (see \refta{tab:para}).
}
\label{fig:Hpsnstau2}
\end{center}
\end{figure}

In \reffis{fig:Hpst1sb1} -- \ref{fig:Hpsnstau2} we show the results for 
the processes $\Hpmdecay$ ($i,j = 1,2$) as a function of 
$\MHp$ and as a function of the relevant complex phases $\phia$. 
These are of particular interest for LHC 
analyses~\cite{stopstophiggs-LHC,Higgsincascades} 
(as emphasized in \refse{sec:intro}). 

\medskip

We start with the decay $\Hpm \to \Stop1 \Sbot1$. 
In the upper plot of \reffi{fig:Hpst1sb1} the first dip (hardly visible) 
at $\MHp = 976\gev$ is an effect due to the threshold 
$\mstop1 + \msbot2 = \MHp$.  The second ``apparently single'' dip is 
in reality two dips at $\MHp \approx 1105, 1108\gev$ coming from the 
thresholds $\mcha1 + \mneu4 = \MHp$ and $\mcha2 + \mneu2 = \MHp$. 
The third dip at $\MHp \approx 1135\gev$ is the threshold 
$\mcha2 + \mneu3 = \MHp$ and the last one is the threshold 
$\mstop2 + \msbot1 = \MHp \approx 1284\gev$.  The size of the 
corrections of the  partial decay widths is especially large very close 
to the production threshold%
\footnote{
  It should be noted that a calculation very close to the production 
  threshold requires the inclusion of additional (nonrelativistic) 
  contributions, which is beyond the scope of this paper. 
  Consequently, very close to the production threshold our calculation 
  (at the tree- and loop-level) does not provide a very accurate 
  description of the decay width.
}
from which on the considered decay mode is kinematically possible. 
Away from this production threshold relative corrections of $\sim +23\%$ 
are found in \Sce\ (see \refta{tab:para}), 
of $\sim +5\%$ in \Scz\ and of $\sim +3\%$ in \Scd.  
The SQCD corrections are slightly larger, \ie the EW
corrections reduce the overall size of the loop corrections by 
$\sim 17\%$.
In the lower plots of \reffi{fig:Hpst1sb1} we show the complex phases
$\phiAtb$ varied at $\MHp = 1000\gev$. 
The tree-level dependence on the two phases is very different.  While for
negative $\At$ a reduction by nearly $50\%$ w.r.t.\ positive $\At$ is
found, negative $\Ab$ leads to an enhancement of about $25\%$.
The full corrections with $\phiAt$ varied are up to $\sim +29\%$
with slightly larger or lower values for the SQCD corrections by up to 
$\sim \pm 4\%$.
The asymmetry depending on $\phiAt$ is rather small.
$\phiAb$ varied can reach $\sim +27\%$ with slightly larger 
values for the SQCD corrections $\sim +31\%$.  Here the $\phiAb$ 
asymmetry is hardly visible.

\medskip

In \reffi{fig:Hpst1sb2} we show the results for $\Hpm \to \Stop1
\Sbot2$. In the upper plot the first ``apparently single'' 
dip is (again) in reality two dips at the thresholds 
$\mcha1 + \mneu4 = \MHp \approx 1105\gev$ and 
$\mcha2 + \mneu2 = \MHp \approx 1108\gev$. 
The second dip at $\MHp \approx 1135\gev$ is (again) the threshold 
$\mcha2 + \mneu3 = \MHp$ and the last (large) one is (again) the 
threshold $\mstop2 + \msbot1 = \MHp \approx 1284\gev$.
Relative corrections of $\sim +8\%$ are found at $\MHp = 1400\gev$ in 
\Scz\ (see \refta{tab:para}) (and $\sim +3\%$ at $\MHp = 1600\gev$ 
in \Scd). The SQCD corrections alone would lead to
an increase of $\sim +21\%$ in \Scz\ ($\sim +13\%$ in \Scd), \ie
they overestimate the full corrections by roughly a factor of three.
In the lower plots of \reffi{fig:Hpst1sb2} the results are shown for
\Scz\ as a function of $\phiAtb$.
One can see that the size of the corrections to the partial decay width
vary substantially with  
the complex phases $\phiAtb$ at $\MHp = 1400\gev$.
At $\phiAt = 180^\circ$ the full corrections reach $\sim +23\%$, 
while the SQCD corrections are much larger $\sim +77\%$.
At $\phiAb = 90^\circ$ the $H^+$ ($H^-$) full corrections reach 
$\sim +55\%$ ($\sim -22\%$), while the SQCD corrections are 
$\sim +38\%$ ($\sim +8\%$).%
\footnote{
  It should be noted that at $\phiAb \approx 180^\circ$ the loop 
  corrections can be larger then the tree results because there 
  the tree level decay width is accidently small, see the 
  lower right plot of \reffi{fig:Hpst1sb2}. 
}

\medskip

Next, in \reffi{fig:Hpst2sb1} the results for $\Hpm \to \Stop2 \Sbot1$
are displayed.  In the upper plot the results are shown as a function 
of $\MHp$.  Relative corrections of $\sim +27\%$ are found at 
$\MHp = 1400\gev$ (see \refta{tab:para}).  In this case the EW 
corrections hardly contribute to the overall one-loop contribution.
In the lower plots the results are displayed as a function of $\phiAtb$
in \Scz.  In the left plot one can see that the size of the corrections 
to the partial decay width vary substantially with the complex phase 
$\phiAt$.  For all $\phiAt$ the full and SQCD corrections are of similar
size and deviate between $+9\%$ and $+27\%$.  The same 
holds for $\phiAb$ with small differences between the full and SQCD
corrections, which vary only between $+25\%$ and $+27\%$.
Here the asymmetries are extremely small and hardly visible.

\medskip

The decay $\Hpm \to \Stop2 \Sbot2$ is shown in \reffi{fig:Hpst2sb2}.
The overall size of this decay width (with real phases) is
(accidentally) very small around $\sim 2 \times 10^{-3} \gev$.
Consequently, the loop corrections, as shown in the upper plot, can be
larger than the tree-level result. The SQCD corrections alone would
overestimate the full result by about $\sim 50\%$.
In the lower plots of \reffi{fig:Hpst2sb2} one can see that the size 
of the tree-level result depends again strongly on the two
phases. Values of $\sim 0.4\ (0.16) \gev$ are reached for negative
$\phiAt\ (\phiAb)$. Again the loop corrections can be substantial.
At $\phiAt = 180^\circ$ the full corrections reach $\sim +63\%$, 
while the SQCD corrections are larger $\sim +72\%$.
At $\phiAb = 180^\circ$ the full corrections reach $\sim +87\%$, 
while the SQCD corrections are up to $\sim +90\%$.  The asymmetries 
are found to be rather small.

\medskip

Now we turn to the charged Higgs boson decays to scalar leptons, 
$\Hpm \to \Snutau \Stau1$ in \reffi{fig:Hpsnstau1} and 
$\Hpm \to \Snutau \Stau2$ in \reffi{fig:Hpsnstau2}. The left plots show
the results as a function of $\MHp$, while the right plots analyze the
dependence on $\phiAtau$ for $\MHp = 1000 \gev$. 
In the left plot of \reffi{fig:Hpsnstau1} and \reffi{fig:Hpsnstau2} 
the first dip (hardly visible) at $\MHp \approx 768\gev$ is an effect 
due to the threshold $\mcha1 + \mneu1 = \MHp$.  The second dip  
stem from the threshold $\mstop1 + \msbot1 = \MHp \approx 907\gev$.
The remaining dips at $\MHp \approx 976,1105,1108,1135,1284\gev$ are the 
same thresholds as in \reffi{fig:Hpst1sb1} (as discussed above).
At $\MHp = 1000\gev$ one-loop corrections of $\sim +19\%$ are found for
$H^\pm \to \Snutau \Stau1$, while for $H^\pm \to \Snutau \Stau2$ they
are only $\sim -1\%$. 
The maximum values of the full one-loop corrections as a function of
$\phiAtau$ reach $\sim +15\%\ (+13\%)$ for 
$\Hpm \to \Snutau \Stau1 (\Snutau \Stau2)$. 
The asymmetries in the decays of a negative charged Higgs w.r.t.\ a
positively charged Higgs are substantial. The size of the respective
loop corrections can nearly be twice as large in one case w.r.t.\ to the
other, depending whether $\phiAtau \le 180^\circ$ or 
$\phiAtau \ge 180^\circ$ is considered.

\clearpage
\newpage

%%%%%%%%%%%%%%%%%%%%%%%%%% F I G U R E %%%%%%%%%%%%%%%%%%%%%%%%%%%%%%%%%%%%%%%%%
%%%%%%%%%%%%%%%%%%%%%%%%%% F I G U R E %%%%%%%%%%%%%%%%%%%%%%%%%%%%%%%%%%%%%%%%%

%\newpage

%%%%%%%%%%%%%%%%%%%%%%%%%% F I G U R E %%%%%%%%%%%%%%%%%%%%%%%%%%%%%%%%%%%%%%%%%
%%%%%%%%%%%%%%%%%%%%%%%%%% F I G U R E %%%%%%%%%%%%%%%%%%%%%%%%%%%%%%%%%%%%%%%%%

%\newpage

%%%%%%%%%%%%%%%%%%%%%%%%%% F I G U R E %%%%%%%%%%%%%%%%%%%%%%%%%%%%%%%%%%%%%%%%%
%%%%%%%%%%%%%%%%%%%%%%%%%% F I G U R E %%%%%%%%%%%%%%%%%%%%%%%%%%%%%%%%%%%%%%%%%

%\newpage

%%%%%%%%%%%%%%%%%%%%%%%%%% F I G U R E %%%%%%%%%%%%%%%%%%%%%%%%%%%%%%%%%%%%%%%%%
%%%%%%%%%%%%%%%%%%%%%%%%%% F I G U R E %%%%%%%%%%%%%%%%%%%%%%%%%%%%%%%%%%%%%%%%%

%\newpage

%%%%%%%%%%%%%%%%%%%%%%%%%% F I G U R E %%%%%%%%%%%%%%%%%%%%%%%%%%%%%%%%%%%%%%%%%
%%%%%%%%%%%%%%%%%%%%%%%%%% F I G U R E %%%%%%%%%%%%%%%%%%%%%%%%%%%%%%%%%%%%%%%%%

\vspace{8mm}

%%%%%%%%%%%%%%%%%%%%%%%%%% F I G U R E %%%%%%%%%%%%%%%%%%%%%%%%%%%%%%%%%%%%%%%%%
%%%%%%%%%%%%%%%%%%%%%%%%%% F I G U R E %%%%%%%%%%%%%%%%%%%%%%%%%%%%%%%%%%%%%%%%%

%%%%%%%%%%%%%%%%%%%%%%%%%%%%%%%%%%%%%%%%%%%%%%%%%%%%%%%%%%%%%%%%%%%%%%%%%%%%%%%

\clearpage
\newpage

\subsubsection{\boldmath{$h_n$} decays into sfermions}
\label{hndecays}

\begin{figure}[htb!]
\begin{center}
\begin{tabular}{c}
\includegraphics[width=0.49\textwidth,height=7.5cm]{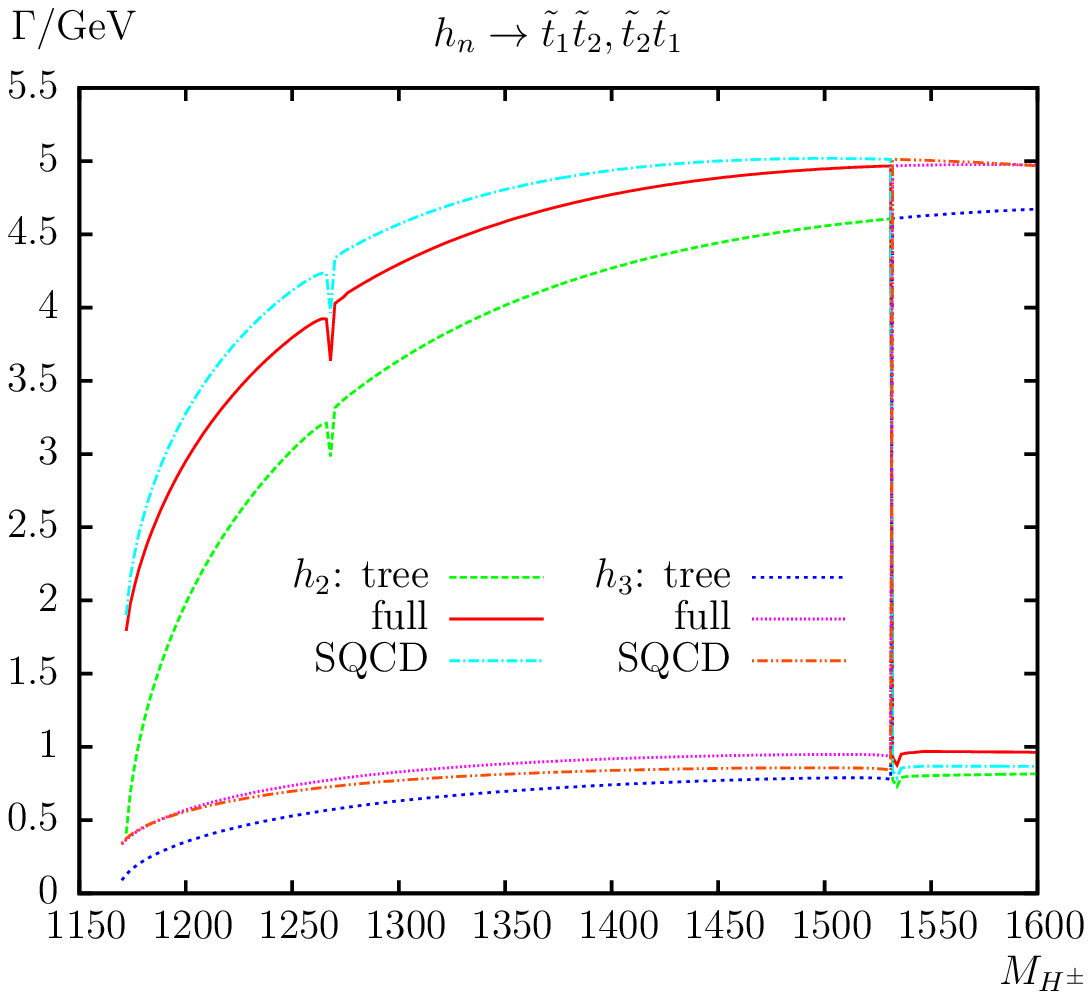}
\\[4em]
\includegraphics[width=0.49\textwidth,height=7.5cm]{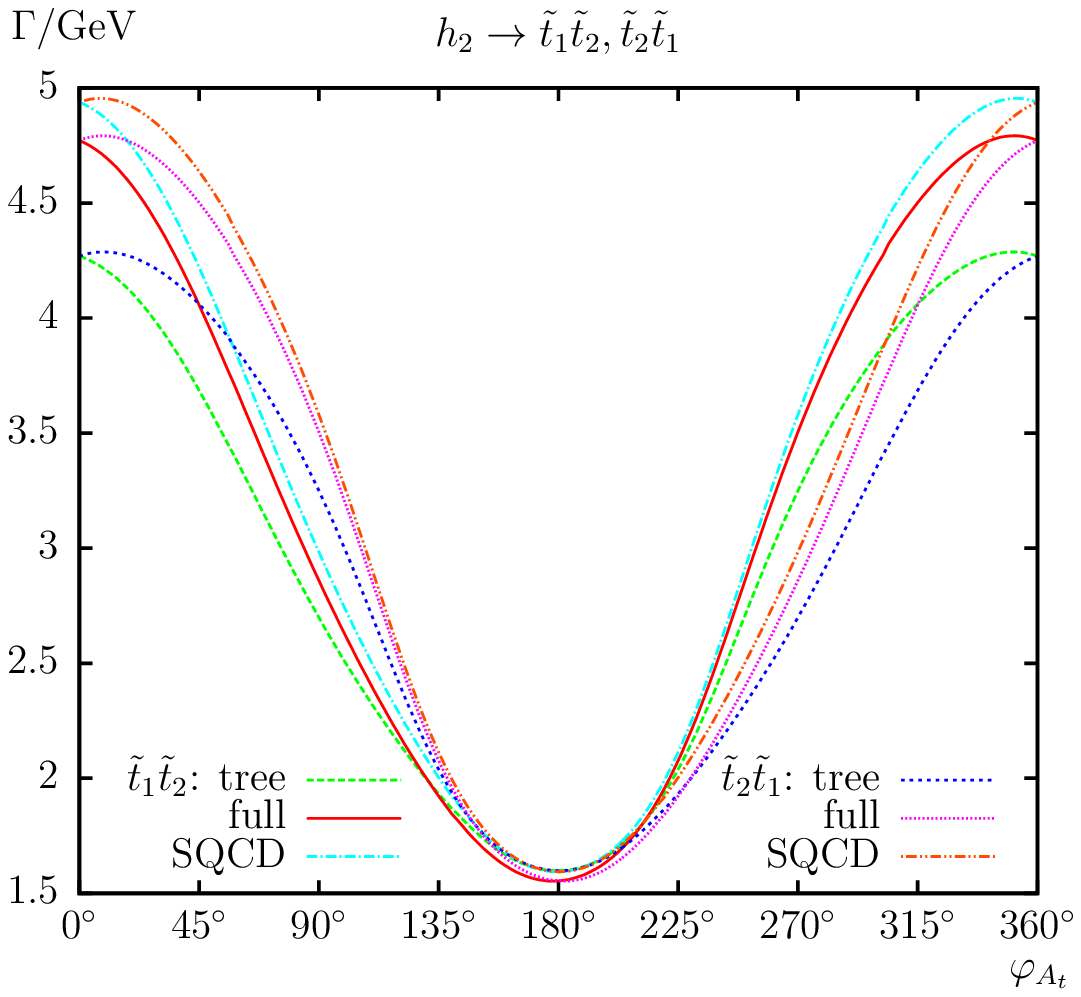}
\hspace{-4mm}
\includegraphics[width=0.49\textwidth,height=7.5cm]{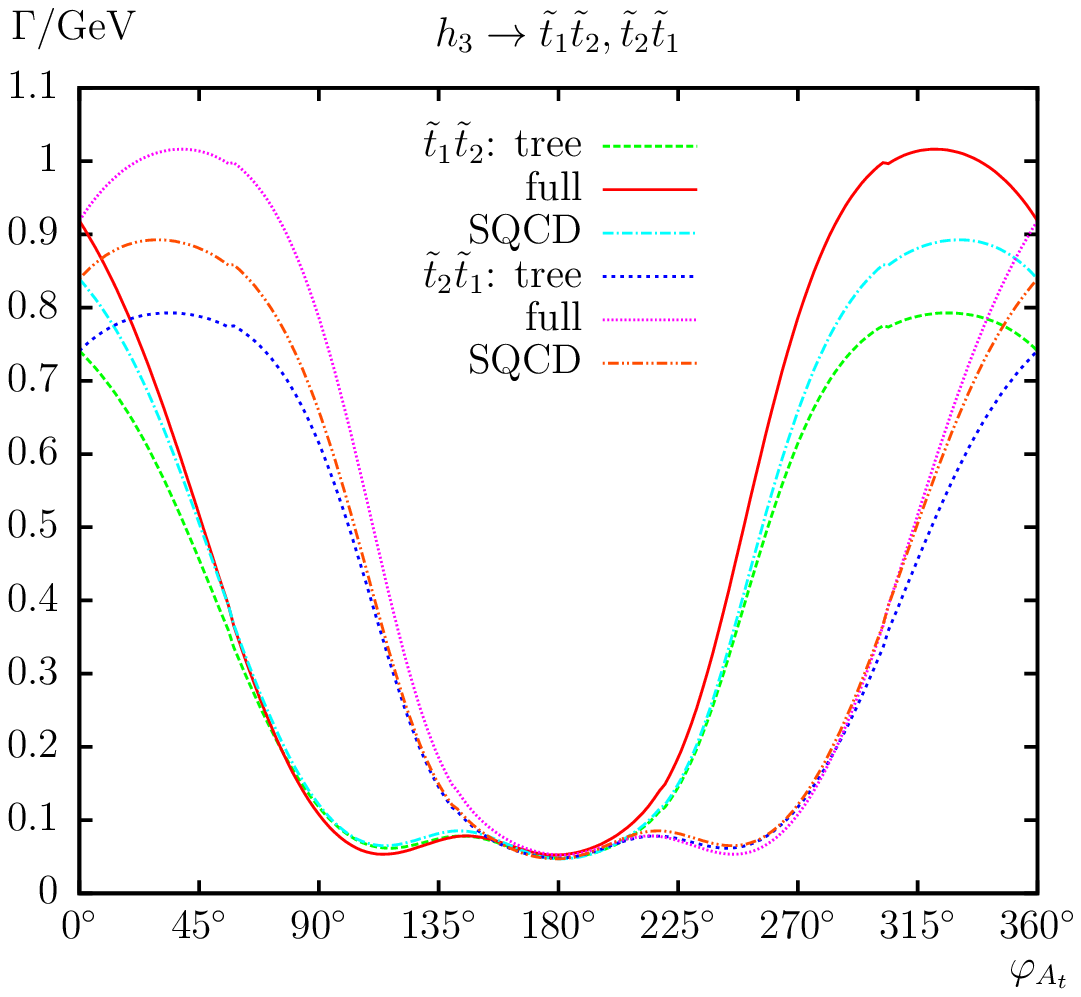}
\end{tabular}
\vspace{1em}
\caption{
  $\Ga(h_n \to \Stop1 \Stop2, \Stop2 \Stop1)$. 
  Tree-level, full and SQCD one-loop corrected partial decay widths are 
  shown.  The upper plot shows the partial decay width with $\MHp$ varied; 
  the lower plots show the complex phase $\phiAt$ varied for $h_2$
  decays (left) and $h_3$ decays (right) with parameters chosen according 
  to \Scz\ (see \refta{tab:para}).
}
\label{fig:hnst1st2}
\end{center}
\end{figure}

\begin{figure}[htb!]
\begin{center}
\begin{tabular}{c}
\includegraphics[width=0.49\textwidth,height=7.5cm]{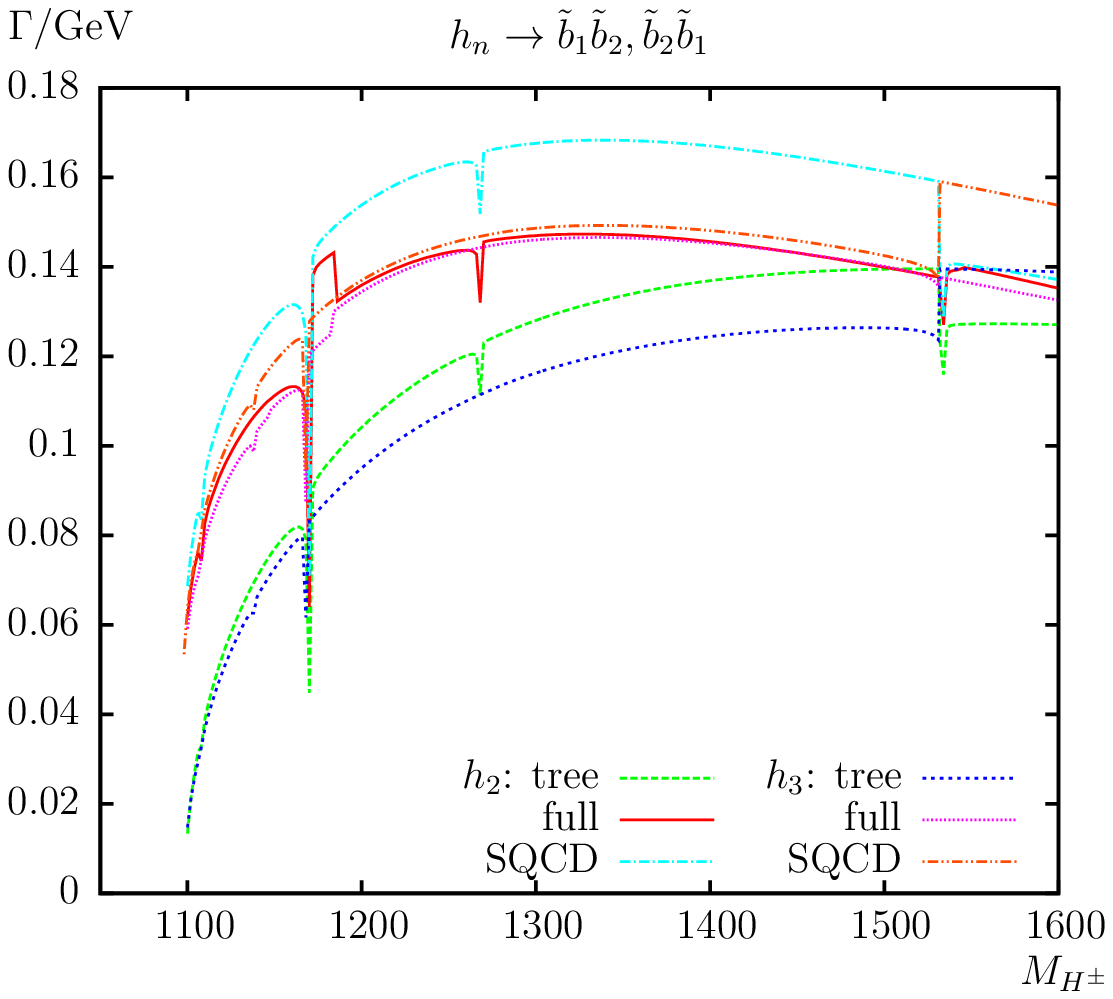}
\\[4em]
\includegraphics[width=0.49\textwidth,height=7.5cm]{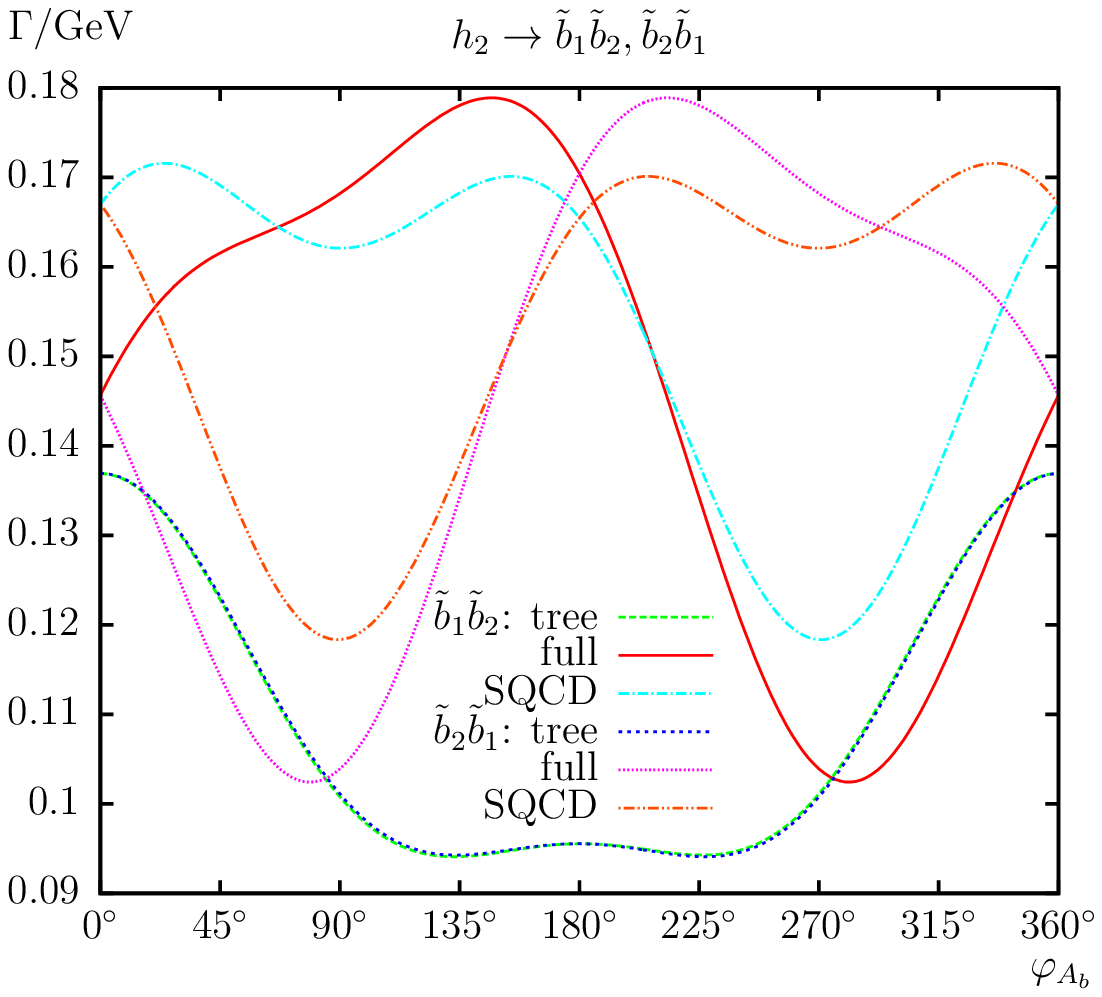}
\hspace{-4mm}
\includegraphics[width=0.49\textwidth,height=7.5cm]{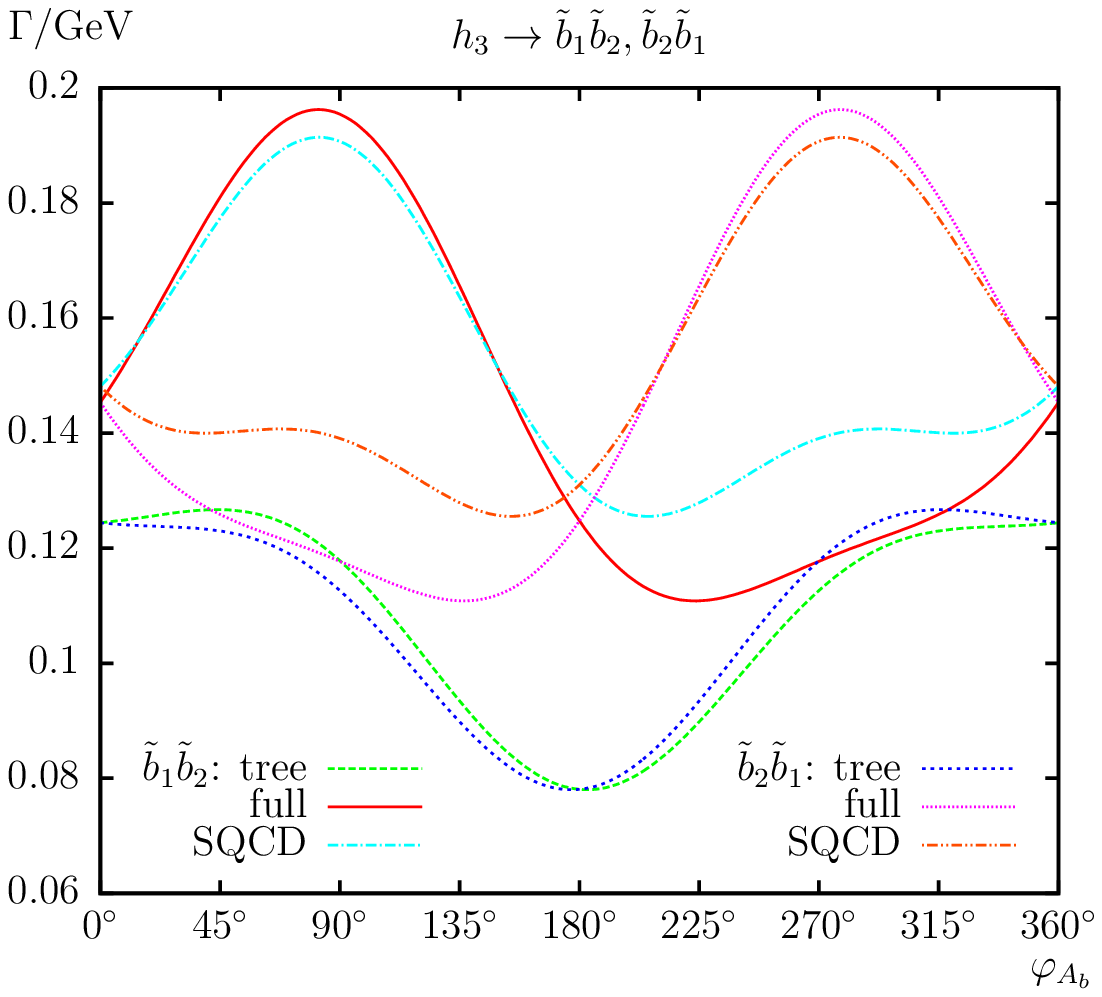}
\end{tabular}
\vspace{1em}
\caption{
  $\Ga(h_n \to \Sbot1 \Sbot2, \Sbot2 \Sbot1)$. 
  Tree-level, full and SQCD one-loop corrected partial decay widths are 
  shown.  The upper plot shows the partial decay width with $\MHp$ varied; 
  the lower plots show the complex phase $\phiAb$ varied for $h_2$
  decays (left) and $h_3$ decays (right) with parameters chosen according 
  to \Scz\ (see \refta{tab:para}). 
}
\label{fig:hnsb1sb2}
\end{center}
\end{figure}

\begin{figure}[htb!]
\begin{center}
\begin{tabular}{c}
\includegraphics[width=0.49\textwidth,height=7.5cm]{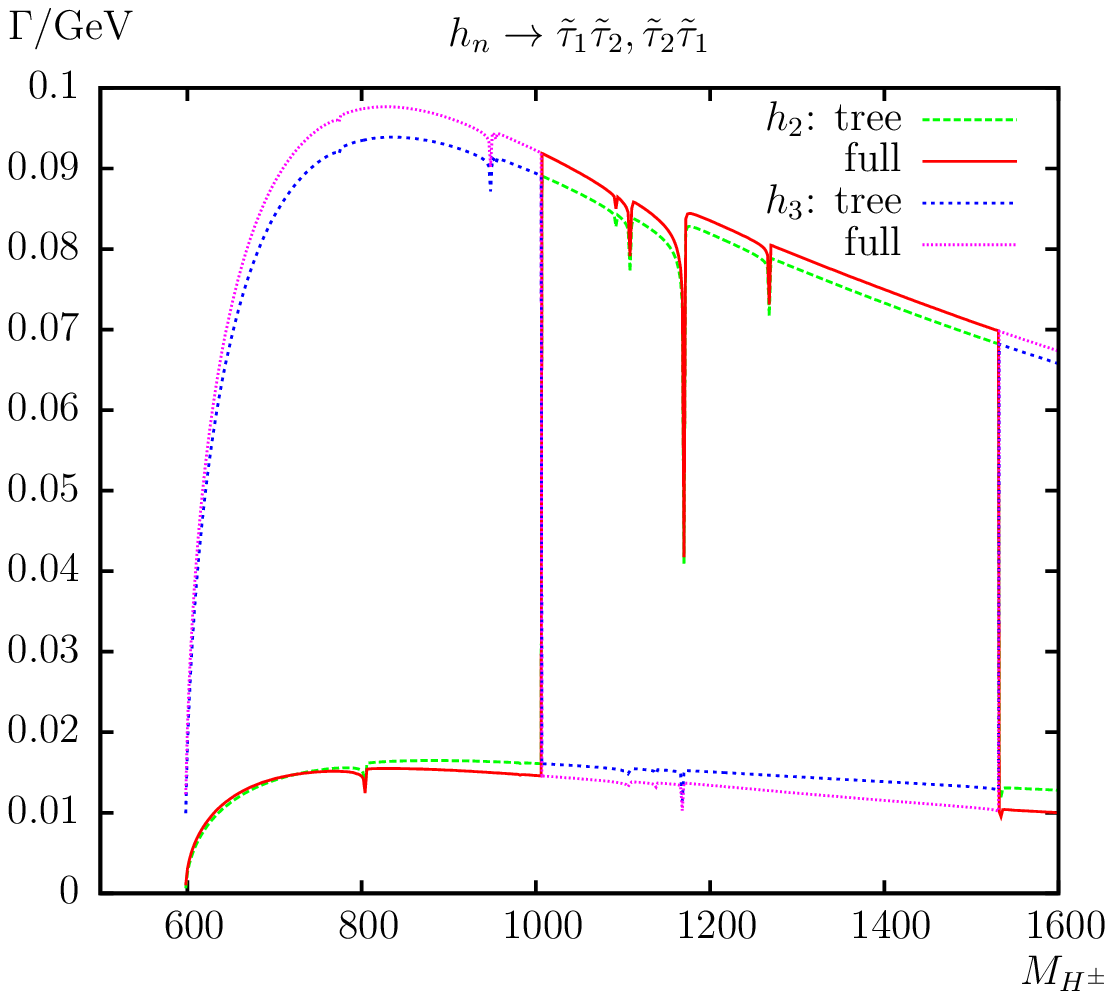}
\\[4em]
\includegraphics[width=0.49\textwidth,height=7.5cm]{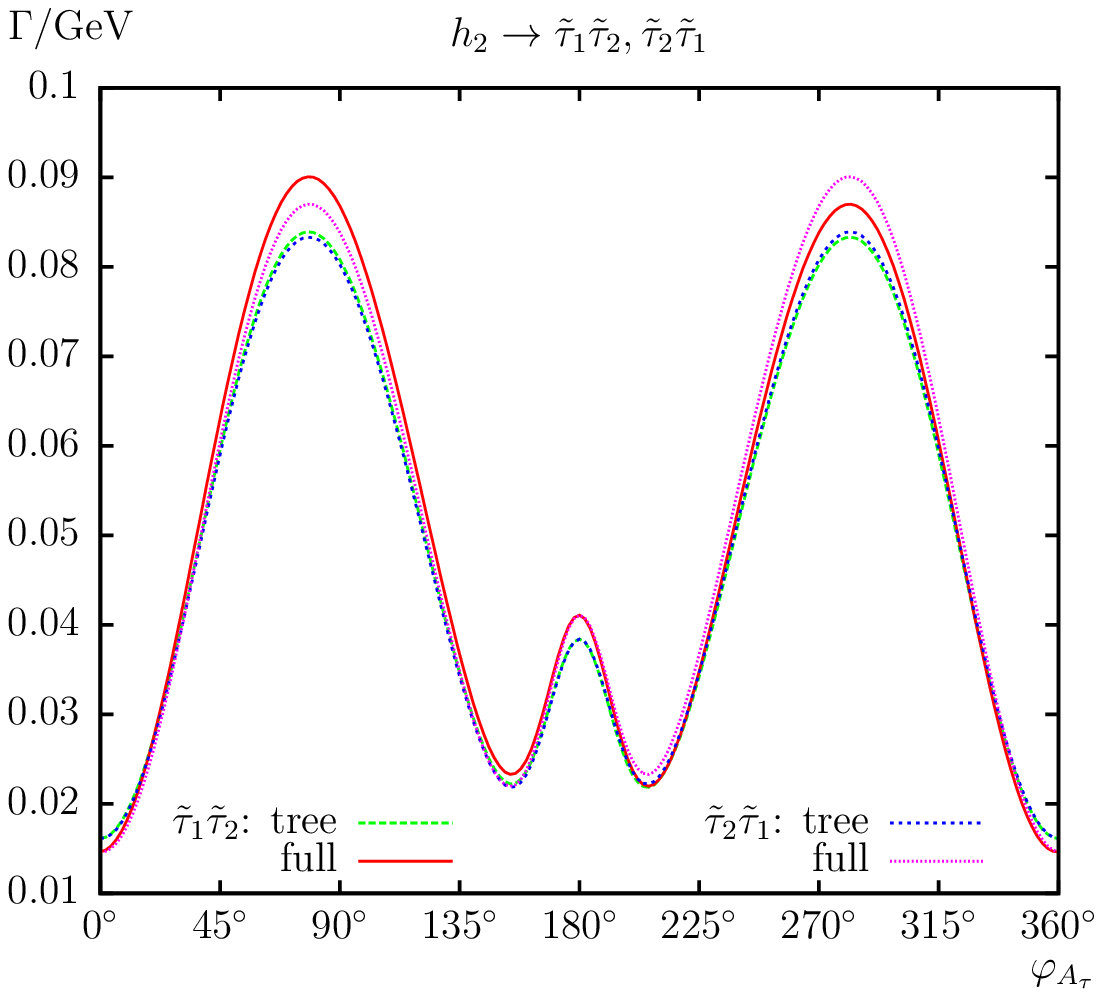}
\hspace{-4mm}
\includegraphics[width=0.49\textwidth,height=7.5cm]{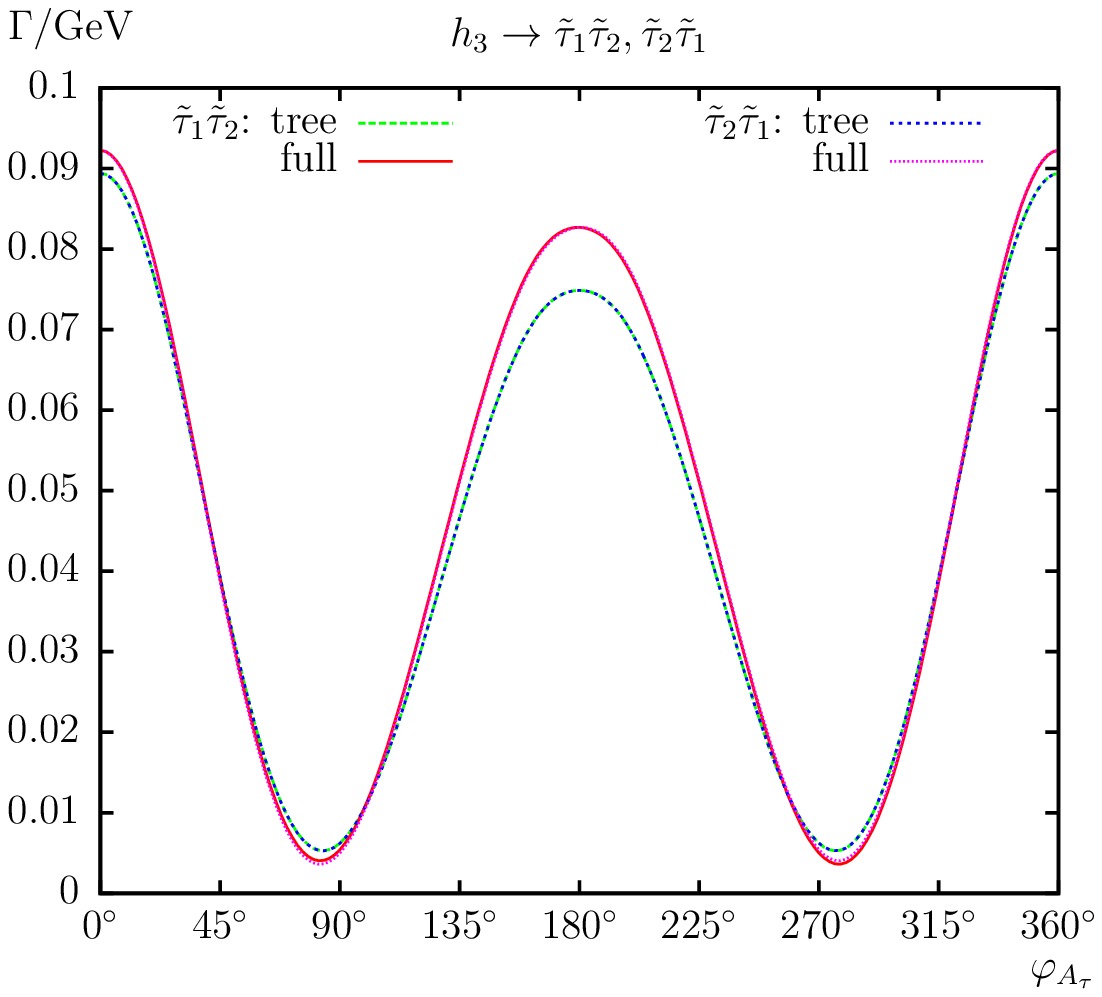}
\end{tabular}
\vspace{1em}
\caption{
  $\Ga(h_n \to \Stau1 \Stau2, \Stau2 \Stau1)$. 
  Tree-level and full one-loop corrected partial decay widths are 
  shown.  The upper plot shows the partial decay width with $\MHp$ varied; 
  the lower plots show the complex phase $\phiAtau$ varied for $h_2$
  decays (left) and $h_3$ decays (right) with parameters chosen according 
  to \Sce\ (see \refta{tab:para}).
}
\label{fig:hnstau1stau2}
\end{center}
\end{figure}

\begin{figure}[htb!]
\begin{center}
\begin{tabular}{c}
\includegraphics[width=0.49\textwidth,height=7.5cm]{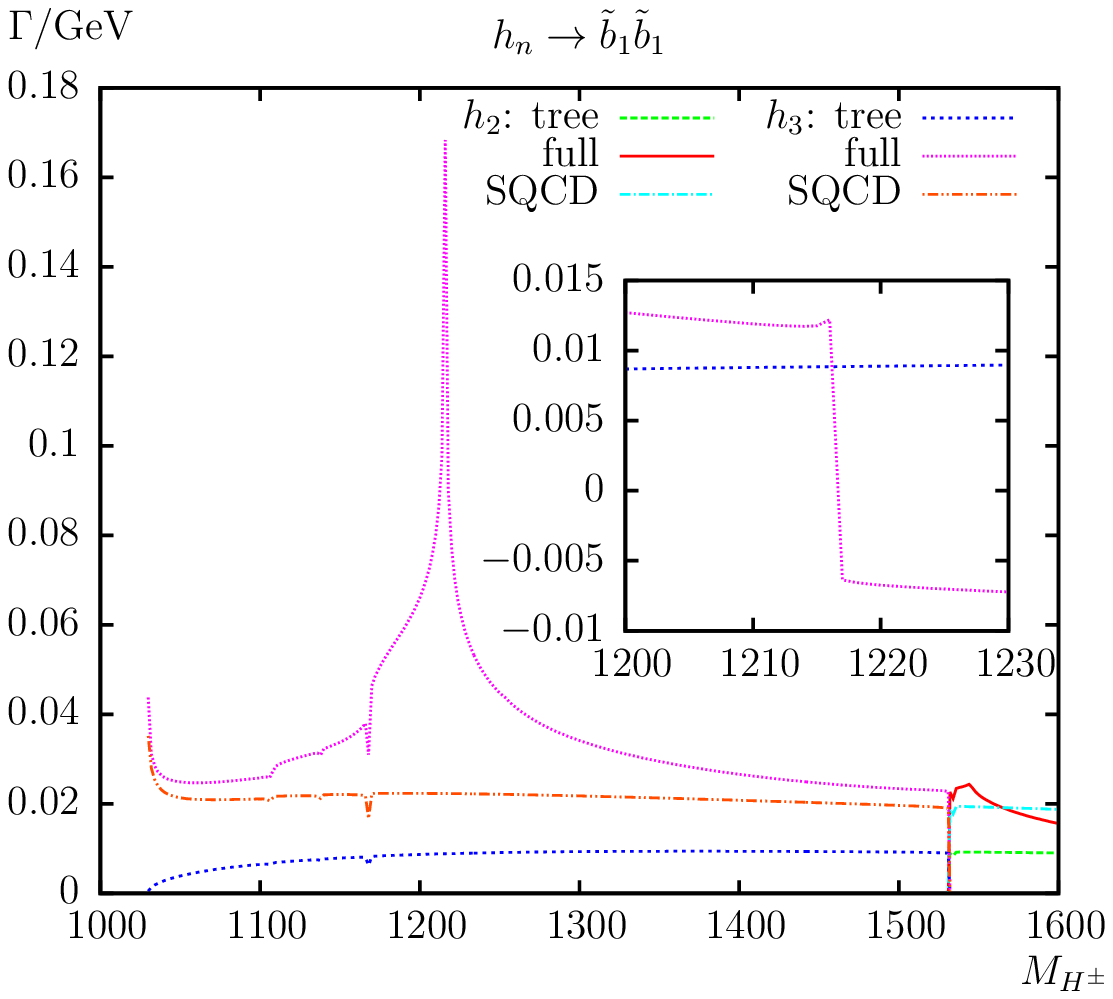}
\\[4em]
\includegraphics[width=0.49\textwidth,height=7.5cm]{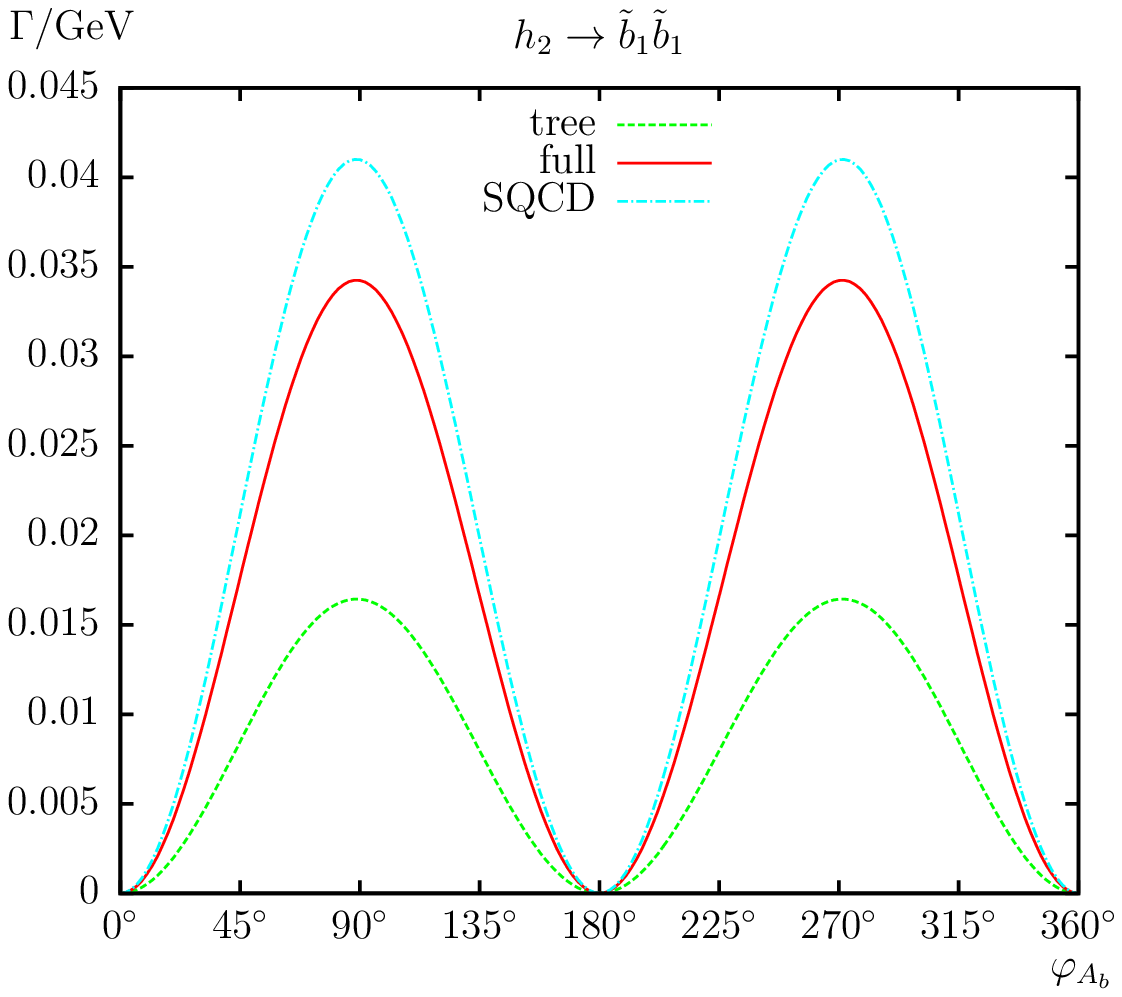}
\hspace{-4mm}
\includegraphics[width=0.49\textwidth,height=7.5cm]{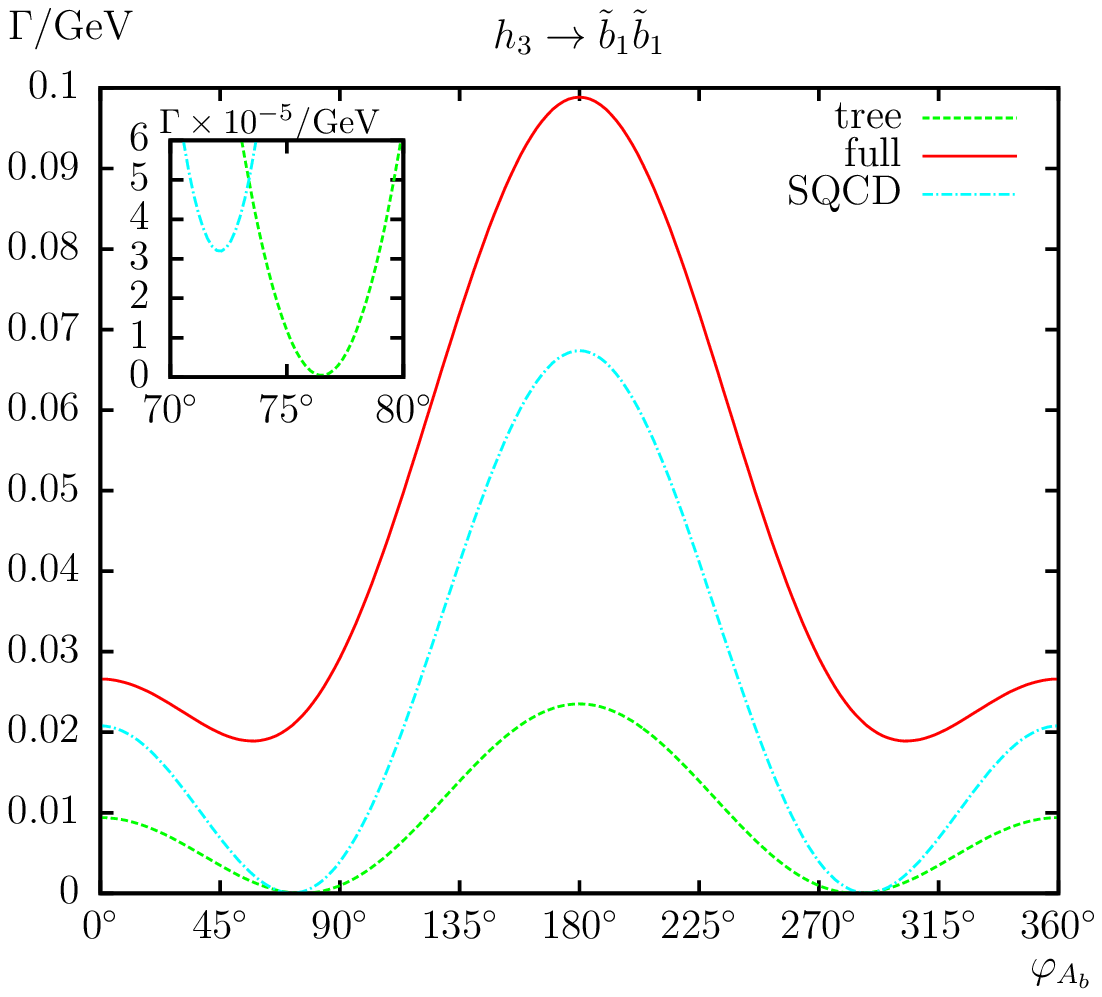}
\end{tabular}
\vspace{1em}
\caption{
  $\Ga(h_n \to \Sbot1 \Sbot1)$. 
  Tree-level, full and SQCD one-loop corrected partial decay widths are 
  shown.  The upper plot shows the partial decay width with $\MHp$ varied; 
  the lower plots show the complex phase $\phiAb$ varied for $h_2$
  decays (left) and $h_3$ decays (right) with parameters chosen according 
  to \Scz\ (see \refta{tab:para}).
}
\label{fig:hnsb1sb1}
\end{center}
\end{figure}

\begin{figure}[htb!]
\begin{center}
\begin{tabular}{c}
\includegraphics[width=0.49\textwidth,height=7.5cm]{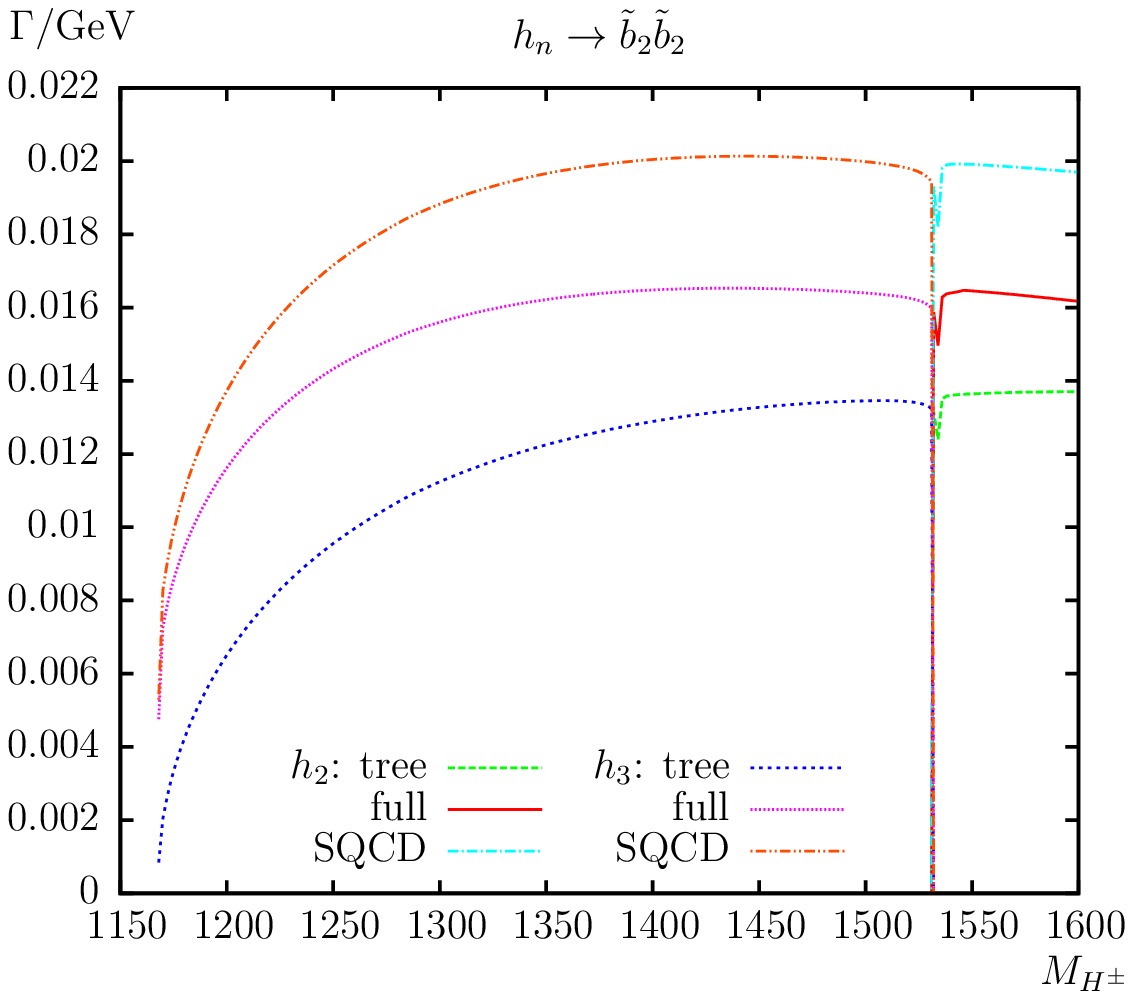}
\\[4em]
\includegraphics[width=0.49\textwidth,height=7.5cm]{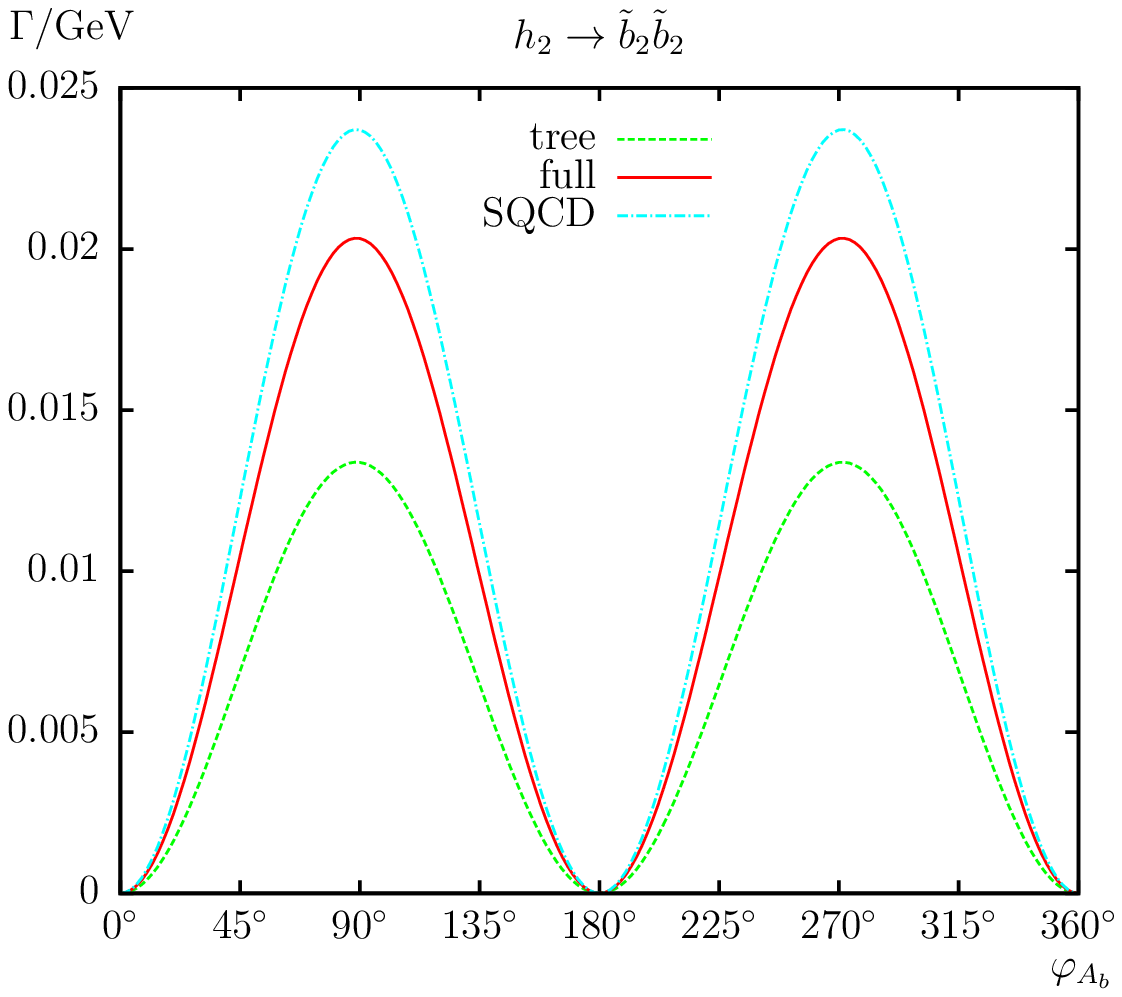}
\hspace{-4mm}
\includegraphics[width=0.49\textwidth,height=7.5cm]{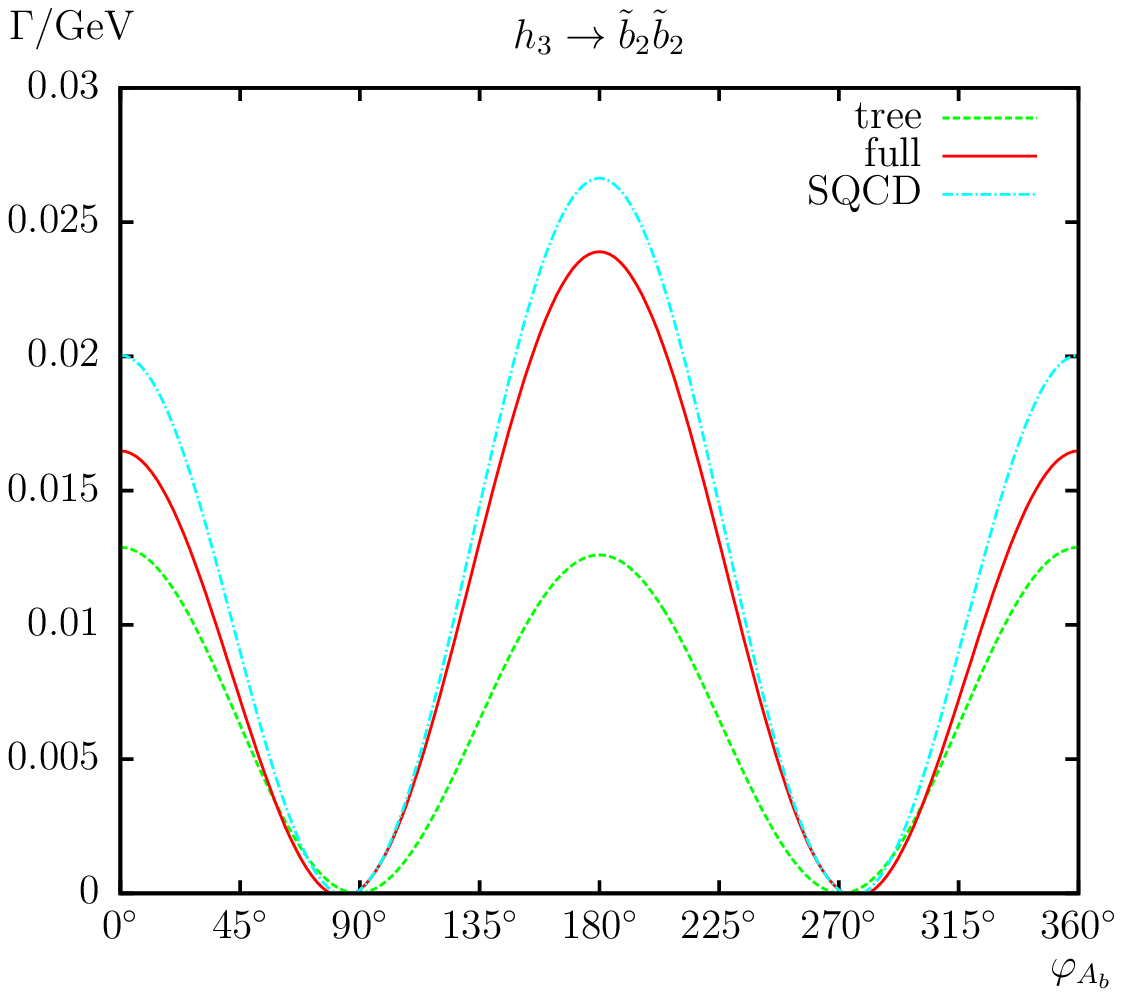}
\end{tabular}
\vspace{1em}
\caption{
  $\Ga(h_n \to \Sbot2 \Sbot2)$. 
  Tree-level, full and SQCD one-loop corrected partial decay widths are 
  shown.  The upper plot shows the partial decay width with $\MHp$ varied; 
  the lower plots show the complex phase $\phiAb$ varied for $h_2$
  decays (left) and $h_3$ decays (right) with parameters chosen according 
  to \Scz\ (see \refta{tab:para}).
}
\label{fig:hnsb2sb2}
\end{center}
\end{figure}

\begin{figure}[htb!]
\begin{center}
\begin{tabular}{c}
\includegraphics[width=0.49\textwidth,height=7.5cm]{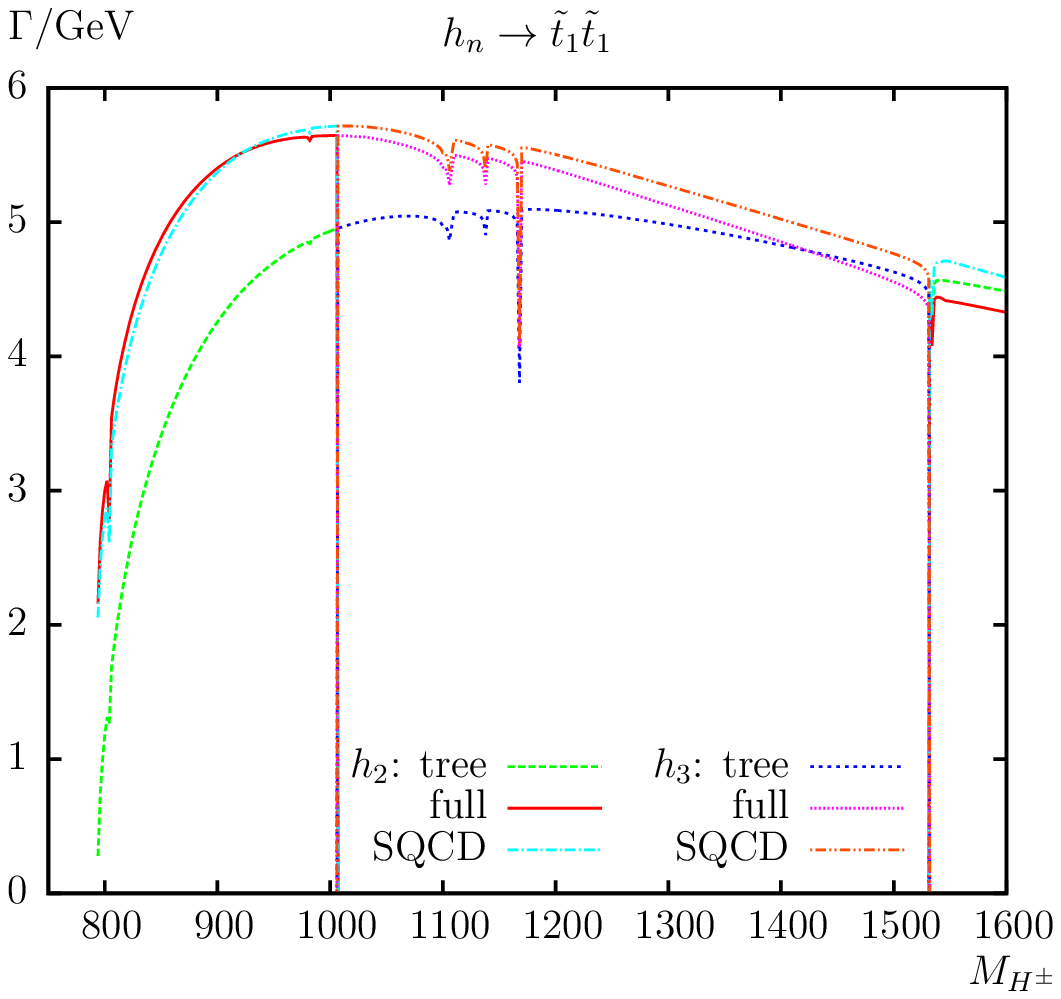}
\\[4em]
\includegraphics[width=0.49\textwidth,height=7.5cm]{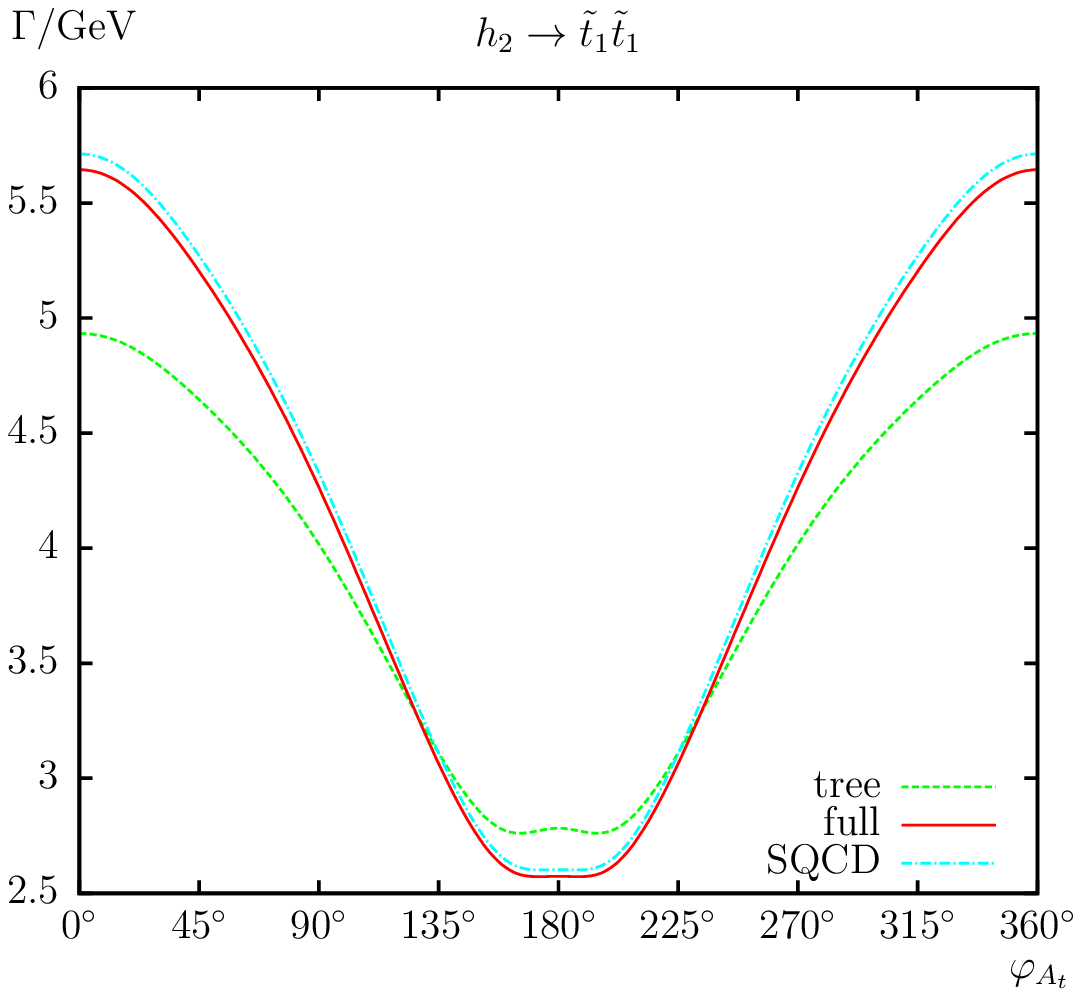}
\hspace{-4mm}
\includegraphics[width=0.49\textwidth,height=7.5cm]{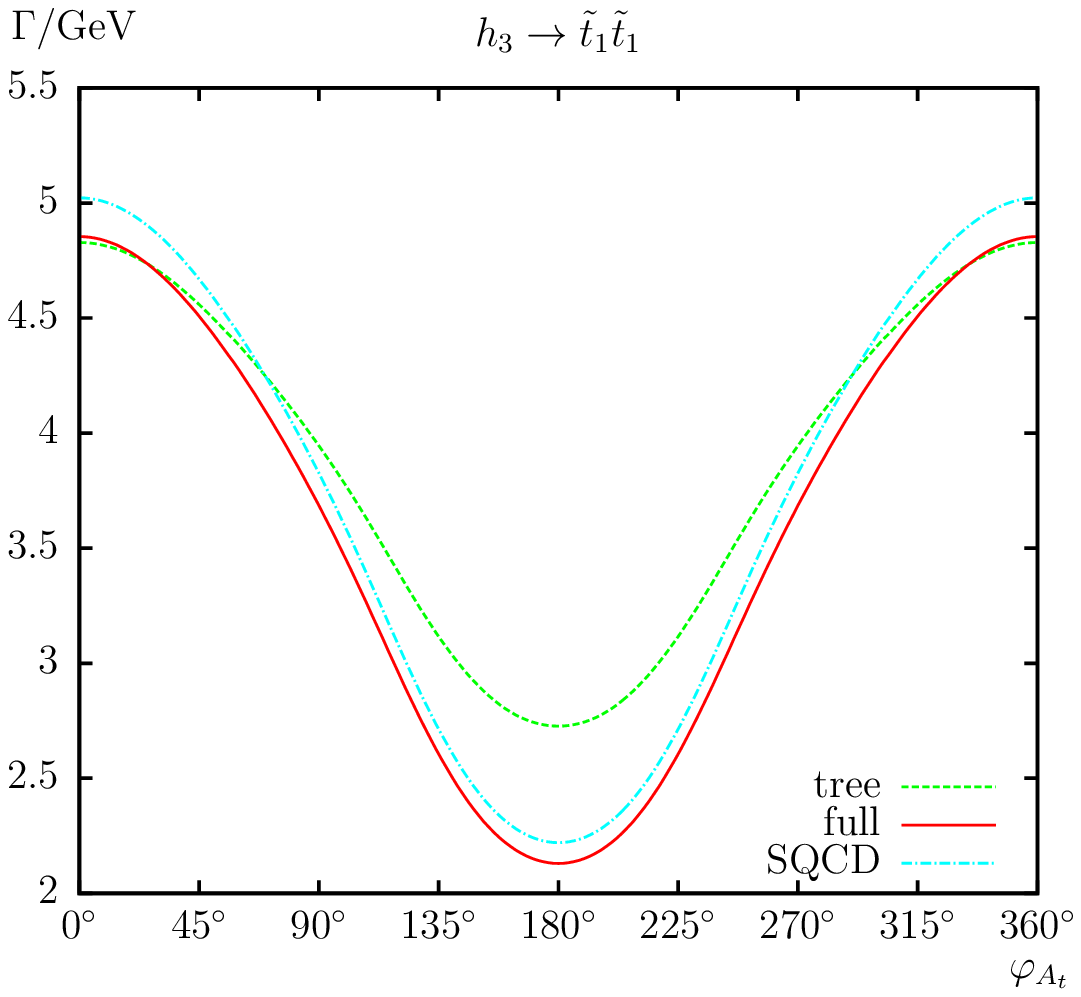}
\end{tabular}
\vspace{1em}
\caption{
  $\Ga(h_n \to \Stop1 \Stop1)$. 
  Tree-level, full and SQCD one-loop corrected partial decay widths are 
  shown.  The upper plot shows the partial decay width with $\MHp$ varied; 
  the lower plots show the complex phase $\phiAb$ varied for $h_2$
  decays (left, \Sce) and $h_3$ decays (right, \Scz).
}
\label{fig:hnst1st1}
\end{center}
\end{figure}

\begin{figure}[htb!]
\begin{center}
\begin{tabular}{c}
\includegraphics[width=0.49\textwidth,height=7.5cm]{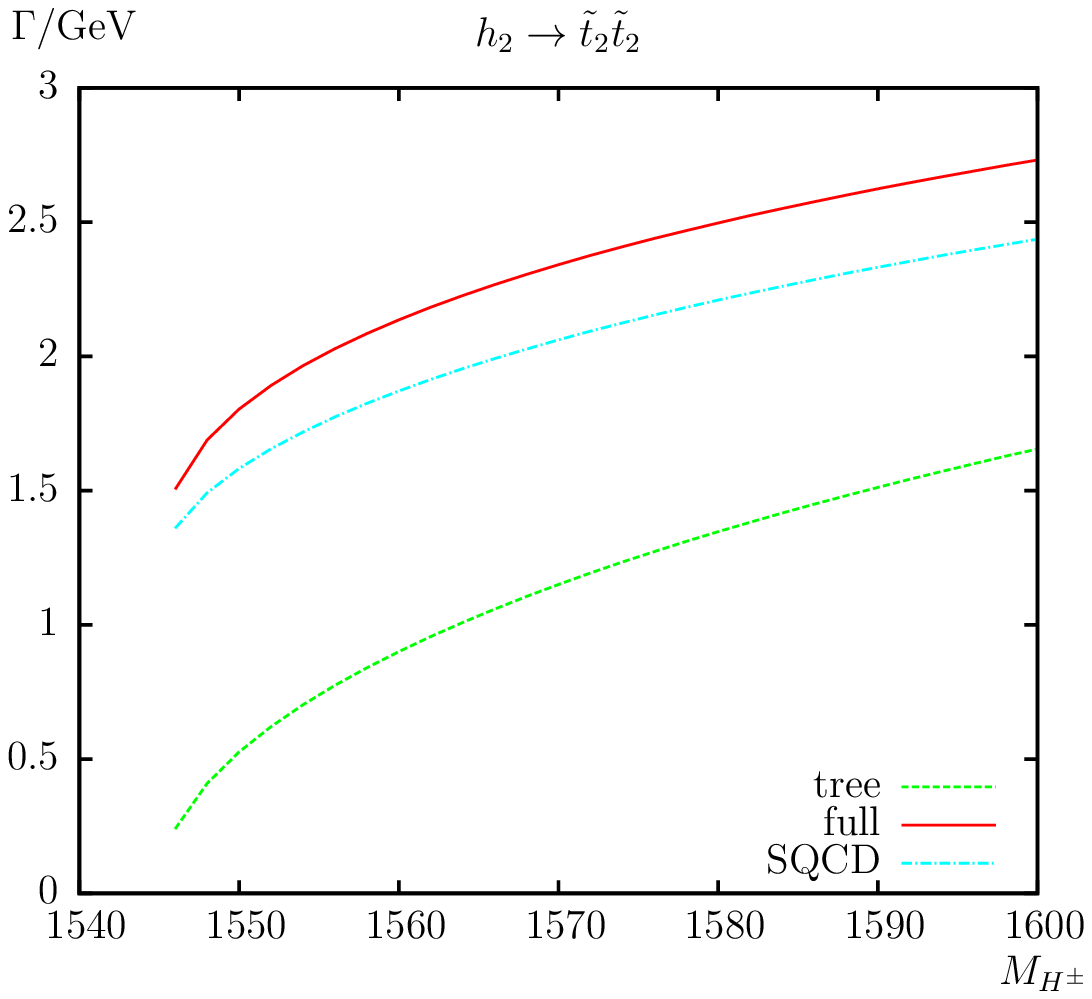}
\\[4em]
\includegraphics[width=0.49\textwidth,height=7.5cm]{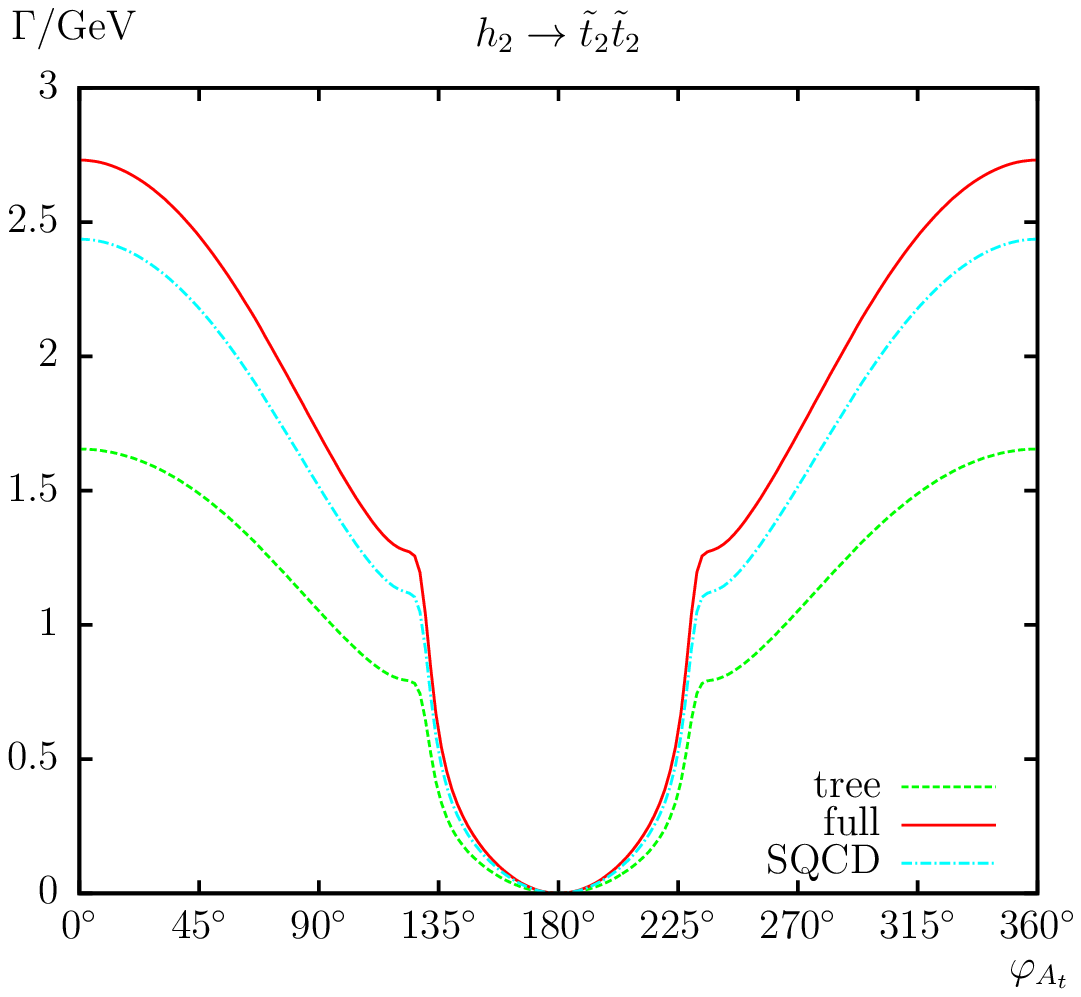}
\hspace{-4mm}
\includegraphics[width=0.49\textwidth,height=7.5cm]{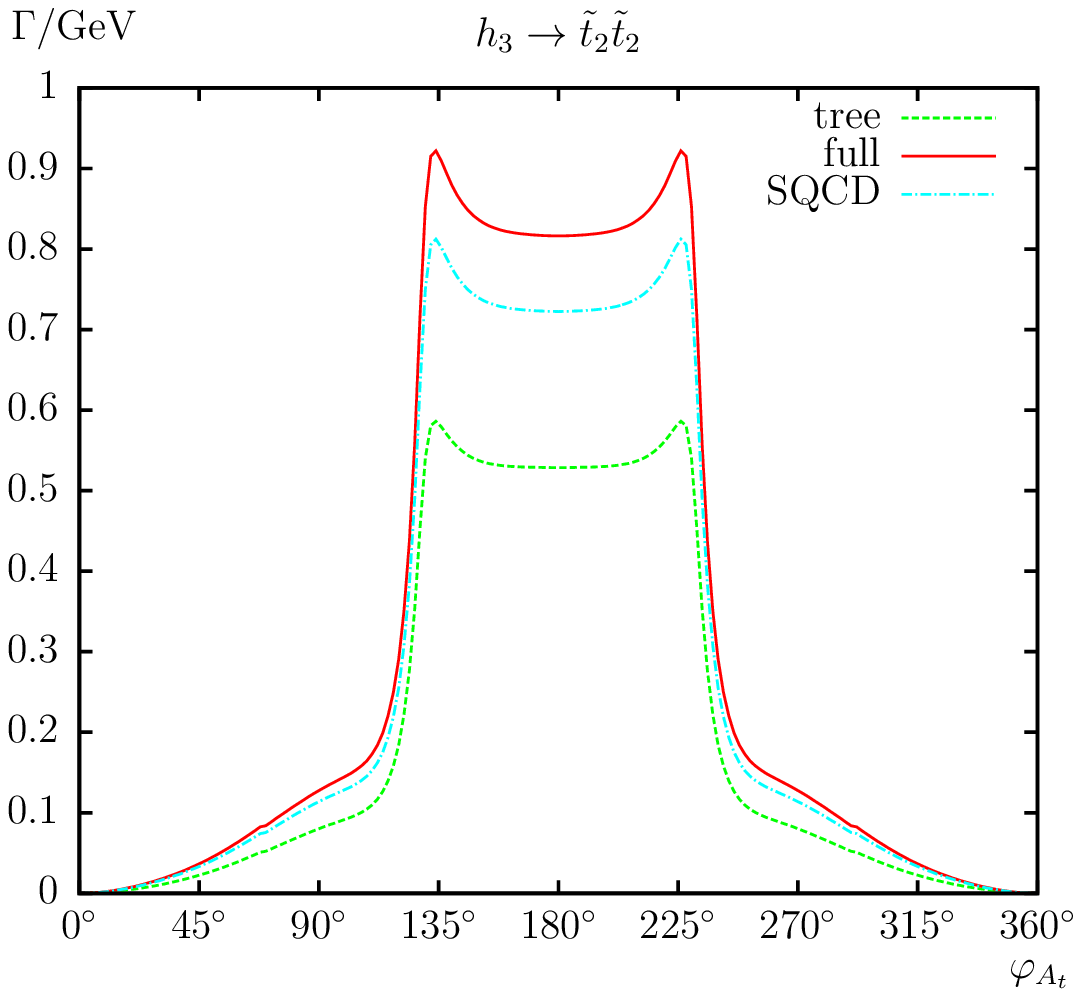}
\end{tabular}
\vspace{1em}
\caption{
  $\Ga(h_n \to \Stop2 \Stop2)$. 
  Tree-level, full and SQCD one-loop corrected partial decay widths are 
  shown.  The upper plot shows the partial decay width with $\MHp$ varied; 
  the lower plots show the complex phase $\phiAt$ varied for $h_2$
  decays (left) and $h_3$ decays (right) with parameters chosen according 
  to \Scd\ (see \refta{tab:para}).
}
\label{fig:hnst2st2}
\end{center}
\end{figure}

\begin{figure}[htb!]
\begin{center}
\begin{tabular}{c}
\includegraphics[width=0.49\textwidth,height=7.5cm]{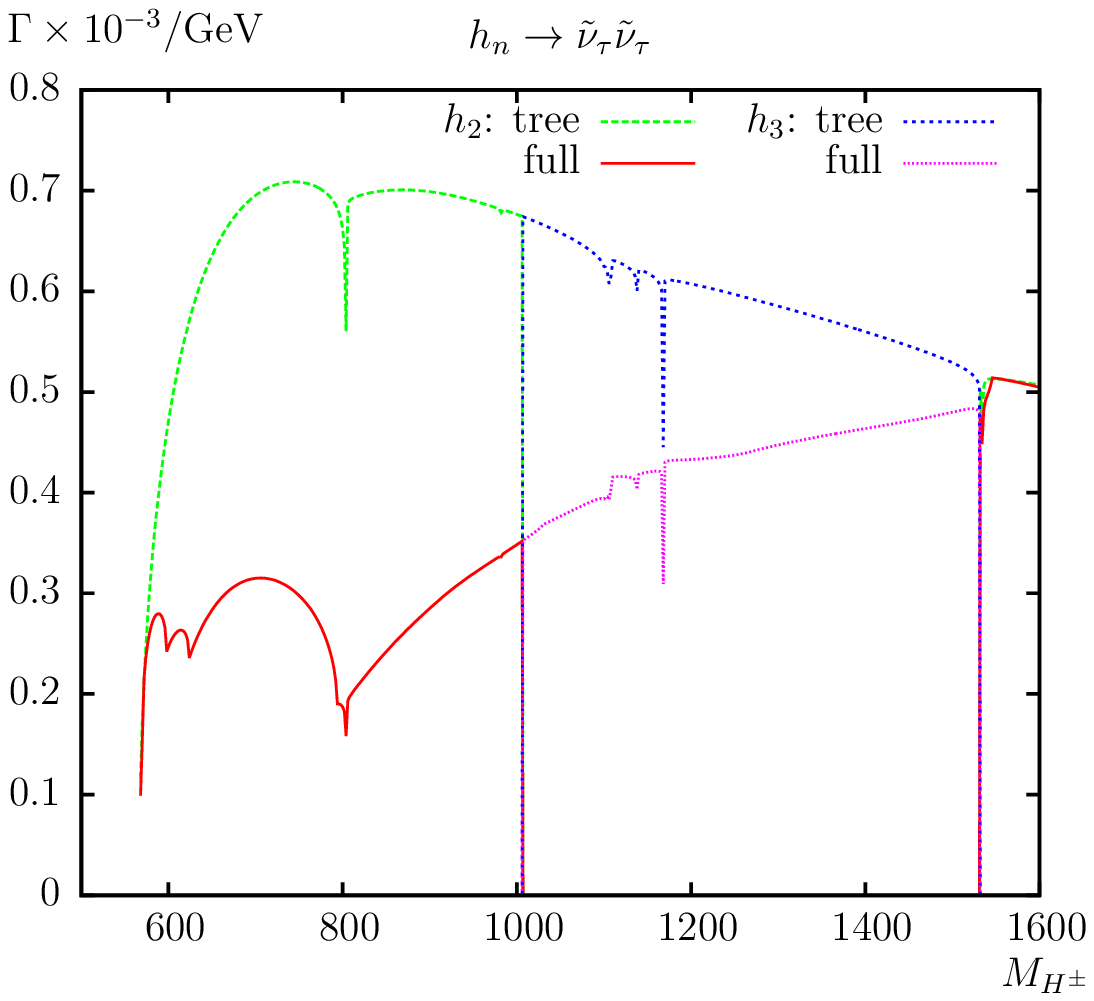}
\\[4em]
\includegraphics[width=0.49\textwidth,height=7.5cm]{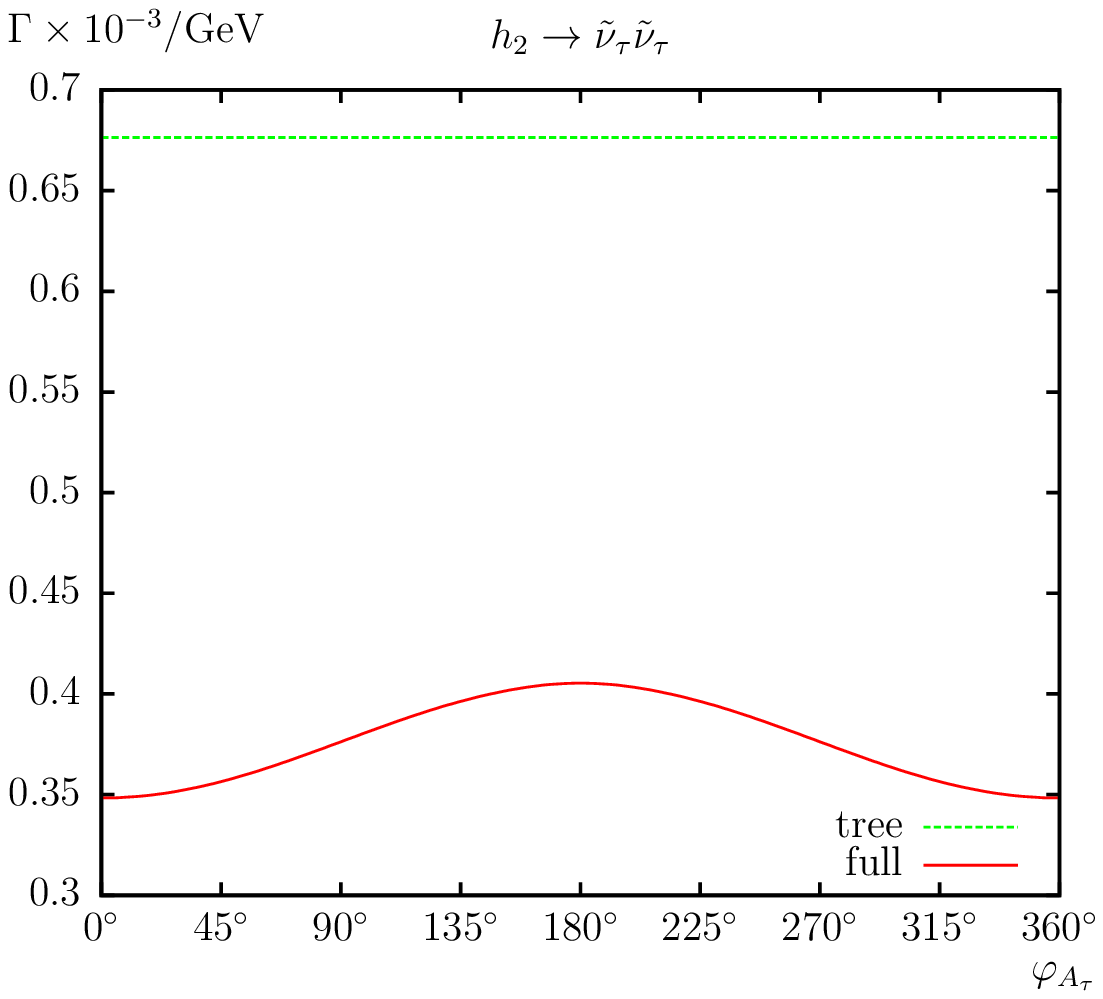}
\hspace{-4mm}
\includegraphics[width=0.49\textwidth,height=7.5cm]{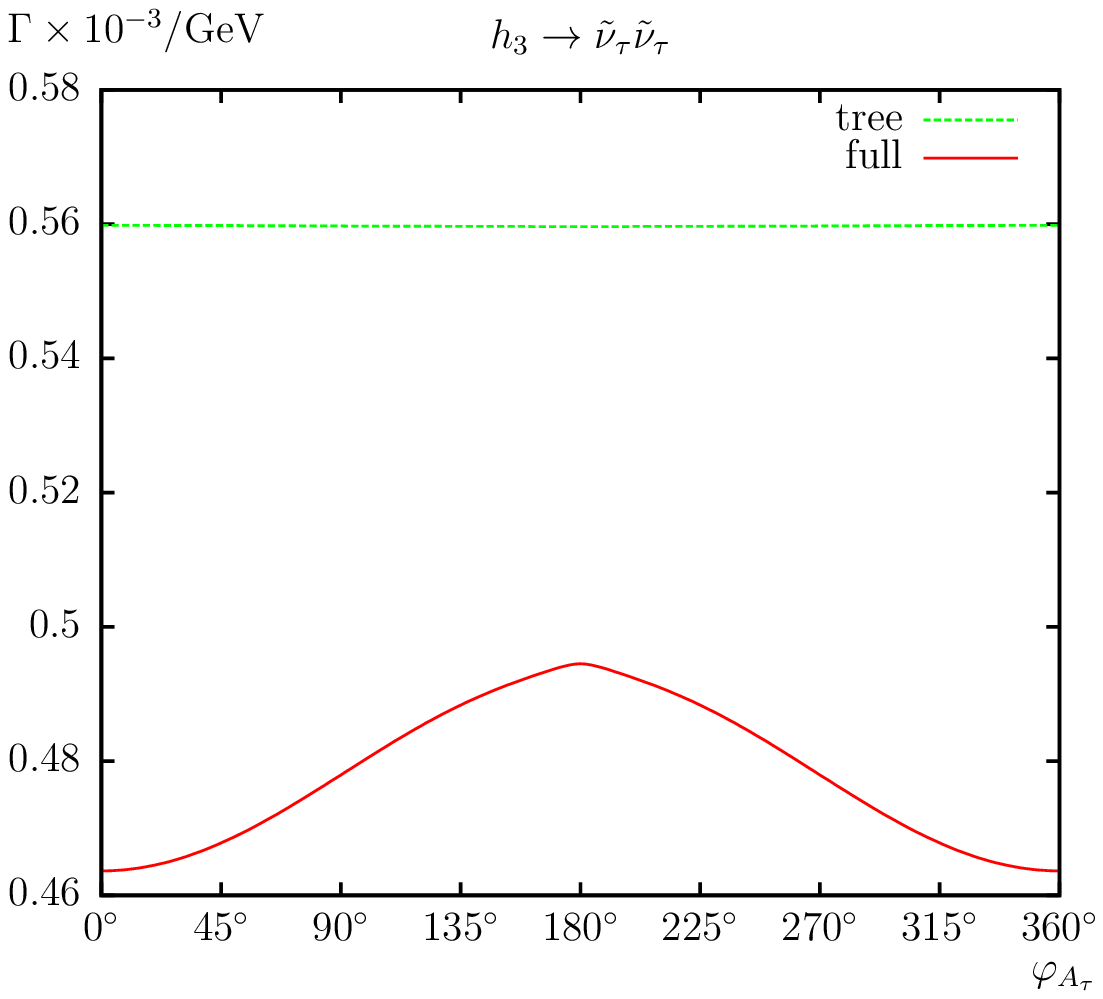}
\end{tabular}
\vspace{1em}
\caption{
  $\Ga(h_n \to \Sn_{\tau} \Sn_{\tau})$. 
  Tree-level and full one-loop corrected partial decay widths are 
  shown.  The upper plot shows the partial decay width with $\MHp$ varied; 
  the lower plots show the complex phase $\phiAtau$ varied for $h_2$
  decays (left, \Sce) and $h_3$ decays (right, \Scz).  
}
\label{fig:hnsnsn}
\end{center}
\end{figure}

\begin{figure}[htb!]
\begin{center}
\begin{tabular}{c}
\includegraphics[width=0.49\textwidth,height=7.5cm]{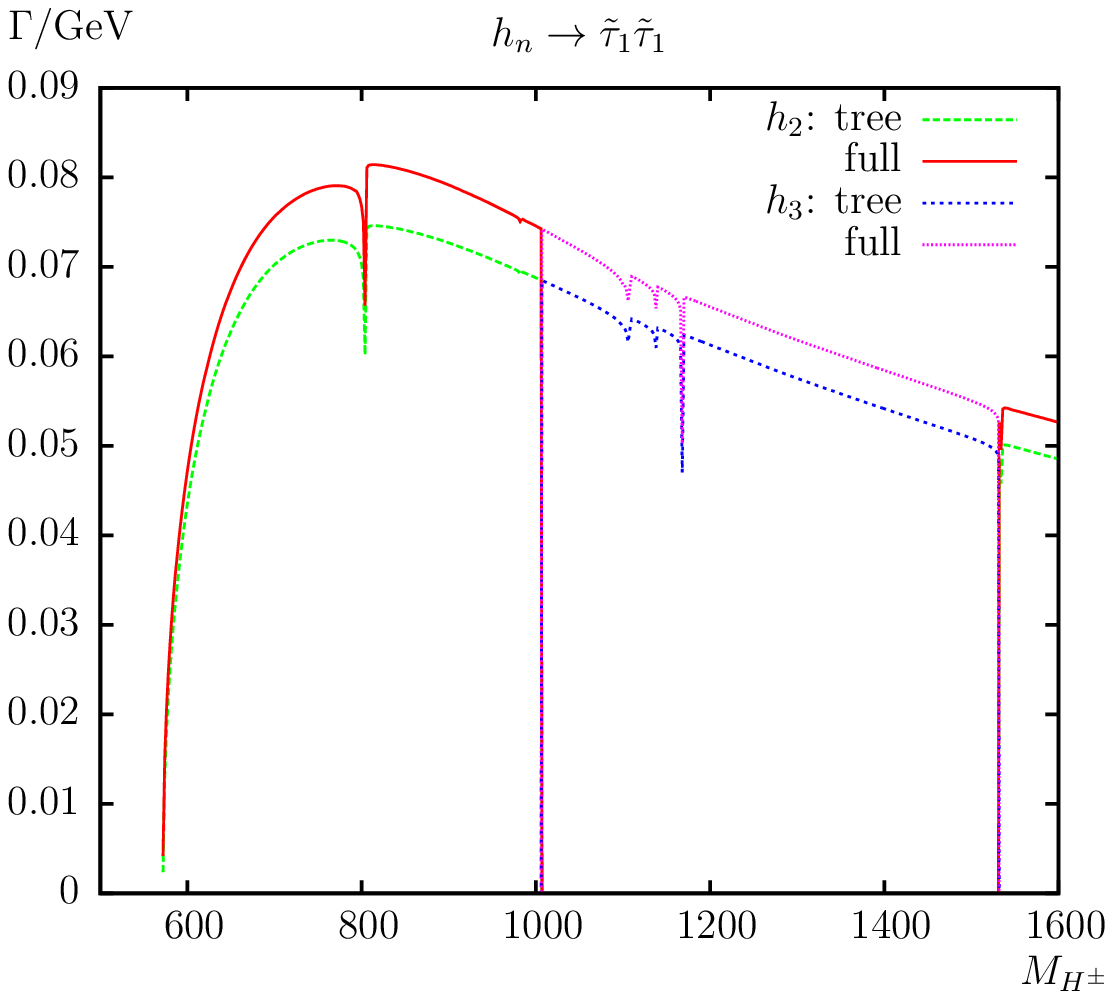}
\\[4em]
\includegraphics[width=0.49\textwidth,height=7.5cm]{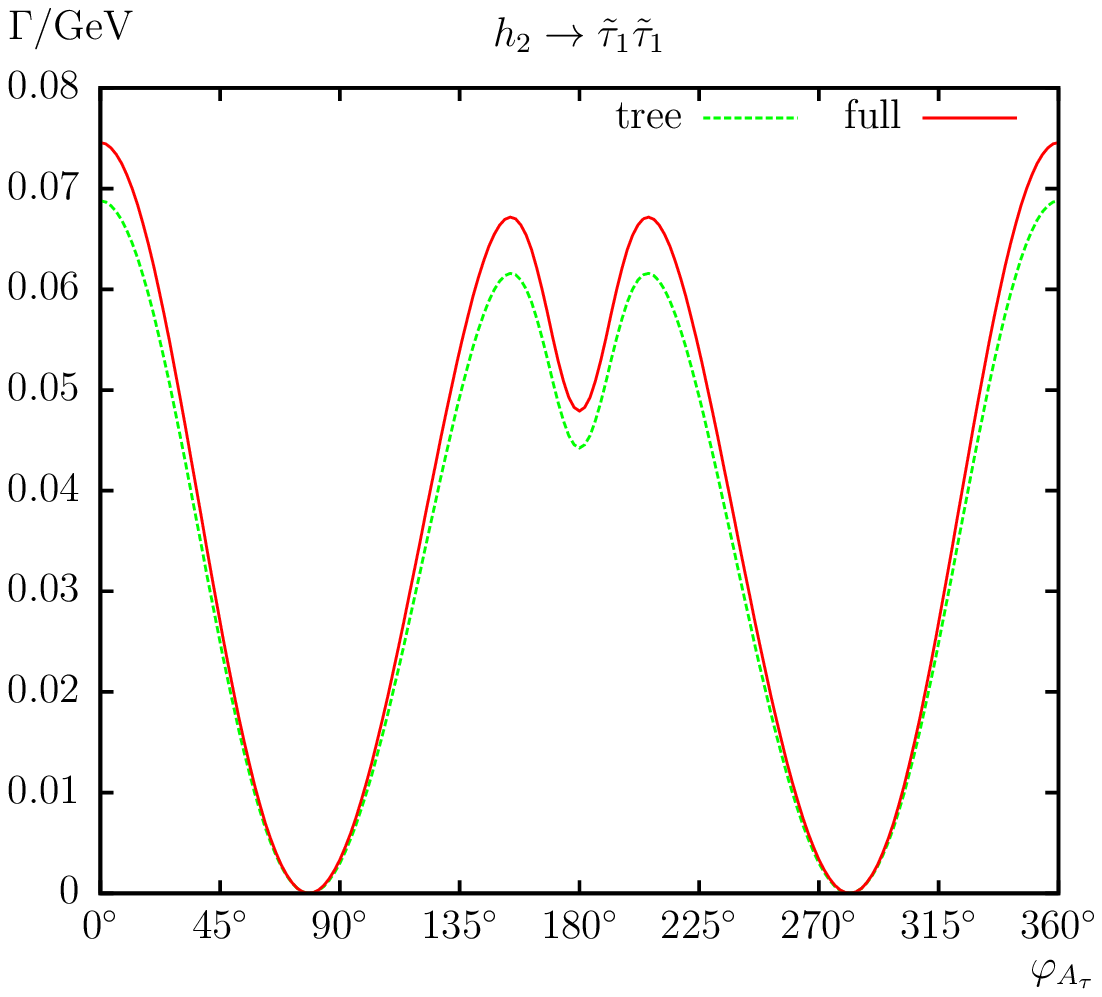}
\hspace{-4mm}
\includegraphics[width=0.49\textwidth,height=7.5cm]{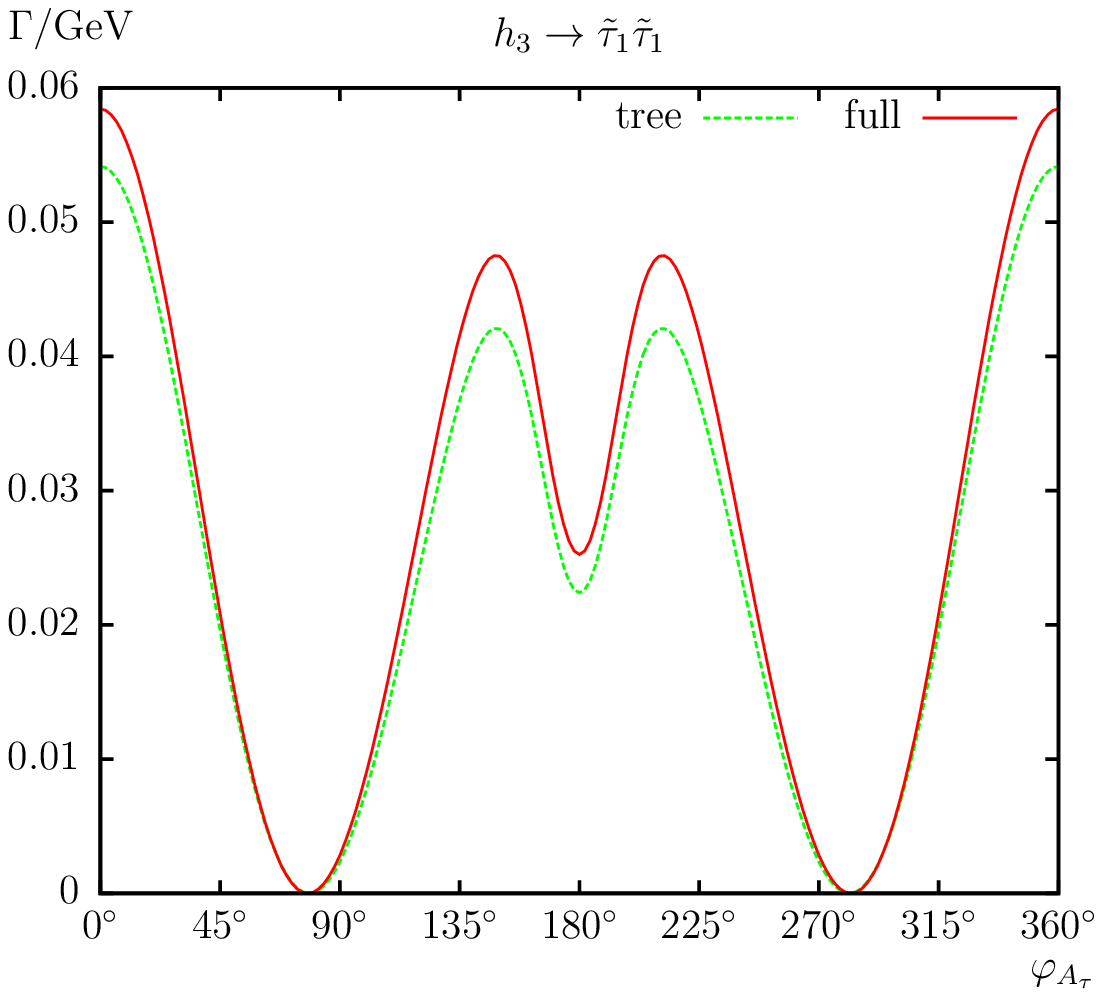}
\end{tabular}
\vspace{1em}
\caption{
  $\Ga(h_n \to \Stau1 \Stau1)$. 
  Tree-level and full one-loop corrected partial decay widths are 
  shown.  The upper plot shows the partial decay width with $\MHp$ varied; 
  the lower plots show the complex phase $\phiAtau$ varied for $h_2$
  decays (left, \Sce) and $h_3$ decays (right, \Scz).
}
\label{fig:hnstau1stau1}
\end{center}
\end{figure}

\begin{figure}[htb!]
\begin{center}
\begin{tabular}{c}
\includegraphics[width=0.49\textwidth,height=7.5cm]{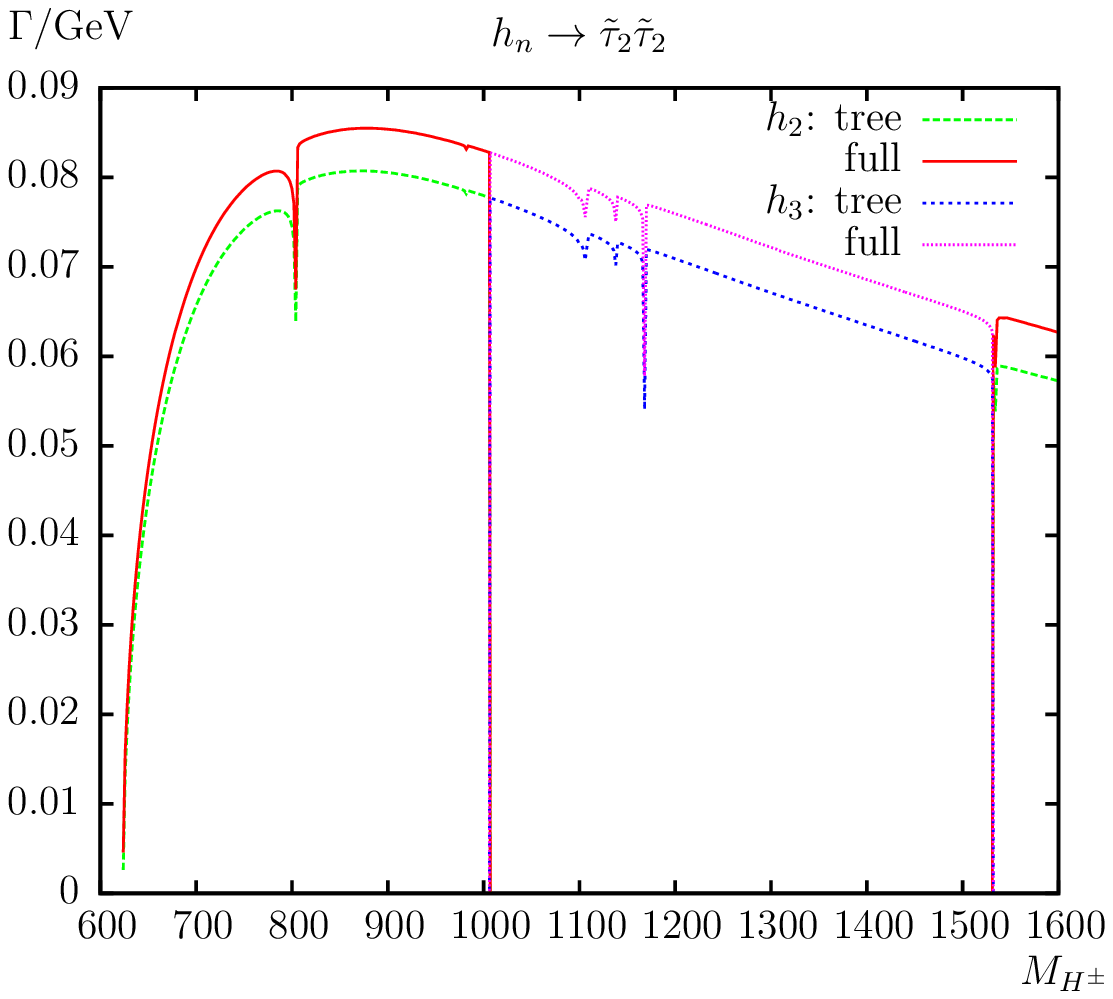}
\\[4em]
\includegraphics[width=0.49\textwidth,height=7.5cm]{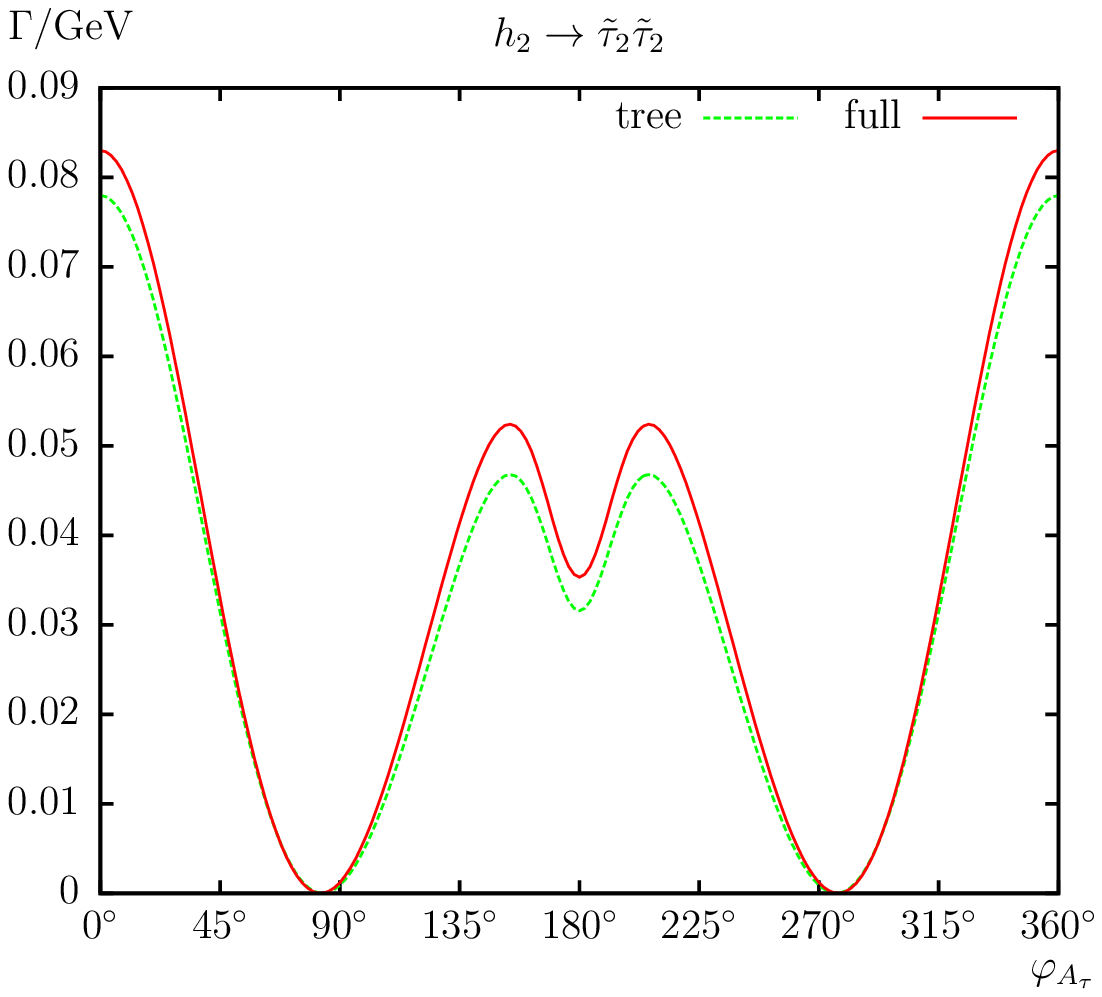}
\hspace{-4mm}
\includegraphics[width=0.49\textwidth,height=7.5cm]{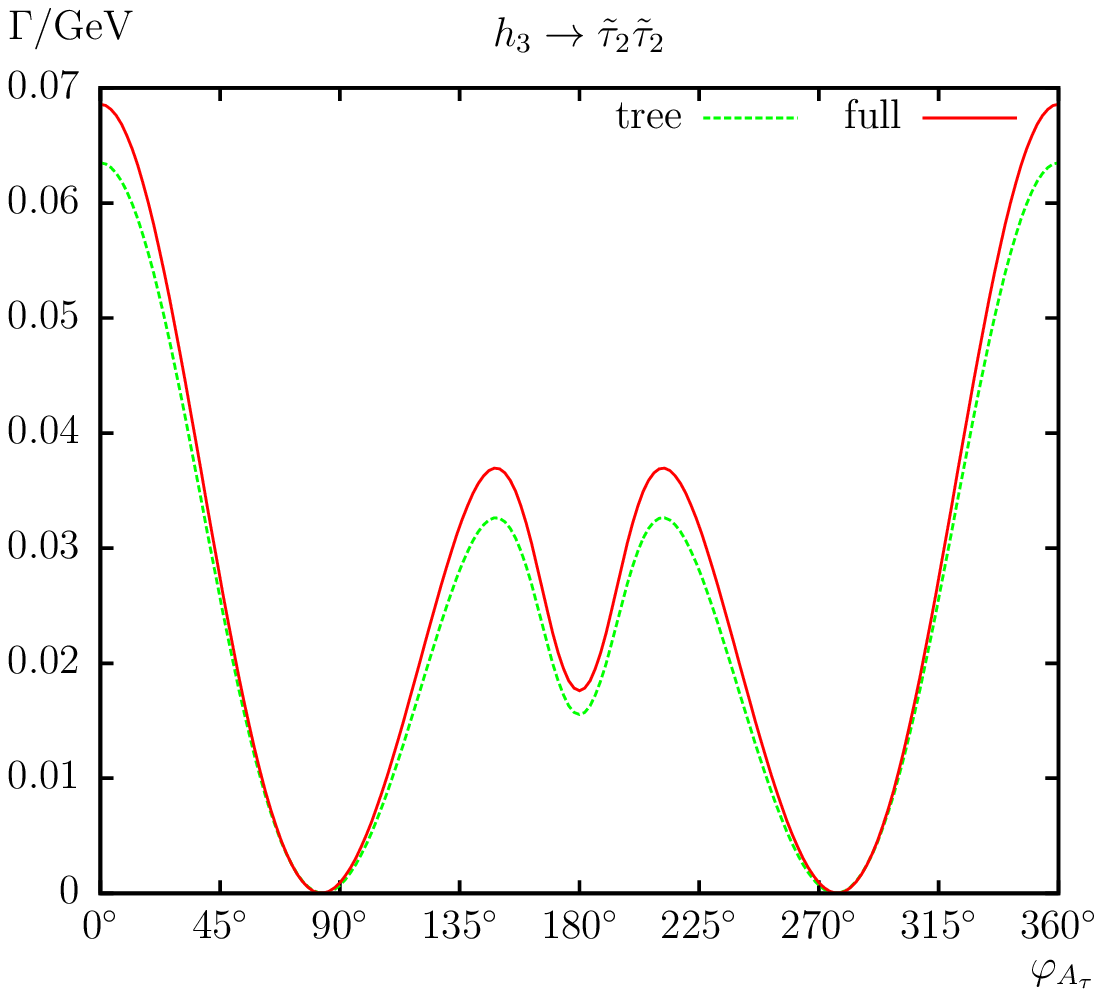}
\end{tabular}
\vspace{1em}
\caption{
  $\Ga(h_n \to \Stau2 \Stau2)$. 
  Tree-level and full one-loop corrected partial decay widths are 
  shown.  The upper plot shows the partial decay width with $\MHp$ varied; 
  the lower plots show the complex phase $\phiAtau$ varied for $h_2$
  decays (left, \Sce) and $h_3$ decays (right, \Scz).  
}
\label{fig:hnstau2stau2}
\end{center}
\end{figure}

We now turn to the decay modes $\hndecay$ ($n = 2,3; i,j = 1,2$). 
Results are shown in the \reffis{fig:hnst1st2} -- \ref{fig:hnstau2stau2}.

Before discussing every figure in detail, it should be noted that 
there is a subtleness concerning the mixture of the $h_n$ bosons.
Depending on the input parameters, the higher-order corrections to the
three neutral Higgs boson masses can vary substantially. 
The mass ordering $\mh1 < \mh2 < \mh3$ (as performed automatically by
\FH), even in the case of real parameters, can yield a heavy $\CP$-even
Higgs mass higher {\em or} lower than the (heavy) $\CP$-odd Higgs mass. 
Such a transition in the mass ordering (or ``mass crossing'') is
accompanied by an abrupt change in the Higgs mixing matrix 
$\matr{\hat Z}$.%
\footnote{In our case the $Z$-factor matrix 
  $\hat{Z}_{ij} \equiv \Code{ZHiggs[\Vi,\,\Vj]}$, see \citere{MSSMCT} 
  (and \citere{mhcMSSMlong}), which contributes at tree level.
  Furthermore $\matr{\hat{Z}}$ is calculated by \FH\ which uses 
  $\mb(\mb)$ and tree level sfermion masses instead of the shifted 
  masses, causing a slight displacement in the threshold position.
}
For our input parameters (see \refta{tab:para}) there are two (possible) 
crossings.  The first (called ``MC1'' below) appears at 
$\MHp \approx 1006\gev$.  Before the crossing we find 
$h_2 \sim H$ ($h_3 \sim A$), whereas after the crossing it changes to 
$h_2 \sim A$ ($h_3 \sim H$).  The second crossing (called ``MC2'') is 
found at $\MHp \approx 1535\gev$, \ie the changing of the 
mixture from $h_2 \sim A$ ($h_3 \sim H$) to $h_2 \sim H$ ($h_3 \sim A$).
Very close to the mass crossings the $\matr{\hat Z}$~matrix can 
yield small numerical instabilities.  As an example, for 
$1534\gev \lesssim \MHp \lesssim 1536\gev$ the $\matr{\hat Z}$~matrix
causes structures appearing similar to ``usual'' dips from thresholds.

\medskip

We start with the decay $\hnstst$ as shown in \reffi{fig:hnst1st2}. 
The upper plot shows the results as a function of $\MHp$, whereas in the 
lower plots we present the decay widths as a function of $\phiAt$ in \Scz. 
We show separately the results for the $h_2$ and $h_3$ decay widths.
In the upper plot of \reffi{fig:hnst1st2} the ``apparently single'' dip 
at $\MHp \approx 1268\gev$ is (again) in reality two dips coming from 
the thresholds $\mcha2 + \mcha2 = \mh2 \approx 1264\gev$ and 
$\mneu4 + \mneu4 = \mh2 \approx 1265\gev$.  
Away from the production threshold relative corrections of $\sim +12\%$ 
are found in \Scz\ (see \refta{tab:para}) for the $h_2$ decay.  
There the SQCD corrections overestimate the full correction by
about $30\%$.  In case of the $h_3$ decay the relative corrections 
are $\sim +24\%$ in \Scz\ (see \refta{tab:para}) and the SQCD corrections 
underestimate the full result by about $50\%$. 
The MC2 can be observed at $\MHp \sim 1535 \gev$ as
described above. Here $h_2$ and $h_3$ change their role. Within the
unrotated scalar top basis the $\CP$-odd Higgs boson 
can only decay as $A \to \StopL\StopR, \StopR\StopL$, but not as 
$A \to \StopL\StopL, \StopR\StopR$, whereas the $\CP$-even Higgs boson
has all four decays possible. Consequently, the decay to 
$\Stop1\Stop2, \Stop2\Stop1$ can depend strongly on the $\CP$ 
nature of the decaying Higgs boson.
While below MC2 we find $\Ga(\hzstst) \gg \Ga(\hdstst)$,
above MC2 we have correspondingly $\Ga(\hzstst) \ll \Ga(\hdstst)$, 
as can be clearly observed in the upper plot of \reffi{fig:hnst1st2}.

We now turn to the phase dependence of the decay width shown 
in \Scz, \ie for $\MHp = 1400 \gev$, where the left (right) plot in 
\reffi{fig:hnst1st2} shows the dependence of $\Ga(\hzstst)$
($\Ga(\hdstst)$) on $\phiAt$.  
In the lower left plot one can observe that already the tree level 
result (green dashed) and the tree of the conjugated process 
(blue short dashed) are asymmetric and depend strongly on the phase.
The assymetry at the tree-level is due to the contribution from 
the $\matr{\hat{Z}}$~matrix, which is not in general unitary and 
depends via the stop contributions to the Higgs boson self-energies on 
$\phiAt$, see \citere{mhcMSSMlong}.  While for $\phiAt \sim 180^\circ$ a 
width of about $1.5 \gev$ is observed, for $\phiAt \sim 0^\circ$ a three 
times higher decay width is found. The full corrections for $\phiAt$ 
varied are $\lesssim +12\%$ for \Scz, while the SQCD corrections 
overestimate the full corrections up to $30\%$.
In the lower right plot of \reffi{fig:hnst1st2}, where we show the
$\phiAt$ dependence of the $h_3$ decay one can see that as for the
$h_2$ case already the tree level results (green dashed) and the tree
of the conjugated  process (blue short dashed) depend strongly on
the phase and exhibit an asymmetry.  The latter is again due to the 
contribution from the $\matr{\hat{Z}}$~matrix.  The relative corrections 
for $\phiAt$ are up to $\sim +29\%$ for \Scz.  The SQCD corrections 
are smaller and would underestimate the full corrections by more
than~$50\%$.  

\medskip

In \reffi{fig:hnsb1sb2} we present the results for the decays
$\hnsbsb$, where in the upper (lower) row we show the
dependence on $\MHp$ ($\phiAb$).  In the upper row plot the first 
``apparently single'' dip in the $h_2$ decay (upper lines) is in 
reality two dips at $\MHp \approx 1108\gev$ and 
$\MHp \approx 1112\gev$ coming from the thresholds 
$\mcha1 + \mcha2 = \mh2 \approx 1105\gev$ and 
$\mneu2 + \mneu4 = \mh2 \approx 1108\gev$.  The second dip at 
$\MHp \approx 1171\gev$ is the threshold 
$\mstop1 + \mstop2 = \mh2 = 1165\gev$.  
The ``step'' at $\MHp \approx 1184\gev$ could be traced back to
the $C$-functions  
$C_{0,1,2}(\mh2^2,\msbot1^2,\msbot2^2,\mstop{i}^2,\mstop{j}^2,\MW^2)$ 
with $i \ne j$ (but without any apparent threshold).
The remaining dip (at $\MHp \approx 1268\gev$) is the same as in 
\reffi{fig:hnst1st2} for the same reasons (see above).  
At $\MHp = 1400\gev$ the full one-loop corrections to the $h_2$ decay 
reach only $\sim +6\%$, while the SQCD corrections would overestimate 
this by a factor of $\sim 2.5$. 
Now we turn to the corresponing $h_3$ decay.
The first three dips and the ``step'' at $\MHp \approx 1182\gev$ 
are the same as for the $h_2$ decay, see above.
For the decay of the $h_3$ at $\MHp = 1400 \gev$ we find full
corrections at the level of $+17\%$, where the SQCD results are 
only slightly larger.  As in \reffi{fig:hnst1st2} one can observe the 
MC2 with an ``interchange'' of $h_2$ and $h_3$.

In the lower left plot of \reffi{fig:hnsb1sb2} we present $\Ga(\hzsbsb)$
as a function of $\phiAb$ in \Scz.
The variation with $\phiAb$ is found to be very large, full relative 
corrections are up to $\sim +90\%$ for \Scz, where the SQCD corrections 
account for about $60\%$ of those. This can partially be attributed to 
the very small tree-level within the region
$60^{\circ} \lesssim \phiAb \lesssim 300^{\circ}$.  Furthermore a very
strong asymmetry between one decay and its complex conjugate can be
observed, reaching up to $60\%$.
In the lower right plot the corresponding results for the $h_3$ decay
are shown.  One can see that again already the tree level result 
(green dashed) and the tree of the conjugated process (blue short dashed) 
are asymmetric, which is caused again by the $\matr{\hat{Z}}$~matrix
contribution, where $\phiAb$ enters via the $\sbot$~contributions to the 
Higgs-boson self-energies.  As in the $h_2$ case the size of the corrections 
shows also a large variation with $\phiAb$.  The full relative corrections 
are up to $\sim +66\%$ for \Scz, where the SQCD corrections are only 
slightly smaller.

\medskip

The third ``mixed case'', the decays $\hnstaustau$, is shown in 
\reffi{fig:hnstau1stau2}. As before, the upper plots depict the result 
as a function of $\MHp$, whereas the lower row presents the $\phiAtau$ 
dependence.  We start with the $h_2$ decay as a function of $\MHp$.
The first dip at $\MHp \approx 805\gev$ is the threshold 
$\mneu1 + \mneu3 = \mh2 \approx 799\gev$.  The second (small) dip at 
$\MHp \approx 1092\gev$ is the threshold 
$\msbot1 + \msbot2 = \mh2 \approx 1086\gev$.%
\footnote{
  It should be noted that the ``squark'' thresholds (in a $h_n$ decay into
  sleptons) enter \textit{into the tree level} only via the
  $\matr{\hat{Z}}$~matrix contribution.  
  Via $2 \Re \{{\cal M}_{\text{tree}}\, {\cal M}_{\text{loop}}\}$ 
  these effects propagate also into the loop corrections.  
  (Of course there are in addition pure loop corrections from the 
  squark-squark-slepton-slepton couplings, see third row, third 
  column of \reffi{fig:hnsfisfj}.)
}
The third ``apparently single'' dip is (again) in reality two dips at 
$\MHp \approx 1108\gev$ and $\MHp \approx 1112\gev$ coming from the 
thresholds $\mcha1 + \mcha2 = \mh2 \approx 1105\gev$ and 
$\mneu2 + \mneu4 = \mh2 \approx 1108\gev$.
The fourth (large) dip at $\MHp \approx 1171\gev$ is (again) the 
threshold $\mstop1 + \mstop2 = \mh2 = 1165\gev$. 
The last dip (at $\MHp \approx 1268\gev$) is (again) the same as in 
\reffi{fig:hnst1st2} (see above). 
At $\MHp = 1000\gev$ the full one-loop corrections reach $\sim -9\%$.
In the same plot we show also the results for the $h_3$ decay.
The first (small) dip in the upper plot at $\MHp \approx 775\gev$ is the 
threshold $\mneu1 + \mneu2 = \mh3 \approx 771\gev$.  The second dip is 
in reality two dips at $\MHp \approx 948\gev$ and $\MHp \approx 954\gev$ 
coming from the thresholds $\mcha1 + \mcha1 = \mh3 \approx 945\gev$ 
and $\mneu2 + \mneu2 = \mh3 \approx 951\gev$.
The third dip is (again) at $\MHp \approx 1108\gev$ coming from the 
threshold $\mcha1 + \mcha2 = \mh3 \approx 1105\gev$.
The fourth dip at $\MHp \approx 1138\gev$ is (again) the threshold 
$\mneu3 + \mneu4 = \mh3 = 1135\gev$.  The fifth (large) dip at 
$\MHp \approx 1168\gev$ is (again) the threshold 
$\mstop1 + \mstop2 = \mh3 \approx 1165\gev$.
At $\MHp = 1000\gev$ the full one-loop corrections reach $\sim +3\%$.
The two mass crossings, MC1 and MC2, can again be observed, 
where $h_2$ and $h_3$ interchange their $\CP$~character.

We now turn to the results for the $h_2\ (h_3)$ decay as a function
of $\phiAtau$ in the lower left (right) plot of \reffi{fig:hnstau1stau2}.
For the $h_2$ decay the relative corrections for 
$\phiAtau = 80^\circ, 180^\circ, 280^\circ$ are up to $\sim +7\%$ in \Sce.
For the $h_3$ decay, on the other hand, the relative corrections for 
$\phiAtau = 82^\circ, 180^\circ$ are up to $\sim -24\%, \sim +10\%$ in \Sce. 
The asymmetry is to small to be visible in the plot.

\bigskip

Next we consider $h_n$ decays into sfermions with equal sfermion 
indices and it should be noted that the $A \Sf_i \Sf_i$ ($i = 1,2$) 
couplings are exactly zero in case of real input parameters. 

In \reffi{fig:hnsb1sb1} we present the results for the decays 
$h_n \to \Sbot1\Sbot1$. The dependence on $\MHp$ is shown in the upper
plot, whereas the dependence on $\phiAb$ for $\MHp = 1400 \gev$ is
given in the lower plots. We start with $\Ga(h_2 \to \Sbot1\Sbot1)$ in
the upper plot. Only above MC2 this decay width becomes non-zero.
The peak at $\MHp \approx 1545\gev$ (red line) 
is the threshold $\mstop2 + \mstop2 = \mh2 = 1542\gev$.
Furthermore the tree level decay width $\Ga(h_2 \to \Sbot1 \Sbot1)$ 
is accidently very small for the parameter set chosen, see
\refta{tab:para}.  Because of this smallness, the relative size of the 
one-loop correction becomes larger then the tree level, and can even 
turn negative.  Therefore in this case we added $|{\cal M}_{\text{loop}}|^2$ 
to the full one-loop result to obtain a positive decay width. 
The full relative corrections are $\sim +73\%$ at $\MHp = 1600\gev$ and 
the SQCD corrections are $\sim +80\%$.  
Also shown in this plot is the decay $h_3 \to \Sbot1\Sbot1$, which is 
non-zero below MC2.
The first dip at $\MHp \approx 1108\gev$ is the threshold 
$\mcha1 + \mcha2 = \mh3 \approx 1105\gev$. 
The second dip (not visible) at $\MHp \approx 1138\gev$ is the 
threshold $\mneu3 + \mneu4 = \mh3 = 1135\gev$. 
The third dip at $\MHp \approx 1168\gev$ is the threshold 
$\mstop1 + \mstop2 = \mh3 = 1165\gev$.  The large ``spike'' at 
$\MHp \approx 1216\gev$ is caused by the addition of the two-loop
contribution $|{\cal M}_{\text{loop}}|^2$ as explained above (formally
it is caused by the $C$-functions 
$C_{0,1,2}(\mh3^2,\msbot1^2,\msbot1^2,\mstop{i}^2,\mstop{j}^2,\MW^2)$ 
with $i \ne j$).
Without the two-loop contribution it appears
as ``step'' (see the inlay in the upper plot of \reffi{fig:hnsb1sb1}) 
similar to \reffi{fig:hnsb1sb2}.
Because of the smallness of the tree, the full relative corrections 
reach $\sim +183\%$ at $\MHp = 1400\gev$.  Here the SQCD corrections 
are smaller with $\sim +92\%$.  

In the lower left plot of \reffi{fig:hnsb1sb1} we show the $h_2$
decay with the complex phase $\phiAb$ varied at $\MHp = 1400\gev$.  
For $\phiAb = 0^\circ, 180^\circ, 360^\circ$, \ie real parameters the 
$h_2$ decay is purely $\CP$-odd, and thus the decay width is zero. 
For complex values of the phase small, but non-zero values are reached.
Here, for the same reasons as in the upper plot the loop corrections 
can be larger then the tree level and reach actually $\sim +108\%$ 
(and $\sim +125\%$ for SQCD) at $\phiAb = 90^{\circ}, 270^\circ$.
In the lower right plot of \reffi{fig:hnsb1sb1} we show the $h_3$
decay with the complex phase $\phiAb$ varied at $\MHp = 1400 \gev$.
Here (for the same reasons as in the upper plot) 
the loop corrections can be larger then the tree level 
(and for consistency with the upper plot we also add 
$|{\cal M}_{\text{loop}}|^2$ here) and reach 
up to $\sim +320\%$ (and $\sim +165\%$ for SQCD) at $\phiAb = 180^\circ$.
It should be noted that the decay width including SQCD corrections does 
{\em not} go to zero due to $|{\cal M}_{\text{loop}}|^2$, but just reaches a
(very) small value of $\Ga(h_3 \to \Sbot1\Sbot1)$ 
(see the inlay in the lower right plot of \reffi{fig:hnsb1sb1}).

\medskip

In \reffi{fig:hnsb2sb2} we present the decay $h_n \to \Sbot2\Sbot2$, 
in full analogy to \reffi{fig:hnsb1sb1}. The same behavior of $h_2$
and $h_3$ concerning MC2 can be observed.
The full relative corrections for the $h_2$ decay are $\sim +18\%$ at 
$\MHp = 1600\gev$, where the pure SQCD corrections would overestimate 
this correction by a factor $\sim 1.8$.
The decay $h_3 \to \Sbot2\Sbot2$ is again non-zero only below MC2.
The dip (hardly visible) at $\MHp \approx 1168\gev$ in the $h_3$ decay
is the threshold $\mstop1 + \mstop2 = \mh3 = 1165\gev$.  The full
relative corrections are $\sim +28\%$ at $\MHp = 1400\gev$ and the SQCD 
corrections are larger by about a factor of 1.6.
In the lower left plot of \reffi{fig:hnsb2sb2} we show the variation of 
$\Ga(h_2 \to \Sbot2\Sbot2)$ with $\phiAb$ at $\MHp = 1400\gev$. 
Here the loop corrections reach $\sim +52\%$ (and $\sim +60\%$ for SQCD) 
at $\phiAb = 90^{\circ}, 270^{\circ}$.
The decay width goes to zero for real $\Ab$ due to the $\CP$-nature of
the $h_2$.
In the lower right plot of 
\reffi{fig:hnsb2sb2} we show $\Ga(h_3 \to \Sbot2\Sbot2)$ with $\phiAb$ 
varied at $\MHp = 1400\gev$.  Here the loop corrections reach $\sim +90\%$ 
at $\phiAb = 180^{\circ}$ and are slightly overestimated in the pure SQCD case.
The decay width goes to zero for $\phiAb \sim 85^\circ, 275^\circ$. 
For these values the relevant diagonal entries of the
$\matr{\hat Z}$~matrix go through zero, \ie the main effect on the 
decay width stems from the $\matr{\hat Z}$~matrix.

\medskip

We now turn to the neutral Higgs decay to scalar top quarks, 
which are shown in full analogy to the decay to scalar bottom 
quarks above.
In \reffi{fig:hnst1st1} we present the decay $h_n \to \Stop1\Stop1$.
In the upper row we show the results as a function of $\MHp$.
The first dip at $\MHp \approx 805\gev$ in the $h_2$ decay is the 
threshold $\mneu1 + \mneu3 = \mh2 \approx 799\gev$.
The second dip at $\MHp \approx 982\gev$ is the threshold 
$\mneu2 + \mneu3 = \mh2 \approx 979\gev$.
The full relative corrections are $\sim +14\%$ at $\MHp = 1000\gev$ 
(\ie \Sce) and the SQCD corrections are $\sim +16\%$.
The decay width turns zero between MC1 and MC2 and reaches non-zero
values below MC1 and above MC2. Reversely, we find non-zero values for 
$\Ga(h_3 \to \Stop1\Stop1)$ only between MC1 and MC2.
The first dip at $\MHp \approx 1108\gev$ in the $h_3$ decay is 
the threshold $\mcha1 + \mcha2 = \mh3 1105\approx \gev$.
The second dip at $\MHp \approx 1138\gev$ is the threshold 
$\mneu3 + \mneu4 = \mh3 \approx 1135\gev$.
The third dip at $\MHp \approx 1168\gev$ is the threshold 
$\mstop1 + \mstop2 = \mh3 = 1165\gev$.
The full relative corrections at $\MHp = 1400\gev$ are accidentally
small and reach only $+1\%$, where the pure SQCD corrections reach
$\sim +4\%$.
In the lower left plot of \reffi{fig:hnst1st1} we show the $h_2$ decay 
with the complex phase $\phiAt$ varied at $\MHp = 1000\gev$.  Here the 
loop corrections can vary between $\sim +14\%$ for 
$\phiAt \sim 0^\circ, 360^\circ$ and $\sim -7\%$ at $\phiAt = 180^{\circ}$, 
where the SQCD corrections in this case are a good approximation to the 
full result.  In the lower right plot of \reffi{fig:hnst1st1} we show 
the $h_3$ decay with $\phiAt$ varied at $\MHp = 1400\gev$.  Here the 
loop corrections are close to zero for real positive values of $\Ab$, 
where the EW corrections compensate the SQCD contributions.  The full 
corrections can reach $\sim -22\%$ (and $\sim -19\%$ for SQCD) at 
$\phiAt = 180^{\circ}$.

\medskip

The final decays involving stops are shown in \reffi{fig:hnst2st2}.
The results as a function of $\MHp$ are given in the upper plot.  Due to
the large values of $\mstop2$ for real parameters only 
$\Ga(h_2 \to \Stop2\Stop2)$ reaches non-zero values.
The full relative corrections for the $h_2$ decay are $\sim +65\%$ at 
$\MHp = 1600\gev$ (\ie S3) and the SQCD corrections reach $\sim +47\%$.  
In the lower left plot of \reffi{fig:hnst2st2} we show 
$\Ga(h_2 \to \Stop2\Stop2)$ with the complex phase $\phiAt$ varied at 
$\MHp = 1600\gev$.  The smooth structure around 
$\phiAt = 130^{\circ}, 230^{\circ}$ is \textit{not} a threshold but a
numerical effect of the $\matr{\hat{Z}}$~matrix contribution.
The loop corrections can reach $\sim +63\%$ (and $\sim +44\%$ for SQCD) 
at $\phiAt = 90^{\circ}, 270^{\circ}$. For $\phiAt = 180^\circ$ the
decay width goes to zero since the relevant diagonal entries in the
$\matr{\hat Z}$~matrix go through zero, see also the discussion of
\reffi{fig:hnsb2sb2}.
The $h_3$ decay is non-zero above MC2 only for complex parameters.
In the lower right plot of \reffi{fig:hnst2st2} we show 
$\Ga(h_3 \to \Stop2\Stop2)$ with $\phiAt$ varied at 
$\MHp = 1600\gev$.  The smooth structure around 
$\phiAt = 135^{\circ}, 225^{\circ}$ is again a numerical effect of 
the $\matr{\hat{Z}}$~matrix contribution.  The loop corrections can 
reach $\sim +54\%$ (and $\sim +37\%$ for SQCD) at $\phiAt = 180^{\circ}$.

\medskip

We finish our numerical analysis with the remaining decays to
scalar leptons. In \reffi{fig:hnsnsn} the decay widths for 
$h_n \to \Snutau\Snutau$ are shown. In the upper plot the results as 
a function of $\MHp$ are given.  The decay $h_2 \to \Snutau\Snutau$ 
for real parameters is non-zero below MC1 and above MC2 due to the
$\CP$-structure of $h_2$.  The first dip in the $h_2$ decay at 
$\MHp \approx 598\gev$ is the threshold 
$\mneu1 + \mneu1 = \mh2 \approx 591\gev$.
The second dip at $\MHp \approx 624\gev$ is the threshold 
$\mstau2 + \mstau2 = \mh2 = 618\gev$.
The third dip at $\MHp \approx 794\gev$ is the threshold 
$\mstop1 + \mstop1 = \mh2 = 788\gev$.
The fourth dip at $\MHp \approx 805\gev$ is the threshold 
$\mneu1 + \mneu3 = \mh2 \approx 799\gev$.
The last dip (hardly visible) at $\MHp \approx 982\gev$ is the threshold 
$\mneu2 + \mneu3 = \mh2 \approx 979\gev$.
The full relative corrections are found to be $\sim -48\%$ at 
$\MHp = 1000\gev$ (\ie \Sce).
Correspondingly, the decay $h_3 \to \Snutau\Snutau$ for real
parameters is non-zero only between MC1 and MC2.
The first dip at $\MHp \approx 1108\gev$ is (again) the threshold 
$\mcha1 + \mcha2 = \mh3 1105\approx \gev$.
The second dip at $\MHp \approx 1138\gev$ is (again) the threshold 
$\mneu3 + \mneu4 = \mh3 \approx 1135\gev$.
The third dip at $\MHp \approx 1168\gev$ is (again) the threshold 
$\mstop1 + \mstop2 = \mh3 = 1165\gev$.
The full relative corrections are $\sim -17\%$ at $\MHp = 1400\gev$ 
(\ie \Scz).
The dependence on $\phiAtau$ is shown for the $h_2\ (h_3)$ decay in
the lower left (right) plot of \reffi{fig:hnsnsn} for 
$\MHp = 1000\ (1400) \gev$.  For the $h_2$ decay the loop corrections 
can reach $\sim -40\%$ at $\phiAtau = 180^{\circ}$.  For the $h_3$ decay 
they can reach $\sim -12\%$ at $\phiAtau = 180^{\circ}$.

\medskip

In \reffi{fig:hnstau1stau1} we present the results for the decays
$h_n \to \Stau1\Stau1$. The upper row shows the decay widths as a
function of $\MHp$. As before, the decay width of $h_2$ is non-zero for
real parameters only below MC1 and above MC2, whereas the $h_3$ decay
width is non-zero between the two mass crossing points.  Starting with 
$h_2$, the first (large) dip at $\MHp \approx 805\gev$ in the $h_2$ decay 
is (again) the threshold $\mneu1 + \mneu3 = \mh2 \approx 799\gev$.  
The second dip (hardly visible) at $\MHp \approx 982\gev$ is (again) 
the threshold $\mneu2 + \mneu3 = \mh2 \approx 979\gev$.  The full 
relative corrections are $\sim +8\%$ at $\MHp = 1000\gev$ (\ie \Sce). 
The three dips of the $h_3$ decay are the same as in the upper plot of 
\reffi{fig:hnsnsn}, see above.  The full relative corrections at 
$\MHp = 1400\gev$ (\ie \Scz) are $\sim +8\%$.
In the lower left plot of \reffi{fig:hnstau1stau1} we show the $h_2$ 
decay with $\phiAtau$ varied at $\MHp = 1000\gev$.  Here the loop 
corrections can reach $\sim +9\%$ around 
$\phiAtau \sim 140^\circ, 220^\circ$.
For $\phiAtau \sim 80^\circ,280^\circ$ we find again the dominant effects
from the $\matr{\hat Z}$~matrix, leading to a vanishing decay width around
these values.
In the lower right plot of \reffi{fig:hnstau1stau1} we show $h_3$ 
results as a function of $\phiAtau$ with $\MHp = 1400 \gev$.
Here the loop corrections can reach 
$\sim +13\%$ at $\phiAtau \sim 140^\circ, 220^\circ$, and the 
$\matr{\hat Z}$~matrix causes the vanishing width around 
$\phiAtau \sim 80^\circ,280^\circ$.

\medskip

Finally, in \reffi{fig:hnstau2stau2} we present the results for 
$\Ga(h_n \to \Stau2\Stau2)$, which are shown in full analogy to 
$\Ga(h_n \to \Stau1\Stau1)$ above. As before, for real parameters, 
the $h_2$ decay width is found non-zero only below MC1 and above MC2, 
while the $h_3$ width is non-zero only between MC1 and MC2.
The $h_n$ decay widths as a function of $\MHp$ exhibit the same dips 
as in \reffi{fig:hnstau1stau1}, see above.  The full relative 
corrections to the $h_2$ width are $\sim +6\%$ at $\MHp = 1000\gev$ 
(\ie \Sce).  The full relative corrections to the $h_3$ decay at 
$\MHp = 1400\gev$ (\ie \Scz) are $\sim +8\%$.  
In the lower left plot of \reffi{fig:hnstau2stau2} we show the $h_2$ 
decay with $\phiAtau$ varied at $\MHp = 1000\gev$.  Here the loop 
corrections can reach $\sim +12\%$ at 
$\varphi_{A_{\tau}} \sim 140^\circ, 180^\circ, 220^\circ$.
The decay width goes to zero in analogy to \reffi{fig:hnstau1stau1}.
In the right plot of \reffi{fig:hnstau2stau2} the corresponding $h_3$ 
results are shown for $\MHp = 1400 \gev$, where we find the same level 
of higher-order corrections, and the dominating effect of the 
$\matr{\hat Z}$~matrix as in \reffi{fig:hnstau1stau1}.

\clearpage
\newpage

%%%%%%%%%%%%%%%%%%%%%%%%%% F I G U R E %%%%%%%%%%%%%%%%%%%%%%%%%%%%%%%%%%%%%%%%%
%%%%%%%%%%%%%%%%%%%%%%%%%% F I G U R E %%%%%%%%%%%%%%%%%%%%%%%%%%%%%%%%%%%%%%%%%

%\newpage

%%%%%%%%%%%%%%%%%%%%%%%%%% F I G U R E %%%%%%%%%%%%%%%%%%%%%%%%%%%%%%%%%%%%%%%%%
%%%%%%%%%%%%%%%%%%%%%%%%%% F I G U R E %%%%%%%%%%%%%%%%%%%%%%%%%%%%%%%%%%%%%%%%%

%\newpage

%%%%%%%%%%%%%%%%%%%%%%%%%% F I G U R E %%%%%%%%%%%%%%%%%%%%%%%%%%%%%%%%%%%%%%%%%
%%%%%%%%%%%%%%%%%%%%%%%%%% F I G U R E %%%%%%%%%%%%%%%%%%%%%%%%%%%%%%%%%%%%%%%%%

%\newpage

%%%%%%%%%%%%%%%%%%%%%%%%%% F I G U R E %%%%%%%%%%%%%%%%%%%%%%%%%%%%%%%%%%%%%%%%%
%%%%%%%%%%%%%%%%%%%%%%%%%% F I G U R E %%%%%%%%%%%%%%%%%%%%%%%%%%%%%%%%%%%%%%%%%

\newpage

%%%%%%%%%%%%%%%%%%%%%%%%%% F I G U R E %%%%%%%%%%%%%%%%%%%%%%%%%%%%%%%%%%%%%%%%%
%%%%%%%%%%%%%%%%%%%%%%%%%% F I G U R E %%%%%%%%%%%%%%%%%%%%%%%%%%%%%%%%%%%%%%%%%

\newpage

%%%%%%%%%%%%%%%%%%%%%%%%%% F I G U R E %%%%%%%%%%%%%%%%%%%%%%%%%%%%%%%%%%%%%%%%%
%%%%%%%%%%%%%%%%%%%%%%%%%% F I G U R E %%%%%%%%%%%%%%%%%%%%%%%%%%%%%%%%%%%%%%%%%

\clearpage
\newpage

%%%%%%%%%%%%%%%%%%%%%%%%%% F I G U R E %%%%%%%%%%%%%%%%%%%%%%%%%%%%%%%%%%%%%%%%%
%%%%%%%%%%%%%%%%%%%%%%%%%% F I G U R E %%%%%%%%%%%%%%%%%%%%%%%%%%%%%%%%%%%%%%%%%

\newpage

%%%%%%%%%%%%%%%%%%%%%%%%%% F I G U R E %%%%%%%%%%%%%%%%%%%%%%%%%%%%%%%%%%%%%%%%%
%%%%%%%%%%%%%%%%%%%%%%%%%% F I G U R E %%%%%%%%%%%%%%%%%%%%%%%%%%%%%%%%%%%%%%%%%

\newpage

%%%%%%%%%%%%%%%%%%%%%%%%%% F I G U R E %%%%%%%%%%%%%%%%%%%%%%%%%%%%%%%%%%%%%%%%%
%%%%%%%%%%%%%%%%%%%%%%%%%% F I G U R E %%%%%%%%%%%%%%%%%%%%%%%%%%%%%%%%%%%%%%%%%

\newpage

%%%%%%%%%%%%%%%%%%%%%%%%%% F I G U R E %%%%%%%%%%%%%%%%%%%%%%%%%%%%%%%%%%%%%%%%%
%%%%%%%%%%%%%%%%%%%%%%%%%% F I G U R E %%%%%%%%%%%%%%%%%%%%%%%%%%%%%%%%%%%%%%%%%

\clearpage
\newpage

%%%%%%%%%%%%%%%%%%%%%%%%%%%%%%%%%%%%%%%%%%%%%%%%%%%%%%%%%%%%%%%%%%%%%%%%%%%%%%%
%%%%%%%%%%%%%%%%%%%%%%%%%%%%%%%%%%%%%%%%%%%%%%%%%%%%%%%%%%%%%%%%%%%%%%%%%%%%%%%

\section{Conclusions}
\label{sec:conclusions}

We evaluated all partial decay widths corresponding to a two-body 
decay of the heavy MSSM Higgs bosons to scalar fermions, allowing 
for complex parameters. 
The decay modes are given in \refeqs{eq:hnsfsf} and (\ref{eq:Hsfsf}).
The evaluation is based on a full one-loop calculation of all decay 
channels, also including hard QED and QCD radiation. 
In the case of a discovery of additional Higgs bosons a subsequent
precision measurement of their properties will be crucial determine
their nature and the underlying (SUSY) parameters. 
In order to yield a sufficient accuracy, one-loop corrections to the 
various Higgs-boson decay modes have to be considered. With the here
presented full one-loop calculation to scalar fermions another step 
in the direction of a complete one-loop evaluation of all possible 
decay modes has been taken.

We first reviewed the one-loop renormalization procedure of the cMSSM,
which is relevant for our calculation. In most cases we follow
\citere{MSSMCT}. However, in the scalar fermion sector, where we differ
from \citere{MSSMCT} all relevant details are given.

We have discussed the calculation of the one-loop diagrams, the
treatment of UV and IR divergences that are canceled by the inclusion
of (hard and soft) QCD and QED radiation. 
We have checked our result against the literature (where loop
corrections so far only for real parameters were available) and in most
cases found good agreement, once our set-up was changed to the one used
in the existing analyses.

While the analytical calculation has been performed for {\em all} decay
modes to sfermions, in the numerical analysis we concentrated on the
decays to the third generation sfermions: scalar tops, bottoms, 
taus and tau neutrinos.
For the analysis we have chosen a parameter set that allows
simultaneously a maximum number of two-body sfermionic decay modes.
In the analysis either the charged Higgs boson mass or the phase of a
relevant trilinear coupling has been varied.
For $\MHp$ we investigated an interval starting at $\MHp = 600\gev$ 
up to $\MHp = 1.6 \tev$, which roughly coincides with the reach of 
the LHC for high-luminosity running as well as an $e^+e^-$ collider
with a center-of-mass energy up to $\sqrt{s} \sim 3 \tev$.

In our numerical scenarios we compared the tree-level partial decay
widths with the full one-loop corrected  partial decay widths. 
In the case of decays to scalar quarks we also included for comparison 
the pure SQCD one-loop corrections. 
We concentrated on the analysis of the decay widths themselves, since
the size of the corresponding branching ratios (and thus the size of 
the one-loop effects) is highly parameter dependent.

We found sizable, roughly \order{15\%}, corrections in all the
channels. The corrections tend to be larger for the decays to scalar
quarks w.r.t.\ decays to scalar leptons.
For some parts of the parameter space (not only close to thresholds) 
also larger corrections up to $30\%$~or~$40\%$ (and in exceptional 
cases even higher) have been observed.  Consequently, the full one-loop 
corrections should be taken into account for the interpretation of the 
searches for scalar fermions as well as for any future precision 
analyses of those decays.

The size of the tree-level decay widths and of the corresponding full
one-loop corrections often depend strongly on the respective complex
phase, \ie $\phiAt, \phiAb$ or $\phiAtau$.
The one-loop contributions often vary by a factor of $2-3$ as a 
function of the complex phases and sometimes can even turn negative.
Neglecting the phase dependence could lead to a wrong impression of 
the relative size of the various decay widths.
Furthermore, for certain values of the phases the relevant diagonal
entries in the $\matr{\hat Z}$~matrix go through zero. Consequently, also 
the decay widths go to zero for these values, where the 
$\matr{\hat Z}$~matrix yields the dominating effect on the widths.

In case of decays to scalar quarks we have also compared with 
the pure SQCD result.  We have found that in most cases the EW 
corrections are of similar size.  Neglecting those can lead, depending 
on the parameter space, to a large over- or underestimate of the full 
one-loop corrections.

In the cases where a decay and its complex conjugate final state 
are possible we have evaluated both decay widths independently.  
The asymmetries, as a byproduct of our calculation, turn out to be 
sizable, in particular for decays into a pair of lighter and heavier 
scalar fermions. 

The numerical results we have shown are, of course, dependent on the choice 
of the SUSY parameters. Nevertheless, they give an idea of the relevance
of the full one-loop corrections. 
Decay channels (and their respective one-loop corrections) that may look 
unobservable due to the smallness of their decay width in our numerical
examples could become important if other channels are kinematically
forbidden. 
Following our analysis it is evident that the full one-loop corrections
are mandatory for a precise prediction of the various branching ratios.
We emphasize again that in many cases it is not sufficient to include
only SQCD corrections, as electroweak corrections can be of comparable size.
The full one-loop corrections should be taken into account in any precise 
determination of (SUSY) parameters from the decay of heavy MSSM Higgs bosons.
The results for the heavy MSSM Higgs decays will be implemented into the
Fortran code \FH.

%%%%%%%%%%%%%%%%%%%%%%%%%%%%%%%%%%%%%%%%%%%%%%%%%%%%%%%%%%%%%%%%%%%%%%%%%%%%%%

\subsection*{Acknowledgements}

We thank T.~Hahn, W.~Hollik, H.~Rzehak, and G.~Weiglein for helpful 
discussions.  The work of S.H.\ is supported in part by CICYT 
(grant FPA 2013-40715-P) and by the Spanish MICINN's Consolider-Ingenio 
2010 Program under grant MultiDark CSD2009-00064.

%%%%%%%%%%%%%%%%%%%%%%%%%%%%%%%%%%%%%%%%%%%%%%%%%%%%%%%%%%%%%%%%%%%%%%%%%%%%%%%
%%%%%%%%%%%%%%%%%%%%%%%%%%%%%%%%%%%%%%%%%%%%%%%%%%%%%%%%%%%%%%%%%%%%%%%%%%%%%%%

%\newpage

%%%%%%%%%%%%%%%%%%%%%%%%%%%%%%%%%%%%%%%%%%%%%%%%%%%%%%%%%%%%%%%%%%%%%%%%%%%%%%%
%%%%%%%%%%%%%%%%%%%%%%%%%%%%%%%%%%%%%%%%%%%%%%%%%%%%%%%%%%%%%%%%%%%%%%%%%%%%%%%

\newcommand\jnl[1]{\textit{\frenchspacing #1}}
\newcommand\vol[1]{\textbf{#1}}

\end{document}

%% file: paperdef.tex
\newcommand\Code[1]{\ensuremath{\texttt{#1}}}
\newcommand\Var[1]{\ensuremath{\mathit{#1}}}
\newcommand\Vi{\Var{i}}
\newcommand\Vj{\Var{j}}
\newcommand\Vt{\Var{t}}
\newcommand\Vg{\Var{g}}
\newcommand\Vs{\Var{s}}

\newcommand\tb{\tan\beta}
\newcommand\TB{t_\beta}

\newcommand\CBB{c_{2\beta}}

\newcommand\LP{\left(}
\newcommand\RP{\right)}
\newcommand\LB{\left[}
\newcommand\RB{\right]}

\renewcommand\Re{\mathop{\mathrm{Re}}}

\newcommand\ReTilde{\mathop{\widetilde{\mathrm{Re}}}}
\newcommand\ReDiag{\mathop{%
  \raise .5pt\hbox{[}%
  \widetilde{\mathrm{Re}}%
  \raise .5pt\hbox{]}}}
\newcommand\ReOffDiag{\mathop{%
  \raise .5pt\hbox{$\llbracket$}%
  \widetilde{\mathrm{Re}}%
  \raise .5pt\hbox{$\rrbracket$}}}
\newcommand\SE[1]{\Sigma_{#1}}
\newcommand\OS{\mathrm{OS}}
\newcommand\DRbar{\ensuremath{\smash{\overline{\mathrm{DR}}}}}
\newcommand\MSbar{\ensuremath{\overline{\mathrm{MS}}}}
\newcommand\matr[1]{\mathbf{#1}}
\newcommand\mati[1]{\bigl(#1\bigr)}

\newcommand\SW{s_\mathrm{w}}

\newcommand\MW{M_W}
\newcommand\MZ{M_Z}

\newcommand\MHp{M_{H^\pm}}
\newcommand\mf[1]{m_{f_{#1}}}
\newcommand\mb{m_b}
\newcommand\mt{m_t}
\newcommand\Ab{A_b}
\newcommand\At{A_t}
\newcommand\Atau{A_\tau}

\newcommand\Sf{{\tilde f}}
\newcommand\Sfp{{\tilde f^\prime}}
\newcommand\msf[1]{m_{\Sf_{#1}}}
\newcommand\msfp[1]{m_{\Sfp_{#1}}}
\newcommand\Sn{{\tilde\nu}}
\newcommand\msn[1]{m_{\Sn_{#1}}}
\newcommand\Se{\mathrm{\tilde e}}
\newcommand\Fe{\mathrm{e}}
\newcommand\mfe[1]{m_{\Fe_{#1}}}
\newcommand\mse[1]{m_{\Se_{#1}}}
\newcommand\Su{\mathrm{\tilde u}}
\newcommand\Fu{\mathrm{u}}
\newcommand\mfu[1]{m_{\Fu_{#1}}}
\newcommand\msu[1]{m_{\Su_{#1}}}
\newcommand\Sd{\mathrm{\tilde d}}
\newcommand\Fd{\mathrm{d}}
\newcommand\mfd[1]{m_{\Fd_{#1}}}
\newcommand\msd[1]{m_{\Sd_{#1}}}

\newcommand\Stau[1]{{\tilde\tau_{#1}}}

\newcommand\Stop[1]{{\tilde t_{#1}}}

\newcommand\Sbot[1]{{\tilde b_{#1}}}
\newcommand\sbot{\tilde b}

\newcommand\gl{{\tilde g}}
\newcommand\mgl{m_\gl}
\newcommand\phigl{\varphi_\gl}
\newcommand\dTB{\delta\TB}
\newcommand\ino[1]{\tilde\chi_{#1}}

\newcommand\chapm[1]{\ino{#1}^\pm}

\newcommand\cha{\chapm}
\newcommand\mcha[1]{m_{\chapm{#1}}}
\newcommand\neu[1]{\ino{#1}^0}
\newcommand\mneu[1]{m_{\neu{#1}}}

\newcommand\refeq[1]{Eq.~(\ref{#1})}
\newcommand\refeqs[1]{Eqs.~(\ref{#1})}
\newcommand\refta[1]{Tab.~\ref{#1}}
\newcommand\refse[1]{Sect.~\ref{#1}}

\newcommand\citere[1]{Ref.~\cite{#1}}
\newcommand\citeres[1]{Refs.~\cite{#1}}

\newcommand\uscore{\symbol{95}}
\newcommand\eg{e.g.\ }
\newcommand\ie{i.e.\ }

\newcommand{\CP}{{\cal CP}}
\newcommand{\os}{\mathrm{os}}

\newcommand{\onel}{one-loop}
\newcommand{\tev}{\,\, \mathrm{TeV}}
\newcommand{\gev}{\,\, \mathrm{GeV}}
\newcommand{\mev}{\,\, \mathrm{MeV}}
\newcommand{\Hpm}{H^\pm}
\newcommand\hndecay{h_n \to \Sf_i \Sf_j}
\newcommand\hzstst{h_2 \to \Stop1 \Stop2, \Stop2 \Stop1}
\newcommand\hdstst{h_3 \to \Stop1 \Stop2, \Stop2 \Stop1}
\newcommand\hnstst{h_n \to \Stop1 \Stop2, \Stop2 \Stop1}
\newcommand\hzsbsb{h_2 \to \Sbot1 \Sbot2, \Sbot2 \Sbot1}
\newcommand\hnsbsb{h_n \to \Sbot1 \Sbot2, \Sbot2 \Sbot1}
\newcommand\hnstaustau{h_n \to \Stau1 \Stau2, \Stau2 \Stau1}
\newcommand\Hpdecay{H^+ \to \Sf_i^{} \Sf_j^{\prime\dagger}}
\newcommand\Hmdecay{H^- \to \Sf_i^\dagger \Sfp_j}
\newcommand\Hpmdecay{H^\pm \to \Sf_i^{} \Sfp_j}

\newcommand\FA{{\tt FeynArts}}
\newcommand\FC{{\tt FormCalc}}
\newcommand\LT{{\tt LoopTools}}
\newcommand\FH{{\tt FeynHiggs}}
\newcommand\Sq{{\tilde q}}
\newcommand\msq[1]{m_{\Sq_{#1}}}
\newcommand\Mt{\tilde{M}}

\newcommand\mh[1]{m_{h_{#1}}}

\newcommand\mstop[1]{m_{\tilde{t}_{#1}}}
\newcommand\msbot[1]{m_{\tilde{b}_{#1}}}

\newcommand\mstau[1]{m_{\tilde{\tau}_{#1}}}
\newcommand\mtausneu{m_{\tilde{\nu}_{\tau}}}

\newcommand{\Sce}{S1}
\newcommand{\Scz}{S2}
\newcommand{\Scd}{S3}

\newcommand{\db}{\Delta_b}

\newcommand{\StopL}{\tilde{t}_L}
\newcommand{\StopR}{\tilde{t}_R}
\newcommand{\Snutau}{\tilde{\nu}_\tau}

\def\order#1{\ensuremath{{\cal O}(#1)}}
\def\reffi#1{\mbox{Fig.~\ref{#1}}}
\def\reffis#1{\mbox{Figs.~\ref{#1}}}
\def\als{\alpha_s}
\def\alt{\alpha_t}

\def\Ga{\Gamma}
\def\ga{\gamma}
\def\de{\delta}
\def\la{\lambda}
\def\phia{\varphi_{A}}
\def\phiAt{\varphi_{\At}}
\def\phiAtb{\varphi_{A_{t,b}}}
\def\phiAb{\varphi_{\Ab}}
\def\phiAtau{\varphi_{\Atau}}
\def\phimu{\varphi_{\mu}}
\def\phiMe{\varphi_{M_1}}
\def\phiMz{\varphi_{M_2}}
\def\phigl{\varphi_{\gl}}

\definecolor{Orange}{named}{Orange}
\definecolor{Purple}{named}{Purple}
\definecolor{Lightblue}{cmyk}{0.9,0.1,0.1,0.3}
\definecolor{dgelborange}{cmyk}{0.,0.3,0.5, 0.}
\definecolor{Lila}{rgb}{0.5,0.,1}

%% file: HiggsDecay.bbl
\begin{thebibliography}{99} 

%1
\bibitem{mssm}
H.~Nilles, 
\jnl{Phys. Rept.} \vol{110} (1984) 1; \\ 
%%CITATION = PRPLC,110,1;%%
R.~Barbieri, 
\jnl{Riv. Nuovo Cim.} \vol{11} (1988) 1. 
%%CITATION = RNCIB,11,1;%%

%2
\bibitem{HaK85}
H.~Haber, G.~Kane,
\jnl{Phys. Rept.} \vol{117} (1985) 75.
%%CITATION = PRPLC,117,75;%%

%3
\bibitem{GuH86}
J.~Gunion, H.~Haber,
\jnl{Nucl. Phys.} \vol{B 272} (1986) 1.

%4
\bibitem{ATLASdiscovery} 
G.~Aad et al.\ [ATLAS Collaboration],
\jnl{Phys. Lett.} \vol{B 716} (2012) 1
[arXiv:1207.7214 [hep-ex]].
%%CITATION = ARXIV:1207.7214;%%

%5
\bibitem{CMSdiscovery} 
S.~Chatrchyan et al.\  [CMS Collaboration],
\jnl{Phys. Lett.} \vol{B 716} (2012) 30
[arXiv:1207.7235 [hep-ex]].
%%CITATION = ARXIV:1207.7235;%%

%6
\bibitem{ILC-TDR}
H.~Baer et al.,
{\it The International Linear Collider Technical Design Report - Volume 2:
Physics},
arXiv:1306.6352 [hep-ph].
%%CITATION = ARXIV:1306.6352;%%

%7
\bibitem{teslatdr} 
TESLA Technical Design Report [TESLA Collaboration] Part~3, 
{\it Physics at an $e^+e^-$ Linear Collider},
arXiv:hep-ph/0106315,
%%CITATION = HEP-PH 0106315;%%
see:\\ {\tt tesla.desy.de/new\_pages/TDR\_CD/start.html};\\
K.~Ackermann et al.,
DESY-PROC-2004-01.

%8
\bibitem{ilc}
J.~Brau et al.\  [ILC Collaboration],
{\it ILC Reference Design Report Volume 1 - Executive Summary},
arXiv:0712.1950 [physics.acc-ph];\\
%%CITATION = ARXIV:0712.1950;%%
G.~Aarons et al.\  [ILC Collaboration],
{\it International Linear Collider Reference Design Report Volume 2:
  Physics at the ILC},
arXiv:0709.1893 [hep-ph].
%%CITATION = ARXIV:0709.1893;%%

%9
\bibitem{CLIC} 
L.~Linssen, A.~Miyamoto, M.~Stanitzki and H.~Weerts,
arXiv:1202.5940 [physics.ins-det];\\
%%CITATION = ARXIV:1202.5940;%%
H.~Abramowicz et al. [CLIC Detector and Physics Study Collaboration],
\textit{Physics at the CLIC $e^+e^-$ Linear Collider -- 
Input to the Snowmass process 2013},
arXiv:1307.5288 [hep-ex].
%%CITATION = ARXIV:1307.5288;%%

%10
\bibitem{lhcilc} 
G.~Weiglein et al.\ [LHC/ILC Study Group],
\jnl{Phys. Rept.} \vol{426} (2006) 47
[arXiv:hep-ph/0410364];\\
%%CITATION = PRPLC,426,47;%%
A.~De Roeck et al.,
\jnl{Eur. Phys.\ J.} \vol{C 66} (2010) 525
[arXiv:0909.3240 [hep-ph]];\\
%%CITATION = ARXIV:0909.3240;%%
A.~De Roeck, J.~Ellis, S.~Heinemeyer,
\jnl{CERN Cour.} \vol{49N10} (2009) 27.

%11
\bibitem{hff} 
K.~Williams, H.~Rzehak, and G.~Weiglein,
\jnl{Eur. Phys. J.} \vol{C 71} (2011) 1669
[arXiv:1103.1335 [hep-ph]].
%%CITATION = ARXIV:1103.1335;%%

%12
\bibitem{hff0} 
S.~Heinemeyer, W.~Hollik and G.~Weiglein, 
\jnl{Eur. Phys. J.} \vol{C 16} (2000) 139
[arXiv:hep-ph/0003022].
%%CITATION = HEP-PH 0003022;%%

%13
\bibitem{db2l}
D.~Noth and M.~Spira,
\jnl{Phys. Rev. Lett.} \vol{101} (2008)  181801
[arXiv:0808.0087 [hep-ph]];
%%CITATION = ARXIV:0808.0087;%%
\jnl{JHEP} \vol{1106} (2011) 084
[arXiv:1001.1935 [hep-ph]].
%%CITATION = ARXIV:1001.1935;%%

%14
\bibitem{deltab} R.~Hempfling,
\jnl{Phys. Rev. D} \vol{49} (1994) 6168;\\
%%CITATION = PHRVA,D49,6168;%%
L.~Hall, R.~Rattazzi and U.~Sarid,
\jnl{ Phys. Rev. D} \vol{50} (1994) 7048
[arXiv:hep-ph/9306309];\\
%%CITATION = HEP-PH 9306309;%%
M.~Carena, M.~Olechowski, S.~Pokorski and C.~Wagner,
\jnl{ Nucl. Phys. B} \vol{426} (1994) 269
[arXiv:hep-ph/9402253];\\
%%CITATION = HEP-PH 9402253;%%
M.~Carena, D.~Garcia, U.~Nierste and C.~Wagner,
\jnl{ Nucl. Phys. B} \vol{577} (2000) 577
[arXiv:hep-ph/9912516].
%%CITATION = HEP-PH 9912516;%%

%15
\bibitem{hAA} 
S.~Heinemeyer and W.~Hollik,
\jnl{Nucl. Phys.} \vol{B 474} (1996) 32
[arXiv:hep-ph/9602318].
%%CITATION = HEP-PH 9602318;%%

%16
\bibitem{prophecy4f} 
A.~Bredenstein, A.~Denner, S.~Dittmaier and M.~Weber, 
\jnl{Phys. Rev.} \vol{D 74} (2006) 013004
[arXiv:hep-ph/0604011];
%%CITATION = HEP-PH/0604011;%%
\jnl{JHEP} \vol{0702} (2007) 080
[arXiv:hep-ph/0611234];\\
%%CITATION = HEP-PH/0611234;%%
A.~Bredenstein, A.~Denner, S.~Dittmaier, A.~M\"uck and M.~Weber, \\
see: {\tt omnibus.uni-freiburg.de/~sd565/programs/prophecy4f/prophecy4f.html}.

%17
\bibitem{mhcMSSMlong}
M.~Frank, T.~Hahn, S.~Heinemeyer, W.~Hollik, R.~Rzehak and G.~Weiglein,
\jnl{JHEP} \vol{0702} (2007) 047
[arXiv:hep-ph/0611326].
%%CITATION = HEP-PH 0611326;%%

%18
\bibitem{mhiggslong} 
S.~Heinemeyer, W.~Hollik and G.~Weiglein,
\jnl{Eur. Phys. J.} \vol{C 9} (1999) 343
[arXiv:hep-ph/9812472].
%%CITATION = HEP-PH 9812472;%%

%19
\bibitem{feynhiggs}
S.~Heinemeyer, W.~Hollik and G.~Weiglein,
\jnl{Comput. Phys. Commun.} \vol{124} (2000) 76
[arXiv:hep-ph/9812320];\\
%%CITATION = HEP-PH 9812320;%%
T.~Hahn, S.~Heinemeyer, W.~Hollik, H.~Rzehak and G.~Weiglein,
\jnl{Comput. Phys. Commun.} \vol{180} (2009) 1426, 
%%CITATION = CPHCB,180,1426;%%
see {\tt www.feynhiggs.de} .

%20
\bibitem{mhiggsAEC}
G.~Degrassi, S.~Heinemeyer, W.~Hollik, P.~Slavich and G.~Weiglein, 
\jnl{Eur. Phys. J.} \vol{C 28} (2003) 133
[arXiv:hep-ph/0212020].
%%CITATION = HEP-PH 0212020;%%

%21
\bibitem{Mh-logresum} 
T.~Hahn, S.~Heinemeyer, W.~Hollik, H.~Rzehak and G.~Weiglein, 
\jnl{Phys. Rev. Lett.} \vol{112} (2014) 141801
[arXiv:1312.4937 [hep-ph]].
%%CITATION = ARXIV:1312.4937;%%

%22
\bibitem{hdecay} 
A.~Djouadi, J.~Kalinowsli and M.~Spira, 
\jnl{Comput. Phys. Commun.} \vol{108} (1998) 56
[arXiv:hep-ph/9704448];\\
%%CITATION = HEP-PH/9704448;%%
M.~Spira,
\jnl{Fortschr. Phys.} \vol{46} (1998) 203
[arXiv:hep-ph/9705337].
%%CITATION = HEP-PH/9705337;%%

%23
\bibitem{hdecay2} 
A.~Djouadi, J.~Kalinowski, M.~M\"uhlleitner and M.~Spira,
arXiv:1003.1643 [hep-ph].
%%CITATION = 1003.1643;%%

%24
\bibitem{YR3} 
S.~Heinemeyer et al.
[LHC Higgs Cross Section Working Group],
arXiv:1307.1347 [hep-ph].
%%CITATION = ARXIV:1307.1347;%%

%25
\bibitem{benchmark4} 
M.~Carena, S.~Heinemeyer, O.~St{\aa}l, C.~Wagner and G.~Weiglein, 
\jnl{Eur. Phys. J.} \vol{C 73} (2013) 2552
[arXiv:1302.7033 [hep-ph]].
%%CITATION = ARXIV:1302.7033;%%

%26
\bibitem{Hpstsb_als} 
A.~Bartl, H.~Eberl, K.~Hidaka, T.~Kon, W.~Majerotto and Y.~Yamada, 
\jnl{Phys. Lett.} \vol{B 373} (1996) 117
[arXiv:hep-ph/9508283].
%%CITATION = HEP-PH/9508283;%%

%27
\bibitem{Phisqsq_als}
A.~Bartl, H.~Eberl, K.~Hidaka, T.~Kon, W.~Majerotto and Y.~Yamada, 
\jnl{Phys. Lett.} \vol{B 402} (1997) 303
[arXiv:hep-ph/9701398].
%%CITATION = HEP-PH/9701398;%%

%28
\bibitem{Phisqsq_als_1} 
H.~Eberl, K.~Hidaka, S.~Kraml, W.~Majerotto and Y.~Yamada, 
\jnl{Phys. Rev.} \vol{D 62} (2000) 055006
[arXiv:hep-ph/9912463].
%%CITATION = HEP-PH/9912463;%%

%29
\bibitem{Asqsq_1L} 
C.~Weber, H.~Eberl and W.~Majerotto,
\jnl{Phys. Lett.} \vol{B 572} (2003) 56
[arXiv:hep-ph/0305250].
%%CITATION = HEP-PH/0305250;%%

%30
\bibitem{Asfsf_1L} 
C.~Weber, H.~Eberl and W.~Majerotto,
\jnl{Phys. Rev.} \vol{D 68} (2003) 093011
[arXiv:hep-ph/0308146].
%%CITATION = HEP-PH/0308146;%%

%31
\bibitem{Phisqsq_1L} 
C.~Weber, K.~Kovarik, H.~Eberl and W.~Majerotto,
\jnl{Nucl. Phys.} \vol{B 776} (2007) 138
[arXiv:hep-ph/0701134].
%%CITATION = HEP-PH/0701134;%%

%32
\bibitem{SbotRen} 
S.~Heinemeyer, H.~Rzehak and C.~Schappacher,
\jnl{Phys. Rev.} \vol{D 82} (2010) 075010
[arXiv:1007.0689 [hep-ph]];
%%CITATION = PHRVA,D82,075010;%%
\jnl{PoSCHARGED} \vol{2010} (2010) 039
[arXiv:1012.4572 [hep-ph]].
%%CITATION = POSCI,CHARGED2010,039;%%

%33
\bibitem{Stop2decay}
T.~Fritzsche, S.~Heinemeyer, H.~Rzehak, C.~Schappacher, 
\jnl{Phys. Rev.} \vol{D 86} (2012) 035014
[arXiv:1111.7289 [hep-ph]].
%%CITATION = ARXIV:1111.7289;%%

%34
\bibitem{Stau2decay} 
S.~Heinemeyer, C.~Schappacher,
\jnl{Eur. Phys. J.} \vol{C 72} (2012) 2136
[arXiv:1204.4001 [hep-ph]].
%%CITATION = ARXIV:1204.4001;%%

%35
\bibitem{Phisqsq_als_2} 
A.~Arhrib, A.~Djouadi, W.~Hollik and C.~J\"unger,
\jnl{Phys. Rev.} \vol{D 57} (1998) 5860
[arXiv:hep-ph/9702426].
%%CITATION = HEP-PH/9702426;%%

%36
\bibitem{Phisqsq_als_3}
E.~Accomando, G.~Chachamis, F.~Fugel, M.~Spira and M.~Walser, 
\jnl{Phys. Rev.} \vol{D 85} (2012) 015004
[arXiv:1103.4283 [hep-ph]].
%%CITATION = ARXIV:1103.4283;%%

%37	
\bibitem{MSSMCT} 
T.~Fritzsche, T.~Hahn, S.~Heinemeyer, F.~von~der~Pahlen, H.~Rzehak and
C.~Schappacher   
\jnl{Comput. Phys. Commun.} \vol{185} (2014) 1529
[arXiv:1309.1692 [hep-ph]].

%38
\bibitem{mhcMSSM2L}
S.~Heinemeyer, W.~Hollik, H.~Rzehak and G.~Weiglein,
\jnl{Phys. Lett.} \vol{B 652} (2007) 300
[arXiv:0705.0746 [hep-ph]].
%%CITATION = PHLTA,B652,300;%%

%39
\bibitem{dissHR} 
H.~Rzehak, PhD thesis:
``Two-loop contributions in the supersymmetric Higgs sector'', 
Technische Universit\"at M\"unchen, 2005; 
see: {\tt nbn-resolving.de/} \\
with {\tt urn}: {\tt nbn:de:bvb:91-diss20050923-0853568146}\,.

%40
\bibitem{squark_q_V_als}
A.~Bartl, H.~Eberl, K.~Hidaka, S.~Kraml, W.~Majerotto, W.~Porod and Y.~Yamada, 
\jnl{Phys. Lett.} \vol{B 419} (1998) 243
[arXiv:hep-ph/9710286].
%%CITATION = PHLTA,B419,243;%%

%41
\bibitem{stopsbot_phi_als}
A.~Bartl, H.~Eberl, K.~Hidaka, S.~Kraml, W.~Majerotto, W.~Porod and Y.~Yamada, 
\jnl{Phys. Rev.} \vol{D 59} (1999) 115007
[arXiv:hep-ph/9806299].
%%CITATION = PHRVA,D59,115007;%%

%42
\bibitem{dr2lA} 
A.~Djouadi, P.~Gambino, S.~Heinemeyer, W.~Hollik, C.~J\"unger and G.~Weiglein,
\jnl{Phys. Rev. Lett.} \vol{78} (1997) 3626
[arXiv:hep-ph/9612363];
%%CITATION = HEP-PH 9612363;%%
\jnl{Phys. Rev.} \vol{D 57} (1998) 4179
[arXiv:hep-ph/9710438].
%%CITATION = HEP-PH 9710438;%%

%43
\bibitem{hr}
W. Hollik and H. Rzehak, 
\jnl{Eur. Phys. J.} \vol{C 32} (2003) 127
[arXiv:hep-ph/0305328].
%%CITATION = HEP-PH 0305328;%%

%44
\bibitem{mhiggsFDalbals} 
S.~Heinemeyer, W.~Hollik, H.~Rzehak and G.~Weiglein,
\jnl{Eur. Phys. J.} \vol{C 39} (2005) 465
[arXiv:hep-ph/0411114].
%%CITATION = HEP-PH 0411114;%%

%45
\bibitem{pdg} 
J.~Beringer et al. [Particle Data Group],
\jnl{Phys. Rev.} \vol{D 86} (2012) 010001
and 2013 partial update for the 2014 edition.

%46
\bibitem{RunDec}
K.~Chetyrkin, J.~K\"uhn and M.~Steinhauser, 
\jnl{Comput. Phys. Commun.} \vol{133} (2000) 43
[arXiv:hep-ph/0004189];\\
%%CITATION = CPHCB,133,43;%%
B.~Schmidt, M.~Steinhauser,
\jnl{Comput. Phys. Commun.} \vol{183} (2012) 1845
[arXiv:1201.6149 [hep-ph]].

%47
\bibitem{deltab2}
M.~Carena, D.~Garcia, U.~Nierste and C.~Wagner,
\jnl{Nucl. Phys.} \vol{B 577} (2000) 577
[arXiv:hep-ph/9912516].
%%CITATION = HEP-PH 9912516;%%

%48
\bibitem{deltab1}
R.~Hempfling,
\jnl{Phys. Rev.} \vol{D 49} (1994) 6168;\\
%%CITATION = PHRVA,D49,6168;%%
L.~Hall, R.~Rattazzi and U.~Sarid,
\jnl{Phys. Rev.} \vol{D 50} (1994) 7048
[arXiv:hep-ph/9306309];\\
%%CITATION = HEP-PH 9306309;%%
M.~Carena, M.~Olechowski, S.~Pokorski and C.~Wagner,
\jnl{Nucl. Phys.} \vol{B 426} (1994) 269
[arXiv:hep-ph/9402253].
%%CITATION = HEP-PH 9402253;%%

%49
\bibitem{deltabc} 
M.~Carena, J.~Ellis, A.~Pilaftsis and C.~Wagner,
\jnl{Nucl. Phys.} \vol{B 586} (2000) 92
[arXiv:hep-ph/0003180].
%%CITATION = HEP-PH/0003180;%%

%50
\bibitem{alsDRbar} 
R.~Harlander, L.~Mihaila and M.~Steinhauser,
\jnl{Phys. Rev.} \vol{D 72} (2005) 095009
[arXiv:hep-ph/0509048];
%%CITATION = PHRVA,D72,095009;%%
\jnl{Phys. Rev.} \vol{D 76} (2007) 055002
[arXiv:0706.2953 [hep-ph]].
%%CITATION = PHRVA,D72,095009;%%                   

%51
\bibitem{complexmassscheme}
A.~Denner, S.~Dittmaier, M.~Roth and D.~Wackeroth,
\jnl{Nucl. Phys. B} \vol{560} (1999) 33
[hep-ph/9904472].
%%CITATION = HEP-PH/9904472;%%

%52
\bibitem{feynarts}
J.~K\"ublbeck, M.~B\"ohm and A.~Denner, 
\jnl{Comput. Phys. Commun.} \vol{60} (1990) 165;\\
%%CITATION = CPHCB,60,165;%%
T.~Hahn, 
\jnl{Comput. Phys. Commun.} \vol{140} (2001) 418
[arXiv:hep-ph/0012260];\\
%%CITATION = HEP-PH 0012260;%%
T.~Hahn and C.~Schappacher, 
\jnl{Comput. Phys. Commun.} \vol{143} (2002) 54
[arXiv:hep-ph/0105349].\\
%%CITATION = HEP-PH 0105349;%%
The program, the user's guide and the MSSM model files
are available via\\ {\tt www.feynarts.de} .

%53
\bibitem{formcalc}
T.~Hahn and M.~P\'erez-Victoria,
\jnl{Comput. Phys. Commun.} \vol{118} (1999) 153
[arXiv:hep-ph/9807565].
%%CITATION = HEP-PH 9807565;%%

%54
\bibitem{cdr}
F.~del Aguila, A.~Culatti, R.~Mu\~noz Tapia and 
M.~P\'erez-Victoria,
\jnl{Nucl. Phys.} \vol{B 537} (1999) 561
[arXiv:hep-ph/9806451].
%%CITATION = HEP-PH 9806451;%%

%55
\bibitem{dred}
W.~Siegel, 
\jnl{Phys. Lett.} \vol{B 84} (1979) 193; \\
D.~Capper, D.~Jones, and P.~van Nieuwenhuizen,
\jnl{Nucl. Phys.} \vol{B 167} (1980) 479. 

%56
\bibitem{dredDS}
D.~St\"ockinger,
\jnl{JHEP} \vol{0503} (2005) 076
[arXiv:hep-ph/0503129].
%%CITATION = HEP-PH 0503129;%%

%57
\bibitem{dredDS2}
W.~Hollik and D.~St\"ockinger,
\jnl{Phys. Lett.} \vol{B 634} (2006) 63
[arXiv:hep-ph/0509298].
%%CITATION = HEP-PH 0509298;%%

%58
\bibitem{denner}
A.~Denner,
\jnl{Fortsch. Phys.} \vol{41} (1993) 307
[arXiv:0709.1075 [hep-ph]].
%%CITATION = FPYKA,41,307;%%

%59
\bibitem{feynarts-mf}
The couplings can be found in the files
{\tt MSSM.ps.gz}, {\tt MSSMQCD.ps.gz} and 
{\tt HMix.ps.gz} as part of the
\FA\ package~\cite{feynarts}.

%60
\bibitem{abdesslam} 
A.~Arhrib,
{\em private communication}, 08.06.2014.

%61
\bibitem{ccb}
J.~Frere, D.~Jones and S.~Raby,
\jnl{Nucl. Phys.} \vol{B 222} (1983) 11;\\
%%CITATION = NUPHA,B222,11;%%
M.~Claudson, L.~Hall and I.~Hinchliffe,
\jnl{Nucl. Phys.} \vol{B 228} (1983) 501;\\
%%CITATION = NUPHA,B228,501;%%
C.~Kounnas, A.~Lahanas, D.~Nanopoulos and M.~Quiros,
\jnl{Nucl. Phys.} \vol{B 236} (1984) 438;\\
%%CITATION = NUPHA,B236,438;%%
J.~Gunion, H.~Haber and M.~Sher,
\jnl{Nucl. Phys.} \vol{B 306} (1988) 1;\\
%%CITATION = NUPHA,B306,1;%%
J.~Casas, A.~Lleyda and C.~Munoz,
\jnl{Nucl. Phys.} \vol{B 471} (1996) 3
[arXiv:hep-ph/9507294];\\
%%CITATION = NUPHA,B471,3;%%
P.~Langacker and N.~Polonsky,
\jnl{Phys. Rev.} \vol{D 50} (1994) 2199
[arXiv:hep-ph/9403306];\\
%%CITATION = PHRVA,D50,2199;%%
A.~Strumia,
\jnl{Nucl. Phys.} \vol{B 482} (1996) 24
[arXiv:hep-ph/9604417].
%%CITATION = NUPHA,B482,24;%%

%62
\bibitem{MSSMcomplphasen}
S.~Dimopoulos and S.~Thomas,
\jnl{Nucl. Phys.} \vol{B 465} (1996) 23
[arXiv:hep-ph/9510220].
%%CITATION = HEP-PH 9510220;%%

%63
\bibitem{SUSYphases}
M.~Dugan, B.~Grinstein and L.~Hall,
\jnl{Nucl. Phys.} \vol{B 255} (1985) 413.
%%CITATION = NUPHA,B255,413;%%

%64
\bibitem{EDMrev2}
D.~Demir, O.~Lebedev, K.~Olive, M.~Pospelov and A.~Ritz,
\jnl{Nucl. Phys.} \vol{B 680} (2004) 339
[arXiv:hep-ph/0311314].
%%CITATION = HEP-PH 0311314;%%

%65
\bibitem{EDMPilaftsis}
D.~Chang, W.~Keung and A.~Pilaftsis,
\jnl{Phys. Rev. Lett.} \vol{82} (1999) 900
[Erratum-ibid.\  \vol{83} (1999) 3972]
[arXiv:hep-ph/9811202];\\
%%CITATION = HEP-PH 9811202;%%
A.~Pilaftsis,
\jnl{Phys. Lett.} \vol{B 471} (1999) 174
[arXiv:hep-ph/9909485].
%%CITATION = HEP-PH 9909485;%%

%66
\bibitem{EDMRitz}
O.~Lebedev, K.~Olive, M.~Pospelov and A.~Ritz,
\jnl{Phys. Rev.} \vol{D 70} (2004) 016003
[arXiv:hep-ph/0402023].
%%CITATION = HEP-PH 0402023;%%

%67
\bibitem{EDMDoink}
W.~Hollik, J.~Illana, S.~Rigolin and D.~St\"ockinger,
\jnl{Phys. Lett.} \vol{B 416} (1998) 345
[arXiv:hep-ph/9707437];
%%CITATION = HEP-PH 9707437;%%
\jnl{Phys. Lett.} \vol{B 425} (1998) 322
[arXiv:hep-ph/9711322].
%%CITATION = HEP-PH 9711322;%%

%68
\bibitem{EDMheavy}
P.~Nath,
\jnl{Phys. Rev. Lett.} \vol{66} (1991) 2565;\\
%%CITATION = PRLTA,66,2565;%%
Y.~Kizukuri and N.~Oshimo,
\jnl{Phys. Rev.} \vol{D 46} (1992) 3025.
%%CITATION = PHRVA,D46,3025;%%

%69
\bibitem{EDMmiracle}
T.~Ibrahim and P.~Nath,
\jnl{Phys. Lett.} \vol{B 418} (1998) 98
[arXiv:hep-ph/9707409];
%%CITATION = HEP-PH 9707409;%%
\jnl{Phys. Rev.} \vol{D 57} (1998) 478 
[Erratum-ibid.\ \vol{D 58} (1998) 019901] 
[Erratum-ibid.\ \vol{D 60} (1998) 079903] 
[Erratum-ibid.\ \vol{D 60} (1999) 119901]
[arXiv:hep-ph/9708456];\\
%%CITATION = HEP-PH 9708456;%%
M.~Brhlik, G.~Good and G.~Kane,
\jnl{Phys. Rev.} \vol{D 59} (1999) 115004
[arXiv:hep-ph/9810457].
%%CITATION = HEP-PH 9810457;%%

%70
\bibitem{EDMrev1}
S.~Abel, S.~Khalil and O.~Lebedev,
\jnl{Nucl. Phys.} \vol{B 606} (2001) 151
[arXiv:hep-ph/0103320].
%%CITATION = HEP-PH 0103320;%%

%71
\bibitem{EDMrev3}
Y.~Li, S.~Profumo and M.~Ramsey-Musolf,
\jnl{JHEP} \vol{1008} (2010) 062
[arXiv:1006.1440 [hep-ph]].
%%CITATION = JHEPA,1008,062;%%

%72
\bibitem{plehnix}
V.~Barger, T.~Falk, T.~Han, J.~Jiang, T.~Li and T.~Plehn,
\jnl{Phys. Rev.} \vol{D 64} (2001) 056007
[arXiv:hep-ph/0101106].
%%CITATION = HEP-PH 0101106;%%

%73
\bibitem{stopstophiggs-LHC} 
H.~Heath, C.~Lynch, S.~Moretti and C.~Shepherd-Themistocleous,
arXiv:0901.1676 [hep-ph].
%%CITATION = ARXIV:0901.1676;%%

%74
\bibitem{Higgsincascades} 
A.~Datta, A.~Djouadi, M.~Guchait and F.~Moortgat,
{\em Nucl.\ Phys.} {\bf B 681} (2004) 31
[arXiv:hep-ph/0303095].


\end{thebibliography}
